\documentclass[]{jfm}

\usepackage{graphicx}
\usepackage{psfrag}
\usepackage{tikz}
\usepackage{amssymb}
\usepackage{siunitx}
\usepackage{newtxtext}
\usepackage{newtxmath}
\usepackage{natbib}
\usepackage{hyperref}
\usepackage{comment}
\usepackage{pstool}
\usepackage{cleveref}
\usepackage{amsfonts}
\usepackage{setspace}
\usepackage{color}
\usepackage{floatrow}
\usepackage{subfig}
\usepackage{scalerel}
\usepackage{multirow} 
\usepackage{booktabs}
\usepackage[dvipsnames]{xcolor}
\hypersetup{
    colorlinks = true,
    urlcolor   = blue,
    citecolor  = black,
}
\definecolor{magenta2}{RGB}{255,0,255}
\crefformat{section}{\S#2#1#3} 
\crefformat{subsection}{\S#2#1#3}
\crefformat{subsubsection}{\S#2#1#3}
\tikzstyle{longdashed}=                  [dash pattern=on 6pt off 2pt]
\tikzstyle{dashdotdot}=              [dash pattern=on 4pt off 2pt on \the\pgflinewidth off 1pt on \the\pgflinewidth off 2pt]
\tikzstyle{dot}=              [dash pattern=on 1pt off 1pt on 1pt off 1pt on 1pt off 1pt]

\newcommand{\RomanNumeralCaps}[1]

\title{Effect of subgrid-scale anisotropy on wall-modeled large-eddy simulation of turbulent flow with smooth-body separation}

\author {Di Zhou \aff{1,2}\corresp{\email{dzhou6@utk.edu}} \and  H. Jane Bae \aff{1}}

\affiliation{\aff{1} Lynn Booth and Kent Kresa Department of Aerospace, California Institute of Technology, Pasadena, CA 91125, USA
\aff{2} Department of Mechanical and Aerospace Engineering, University of Tennessee, Knoxville, TN 37996, USA
}

\begin{document}
\maketitle

\begin{abstract}
We examine the role of anisotropic subgrid-scale (SGS) stress in wall-modeled large-eddy simulation (WMLES) of flow over a spanwise-uniform Gaussian-shaped bump, with emphasis on predicting flow separation. The simulations show that eddy-viscosity-based SGS models often yield non-monotonic predictions of the mean separation bubble size on the leeward side under grid refinement, whereas models incorporating anisotropic SGS stress produce more consistent results. To identify where SGS anisotropy is most critical, we introduce anisotropic SGS stress in selected regions of the domain. The results reveal that the windward side, where a strong favorable pressure gradient (FPG) occurs, is crucial in determining downstream separation. Analysis of the Reynolds stress transport equation shows that fluctuations of anisotropic SGS stress modify SGS dissipation and diffusion in this region, thereby altering the Reynolds stress and the onset of separation. Examination of the mean streamwise momentum equation indicates that at coarse resolutions, the mean SGS shear stress dominates, and the differences between the eddy-viscosity-based and anisotropic models remain minor. With grid refinement, resolved Reynolds stresses increasingly govern the near-wall momentum transport, and the influence of SGS stress fluctuations grows as they determine the SGS dissipation and diffusion of Reynolds stresses. Component-wise analysis of the SGS stress tensor further shows that the improvement arises mainly from including significant normal stress contributions. An \emph{a priori} study using filtered direct numerical simulation of turbulent Couette-Poiseuille flow confirms that wall-bounded turbulence under FPG is highly anisotropic and that anisotropic SGS models provide a more realistic SGS stress representation than eddy-viscosity-based models.
\end{abstract}

{\bf MSC Codes }  76F65, 76F40 

\section{Introduction}
\label{sec:intro}

{Complex turbulent flows with separation are common in aerodynamic and hydrodynamic applications and strongly affect performance and stall behaviour. Accurate prediction of such flows is therefore essential. Wall-modeled large-eddy simulation (WMLES) has emerged as a promising approach {(see reviews in \citep{larsson2016large, bose2018wall})}, as it resolves the energy-containing turbulent motions away from the wall while modeling near-wall effects. Compared with wall-resolved LES (WRLES), WMLES significantly reduces grid requirements and computational cost, often by one to two orders of magnitude in practical configurations \citep{choi2012grid, yang2021grid}. Owing to these advantages, WMLES is considered a viable tool for high-fidelity simulations of realistic flows, and has been successfully applied to complex engineering problems \citep{park2017wall, lehmkuhl2018large, goc2021large, goc2024wind}.}

{Despite this progress, achieving robust and accurate WMLES predictions across a wide range of flow conditions remains challenging. Many wall models rely on equilibrium assumptions that are not valid in separated or strongly non-equilibrium flows. While improved formulations, such as models based on thin boundary-layer equations \citep{wang2002dynamic,kawai2013dynamic,park2014improved} or dynamic slip approaches \citep{bose2014dynamic, bae2019dynamic}, have been developed, recent studies \citep{lozano2019error, rezaeiravesh2019systematic, zhou2024sensitivity} show that limitations in subgrid-scale (SGS) modeling play an equally important, if not dominant, role in WMLES performance. In particular, the sensitivity of separated-flow predictions to SGS models has been shown to be substantial \citep{zhou2023large, zhou2024sensitivity}.}

{In LES, the effects of unresolved small-scale motions are represented by SGS models, which are often based on the assumption of isotropic turbulence. This assumption underlies classical eddy-viscosity models, originally developed for well-resolved simulations. Although these models provide reasonable energy dissipation, they show weak correlation with the exact SGS stress from direct numerical simulation (DNS) and may fail to capture important statistical properties of turbulent flows \citep{clark1979evaluation, kerr1996small, domaradzki1997subgrid, moser2021statistical}. Their limitations become more pronounced in WMLES, where coarse grids require SGS models to represent more than energy dissipation alone. As a result, eddy-viscosity models often exhibit poor or non-monotonic convergence in separation bubble prediction and require sufficiently fine resolution to remain reliable \citep{whitmore2021large, agrawal2022non, zhou2024sensitivity}. This reliance on fine resolution limits their practical applicability and underscores the need for SGS models suited for WMLES that can accurately predict complex turbulent flows.}

{At coarse resolutions typical of WMLES, SGS models must represent not only energy dissipation but also the contribution of subgrid motions to mean momentum and energy transport. In this regime, the effects of SGS anisotropy become important, particularly at the smallest resolved scales that remain dynamically active. However, classical eddy-viscosity models, with their single degree of freedom, cannot simultaneously represent both SGS stress and dissipation. These limitations highlight the need for SGS models that account for anisotropy and can capture both effects at coarse resolutions. To this end, various anisotropic and nonlinear SGS models have been proposed, including mixed similarity models \citep{zang1993dynamic, liu1994properties, vreman1994formulation, meneveau2000scale, iyer2024efficient}, algebraic models \citep{gatski2000nonlinear, marstorp2009explicit, rasam2017improving}, and other nonlinear formulations \citep{abe2014investigation, vollant2016dynamic, agrawal2022non, uzun2025application}. While these models show improved performance in canonical turbulent flows such as channel flows and turbulent boundary layers (TBLs), their behaviour in complex flow configurations remains insufficiently understood, and the role of SGS anisotropy is not yet fully characterized.}

{Recent \emph{a priori} analyses based on filtered DNS data indicate that SGS anisotropy strongly influences turbulence dynamics, including the evolution of Reynolds stress, vorticity, and enstrophy \citep{horiuti2003roles, abe2019notable, cimarelli2019resolved, inagaki2023analysis}. However, inconsistencies between \emph{a priori} and \emph{a posteriori} evaluations have been widely reported \citep{vreman1997large, park2005toward, duraisamy2021perspectives, choi2025perspective}. In particular, models that perform poorly in \emph{a priori} tests may yield good \emph{a posteriori} results, and vice versa, highlighting a key limitation of \emph{a priori} analysis. To address this gap, the present study conducts a systematic \emph{a posteriori} investigation of anisotropic SGS stress effects in WMLES of separated turbulent flows.}

{In particular, this \emph{a posteriori} investigation focuses on the flow over a Gaussian-shaped bump at a relatively high Reynolds number (see figure~\ref{set_up}) \citep{slotnick2019integrated}. This configuration mimics smooth junctions between an aircraft wing and fuselage, where smooth-body separation of a TBL occurs under combined pressure gradients and surface curvature. As a canonical benchmark, it has been extensively studied experimentally \citep{williams2020experimental, gray2022benchmark, gluzman2022simplified} and computationally \citep{balin2021direct,uzun2022high,arranz2023wall,zhou2024wall,agrawal2024reynolds, iyer2025insights}, with consistent emphasis on the challenge of accurately predicting the extent and location of leeward-side separation. Recent WMLES studies have demonstrated that separation prediction is highly sensitive to the SGS model, and that anisotropic SGS models markedly improve mean velocity field predictions \citep{iyer2022wall, agrawal2022non, zhou2024sensitivity, iyer2024efficient, uzun2025application}, making this an ideal test case for the present analysis. To avoid complexities from spanwise variations, this study focuses on a spanwise-uniform bump with periodic spanwise boundary conditions, following the hybrid DNS-WRLES study of \citet{uzun2022high}, which also provides high-fidelity reference data.}

{Through this study, we aim to improve understanding of how anisotropic SGS models influence the statistics and dynamics of separated flow in WMLES, characterize the properties of anisotropic SGS stress, and identify key modeling features required for accurate and robust predictions in complex separated flows — ultimately providing guidance for future SGS model development.}

The remainder of the paper is organized as follows. {\Cref{sec:Computation_method} describes the numerical approach, flow configuration, and simulation setup. In \Cref{sec:flow_stats}, the sensitivity of mean flow separation to SGS models and mesh resolution is examined, followed by a numerical experiment in which anisotropic SGS stress is introduced in different regions of the computational domain to identify where SGS anisotropy is most critical. \Cref{sec:budget} explains the mechanisms underlying the sensitivity of separation prediction through budgets of the mean streamwise momentum and Reynolds stress transport equations. In \Cref{sec:SGS_properties}, the SGS stress properties are investigated and compared with \emph{a priori} results from filtered DNS of a turbulent Couette--Poiseuille flow.} Finally, \Cref{sec:conclusion} summarizes the key findings and provides insights for further improvement of WMLES.

\section{Computational methodology}
\label{sec:Computation_method}
\subsection{Numerical approach}
\label{sec:numerical_method}

Flow simulations are conducted employing a finite-volume, unstructured-mesh LES code \citep{you2008discrete}, which has been validated in various turbulent flow configurations \citep{yang2013boundary, zhou2020large, zhou2024rotor}. The spatially-filtered incompressible Navier-Stokes equations are solved with second-order accuracy using cell-based, low-dissipative, and energy-conservative spatial discretization and a fully implicit, fractional-step time-advancement method with the Crank--Nicolson scheme. The Poisson equation for pressure is solved using the bi-conjugate gradient stabilized method \citep{van1992bi}. The governing equations are
\begin{equation}\label{continuity_EQ}
   \frac{\partial \tilde{u}_i}{\partial x_i} = 0\hspace{3pt},
\end{equation}
and
\begin{equation}\label{momentum_EQ}
\frac{\partial \tilde{u}_i }{\partial  t} +
\tilde{u}_j  \frac{\partial \tilde{u}_i}{\partial x_j}  = -\frac{1}{\rho}\frac{\partial \tilde{ p}}{\partial x_i} +\nu \frac{\partial^2 \tilde{u}_i}{\partial x_j \partial x_j}  - \frac{\partial}{\partial x_j} {\tau}^\text{sgs}_{ij}\hspace{3pt},
\end{equation}
where $u_i$ is the instantaneous flow velocity, $p$ is the instantaneous static pressure, $\rho$ is fluid density, $\nu$ is fluid kinematic viscosity and $\widetilde{(\cdot)}$ denotes grid-filtering operation. The SGS stress tensor is $\tau^\text{sgs}_{ij} = \widetilde{u_i u_j}-\tilde{ u}_i \hspace{1pt}\tilde{ u}_j$, whose deviatoric part is modeled using an SGS model for closure of the equations, and the isotropic component of the SGS stress is absorbed into pressure. In what follows, the tilde is omitted and $u_i$, $p$ denote resolved-field quantities directly for brevity.

Two SGS models are investigated. The first is the classical Smagorinsky model (SM) \citep{smagorinsky1963general},
\begin{equation}\label{SM_EQ}
{\tau}^\text{sgs}_{ij}= {\tau}^\text{SM}_{ij} = -2\nu_tS_{ij}= -2(C_s\Delta)^2\lvert S\rvert S_{ij}\hspace{3pt},
\end{equation} 
where {$\nu_t$ is the eddy viscosity}, $S_{ij}$ is the strain-rate tensor, $\lvert S\rvert=(2S_{ij}S_{ij})^{1/2}$, $\Delta$ is the grid filter width, and $C_s=0.16$ is calibrated for homogeneous isotropic turbulence. The second is a modified SM (MSM), which augments the isotropic SM with an anisotropic SGS stress term,
\begin{equation}\label{aniso_EQ}
{\tau}^\text{sgs}_{ij}= {\tau}^\text{SM}_{ij} + \mathbf{\tau}^\text{ani}_{ij}\hspace{3pt},
\end{equation}
where ${\tau}^\text{ani}_{ij}=C_a\Delta^2(S_{ik}R_{kj}-R_{ik}S_{kj})$ and $R_{ij}$ is the rotation-rate tensor. This term is one of the six independent terms arising from the expansion of SGS stress in strain-rate and rotation-rate tensors \citep{lund1992parameterization, gatski2000nonlinear}, and is explicitly incorporated in several recently developed anisotropic SGS models that have shown promising results for flow over a Gaussian bump \citep{agrawal2022non, uzun2025application}. Importantly, this term contributes no SGS kinetic energy dissipation, i.e.\ ${\tau}^\text{ani}_{ij}S_{ij}=0$ \citep{lund1992parameterization, silvis2019nonlinear, inagaki2023analysis}, allowing the physical role of anisotropic SGS stress to be studied independently of energy transfer. To minimize differences in kinetic energy dissipation between the two models, $C_s=0.16$ is retained in the MSM. For $C_a$, tests over the range $-1/30$ to $-1/6$ (consistent with values used in related anisotropic models \citep{bardina1983improved, kosovic1997subgrid, sarghini1999scale, wang2005dynamic, marstorp2009explicit, silvis2019nonlinear}) showed no significant change in the predicted separation bubble; $C_a=-1/30$ is therefore adopted as the most conservative choice. No near-wall damping is applied to the eddy viscosity in either model, consistent with the coarse-mesh WMLES focus of this study. 

\subsection {Flow configuration and simulation set-up}
\label{sec:setup}

\begin{figure}
\centering
\includegraphics[width=.86\textwidth,trim={0cm 0.1cm 0.0cm 0.1cm},clip]{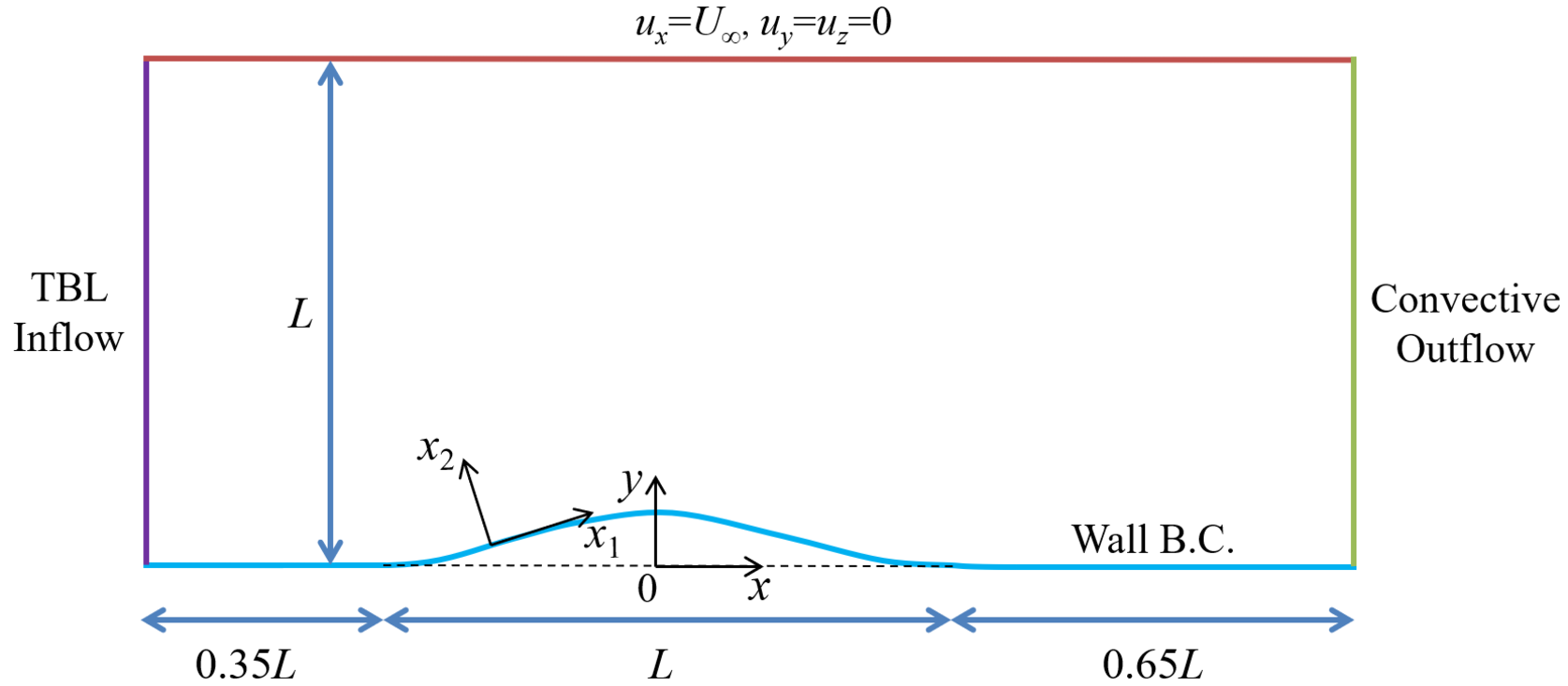}
\caption{Simulation set-up for flow over a {Gaussian}-shaped bump.}
\label{set_up}
\end{figure}

The physical conditions follow the hybrid DNS-WRLES of \citet{uzun2022high}. The bump geometry is $y=f_b(x)=h\exp{[-(x/x_0)^2]}$, with maximum height $h=0.085L$, $x_0=0.195L$, and $L$ the bump width. The Reynolds number is $Re_L=U_\infty L/\nu=2\times10^6$, where $L$ is the bump width and $U_\infty$ is the free-stream velocity, matching the reference DNS \citep{uzun2022high}. The flow configuration and boundary conditions are shown schematically in figure~\ref{set_up}.

Simulations are conducted in a rectangular domain of length $2L$, height $L$, and spanwise depth $0.08L$, matching the DNS dimensions in the vertical and spanwise directions \citep{uzun2022high}. A Cartesian coordinate system ($x$-$y$-$z$) with velocity components {($u_x$, $u_y$, $u_z$)} serves as the global reference, with origin at the base of the bump peak, located $0.85L$ downstream from the inlet. A surface-fitted local coordinate system ($x_1$-$x_2$-$x_3$) is also used with velocity components ($u_1$, $u_2$, $u_3$), where $x_1$ is tangential to the bump surface in the flow direction, $x_2$ is surface-normal, and $x_3$ is spanwise.

Boundary conditions consist of a TBL inflow at the inlet, free-stream condition on the top boundary, convective outflow at the exit, and periodic conditions in the spanwise direction. The TBL inflow is generated by a separate flat-plate LES using the rescale-and-recycle method of \citet{lund1998generation}. We note that the inflow-generation method used in the study of \citet{uzun2022high} is different from the current method. The friction Reynolds number is $Re_\tau \approx 620$ and the TBL thickness is $\delta_{\text{in}}/L=0.0061$, which is approximately 10\% larger than in the DNS. The momentum thickness Reynolds number of the inflow is $Re_{\theta} \approx 1074$, compared to their slightly smaller value of approximately 1035. Further details are given in \citet{zhou2024sensitivity}. To isolate SGS model effects from wall-model interactions, the no-slip condition is replaced by an idealized Neumann boundary condition prescribing the mean wall-shear stress from DNS,
\begin{equation}
\left.\left(\frac{\partial u_1}{\partial x_2}\right)\right|_w=\frac{\tau_{w,1}^{\text{DNS}}}{\rho\nu} \hspace{3pt},
\label{Neumann}
\end{equation}
where $\tau_{w,1}^{\text{DNS}}$ is the mean wall-shear stress taken directly from \citet{uzun2022high}, with a no-penetration condition enforced for $u_2$. {Since the spanwise mean wall-shear stress is zero in the DNS, the corresponding boundary condition for $\left({\partial u_3}/{\partial x_2}\right)$ at the wall is set to zero.}

The computational mesh combines structured-mesh blocks near the bottom surface with unstructured-mesh blocks in the outer region, using isotropic cells to avoid resolution-induced anisotropy \citep{haering2019resolution}. Mesh parameters are listed in table~\ref{tab:table1}. Based on the TBL thickness at $x/L=-0.65$ from the DNS \citep{uzun2022high}, the TBL is resolved by approximately 5, 9, 18, and 36 cells {for the coarsest, coarse, medium, and fine meshes}, respectively. The fine mesh resolution ranges from 10 to 30 wall units in attached-flow regions, comparable to standard WRLES in the streamwise and spanwise directions, but an order of magnitude coarser in the wall-normal direction near the wall.

\begin{table}
  \begin{center}
\def~{\hphantom{0}}
      \noindent\rule{\textwidth}{0.4pt}
\small
  \begin{tabular}{lccc}
      \\ Mesh label & $\Delta_{\text{c}}/L$ & Cell number \\[3pt]
      Coarsest mesh & $1.90\times 10^{-3}$ & $1050 \times 44 \times 42 \approx 1.94$~million \\
      Coarse mesh & $9.52\times 10^{-4}$ & $2100 \times 88 \times 84 \approx 15.5$~million \\
      Medium mesh&$4.76\times 10^{-4}$ & $4200 \times 176 \times 168 \approx 124$~million \\
      Fine mesh &$2.38\times 10^{-4}$ & $ 8400 \times 352 \times 336 \approx 993$~million \\
      \\
  \end{tabular}
  \caption{Parameters of the computational meshes utilizing isotropic cells.}
  \noindent\rule{\textwidth}{0.4pt}
  \label{tab:table1}
  \end{center}
\end{table}

A maximum Courant--Friedrichs--Lewy number of 1.0 is used throughout. Simulations are run for two flow-through times ($4L/U_\infty$) to remove initial transients, followed by three flow-through times ($6L/U_\infty$) for statistics collection.

\section{Sensitivity of mean flow separation prediction to SGS model}\label{sec:flow_stats}

\subsection{Separation prediction and grid convergence test}

\begin{figure}
\centering
\includegraphics[width=.7\textwidth,trim={0.0cm 0.1cm 0.0cm 0.0cm},clip]{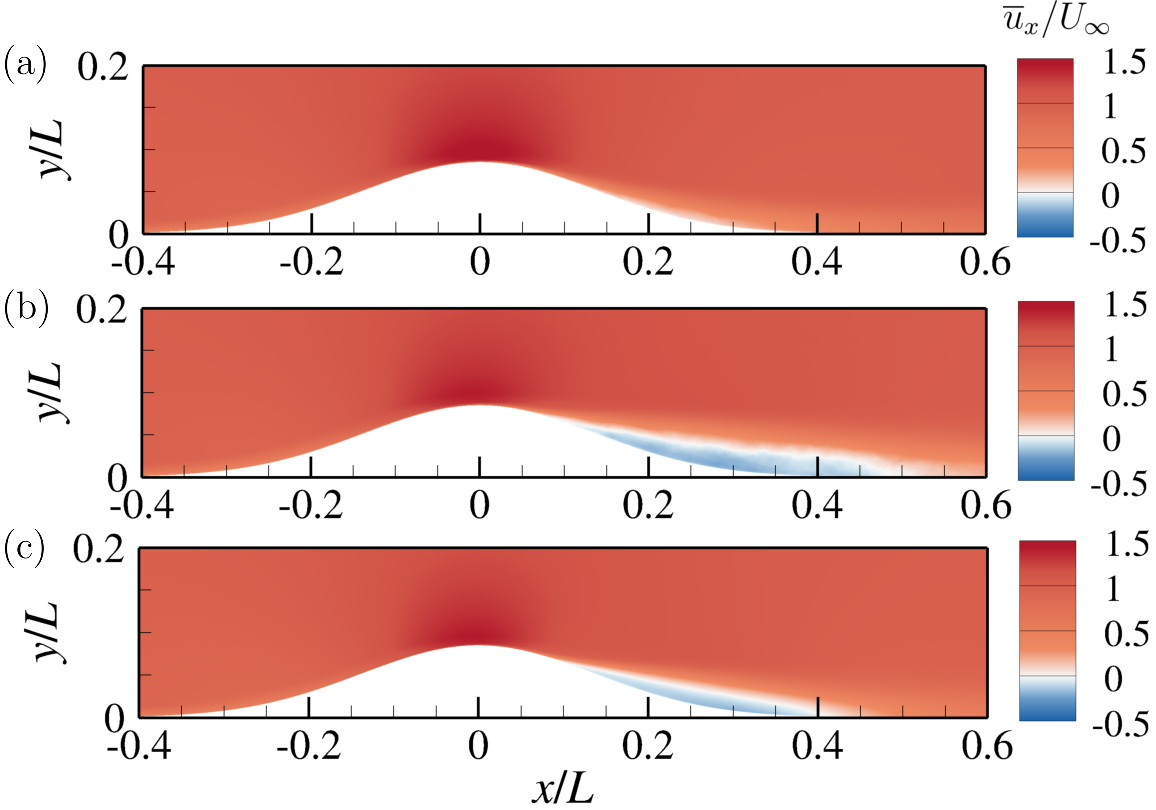}
\caption{Isocontours of mean velocity $\overline{u}_x/U_{\infty}$ from the medium-mesh simulations with the SM (a) and MSM (b) and from the reference DNS \citep{uzun2022high} (c).}
\label{ux_mean}
\end{figure}

The flow field around the Gaussian-shaped bump, obtained using the medium mesh for the two SGS models, is shown in figure~\ref{ux_mean}. Here, $\overline{(\cdot)}$ denotes both temporal averaging and spatial averaging along the homogeneous directions. With this definition, an instantaneous quantity $\varphi$ can be decomposed as $\varphi = \overline{\varphi} + \varphi'$. {Following the convention introduced in \Cref{sec:numerical_method}, the tilde denoting grid filtering is omitted throughout, so that all quantities including $\varphi$, its mean $\overline{\varphi}$ and the fluctuation $\varphi'$ refer to the resolved (filtered) field.} For reference, the DNS results from \citet{uzun2022high} are also included in the figure.

In the DNS flow field, the incoming TBL accelerates upstream of the bump peak and then decelerates downstream under the influence of an adverse pressure gradient (APG), leading to rapid thickening of the boundary layer on the leeward side. Farther downstream, a pronounced separation bubble forms. In contrast, for the present medium-mesh WMLES, the predicted separation behaviour is highly sensitive to the SGS model. The simulation using the SM does not predict any flow separation, whereas the simulation using the MSM exhibits a separation bubble that is larger than the one observed in the DNS. 

\begin{figure}
\centering
\includegraphics[width=.7\textwidth,trim={0.0cm 0.1cm 0.0cm 0.0cm},clip]{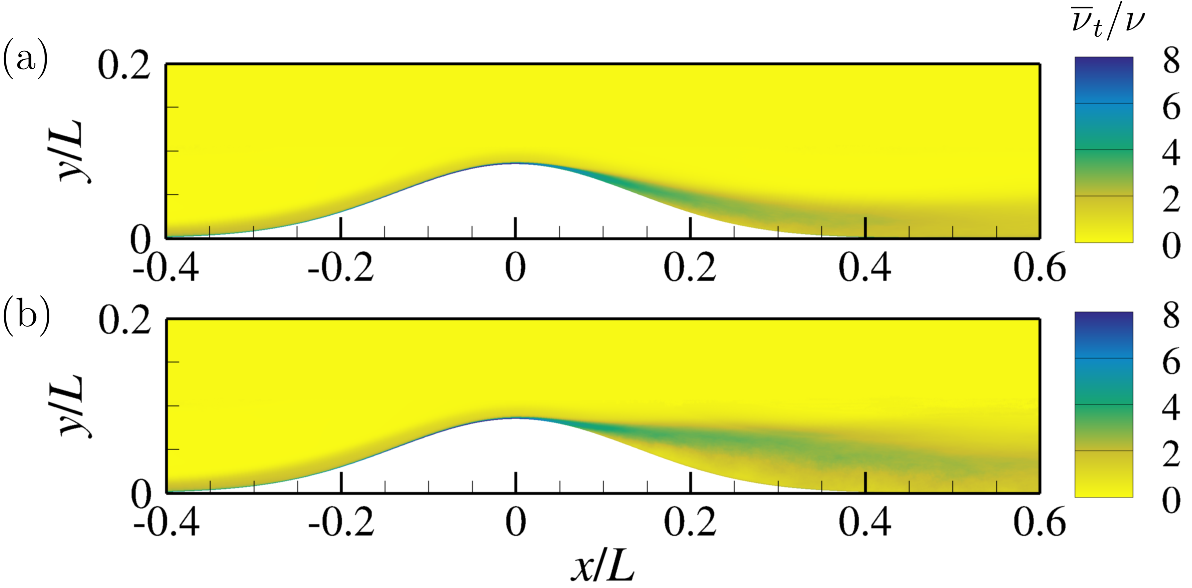}
\caption{{Isocontours of mean eddy viscosity $\overline{\nu}_t/\nu$ from the medium-mesh simulations with the SM (a) and MSM (b).}}
\label{nut_mean}
\end{figure}

The contours of the mean eddy viscosity {of} SGS model, $\overline{ \nu}_t /\nu$, from the medium-mesh simulations in an $x$\nobreakdash-$y$ plane are shown in figure~\ref{nut_mean}. For the MSM, the eddy viscosity arises solely through the isotropic stress term, as defined in equation~\eqref{aniso_EQ}. The magnitude of the eddy viscosity within the TBL is on the same order of magnitude as the fluid viscosity. {Upstream of the bump peak, the eddy viscosity produced by the two SGS models is similar in magnitude, indicating that the mean kinetic energy dissipation is similar for the two models in this region.} However, farther downstream, noticeable differences emerge: although the MSM uses the same model coefficient $C_s$ as the SM, the eddy viscosity distribution on the leeward side is significantly modified by the additional anisotropic stress term, indicating that the SGS dissipation of kinetic energy introduced by the MSM differs substantially from that of the SM in this region.

\begin{figure}
\centering
{\psfrag{a}[][]{{$x/L$}}
\psfrag{b}[][]{{$C_p$}}
\includegraphics[width=.65\textwidth,trim={0.1cm 6cm 0.1cm 0.1cm},clip]{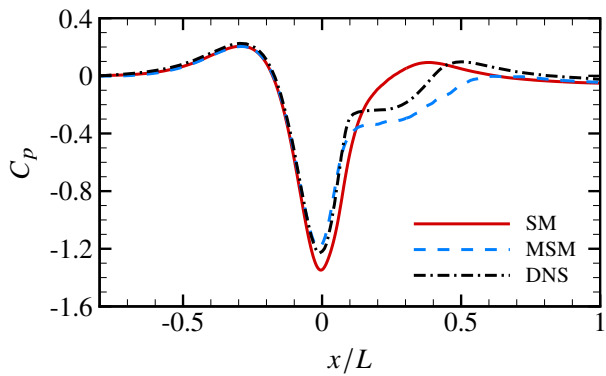}}
\caption{Mean pressure coefficient on the bottom surface from the medium-mesh simulations with the SM and MSM along with the reference DNS \citep{uzun2022high}.}
\label{Cp}
\end{figure}

In figure~\ref{Cp}, the distribution of the mean pressure coefficient, $C_p=(\overline{p}_{\text{w}}-P_\infty)/(\frac{1}{2}\rho U_\infty^2)$, on the bottom surface is compared with the DNS data \citep{uzun2022high}. Here, $p_{\text{w}}$ denotes the instantaneous static pressure at the wall, and the reference pressure $P_\infty$ is taken near the top boundary at the inlet. The $C_p$ distributions show a strong favorable pressure gradient (FPG) on the windward side near the bump peak. Downstream of the peak, the flow experiences a strong APG, followed by a milder APG over the majority of the leeward side. The two medium-mesh simulations agree reasonably well upstream of the bump peak and in the flat region downstream of the bump, but clear differences appear near the peak and along the leeward side. In particular, the MSM predicts a plateau in $C_p$ on the leeward side corresponding to the presence of a separation bubble, in better agreement with the DNS.

\begin{figure}
\centering

\sidesubfloat[]{
{\psfrag{a}[][]{{$\overline{u}_1/U_\infty$}}
\psfrag{b}[][]{{$x_2/L$}}
\includegraphics[width=0.22\textwidth,trim={1.8cm 0.6cm 7.6cm 0.5cm},clip]{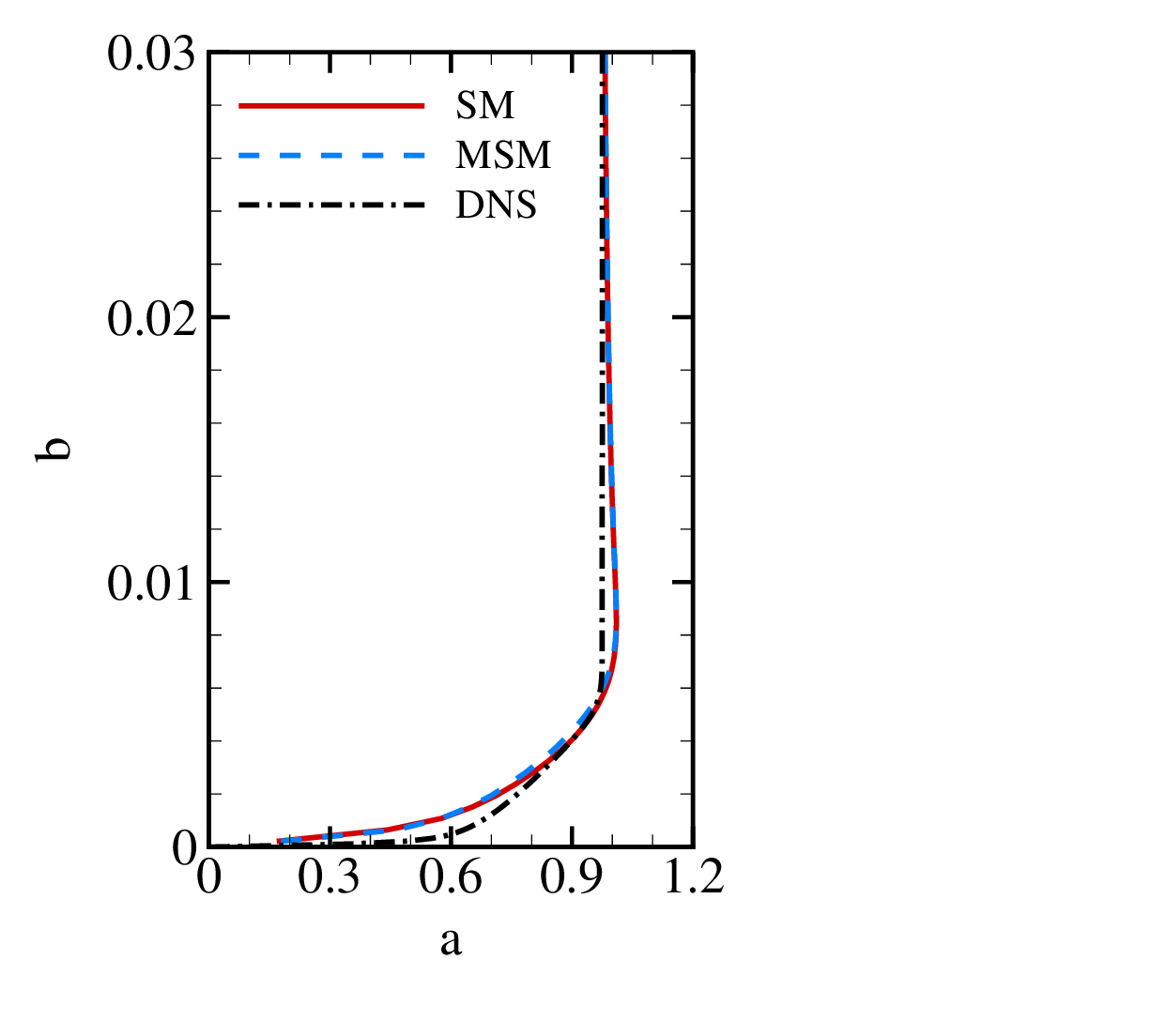}}}
\sidesubfloat[]{
{\psfrag{a}[][]{{$\overline{u}_1/U_\infty$}}
\psfrag{b}[][]{{$x_2/L$}}
\includegraphics[width=0.22\textwidth,trim={1.7cm 0.5cm 7.7cm 0.6cm},clip]{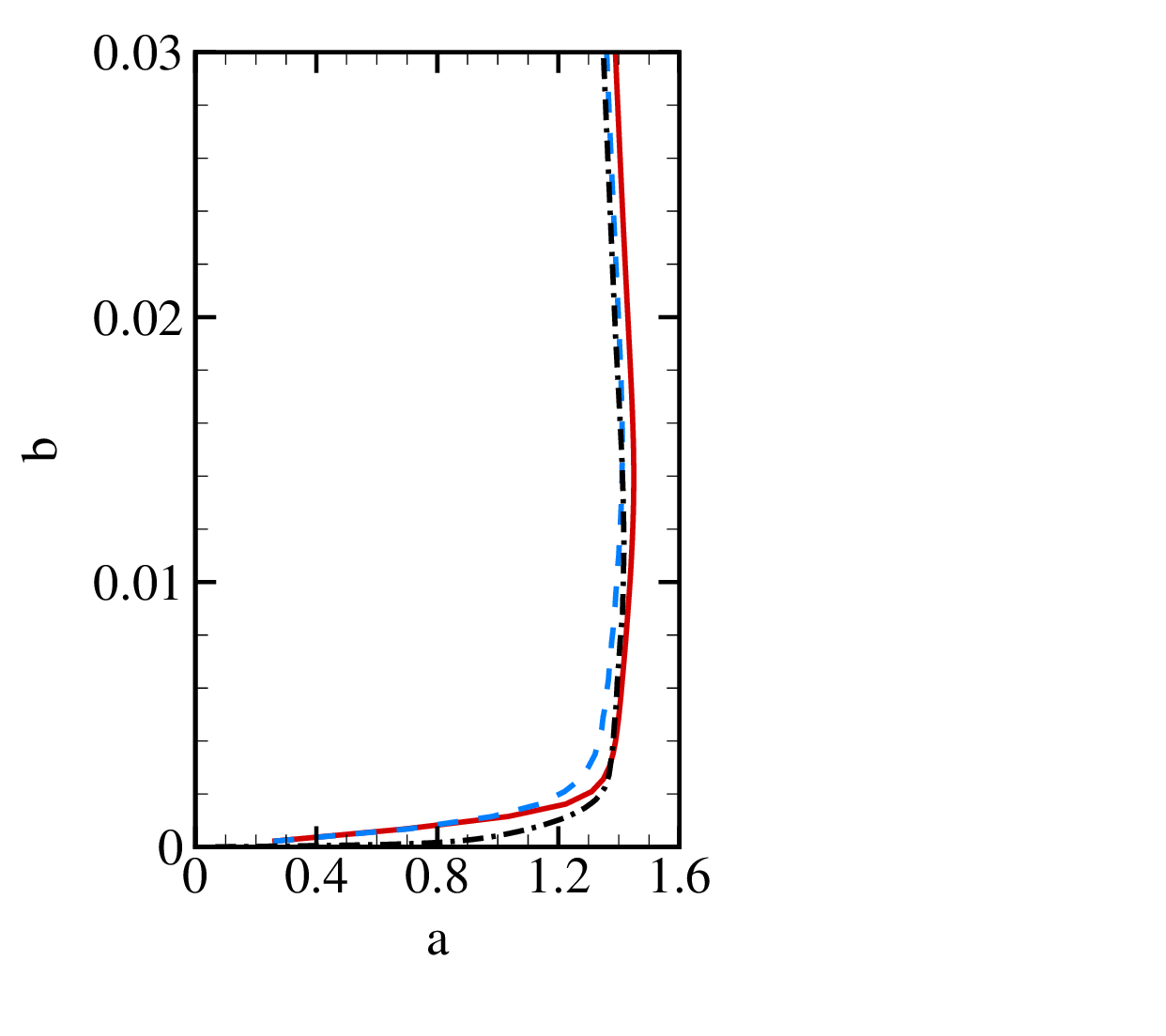}}}
\sidesubfloat[]{
{\psfrag{a}[][]{{$\overline{u}_1/U_\infty$}}
\psfrag{b}[][]{{$x_2/L$}}
\includegraphics[width=0.22\textwidth,trim={1.7cm 0.5cm 7.7cm 0.6cm},clip]{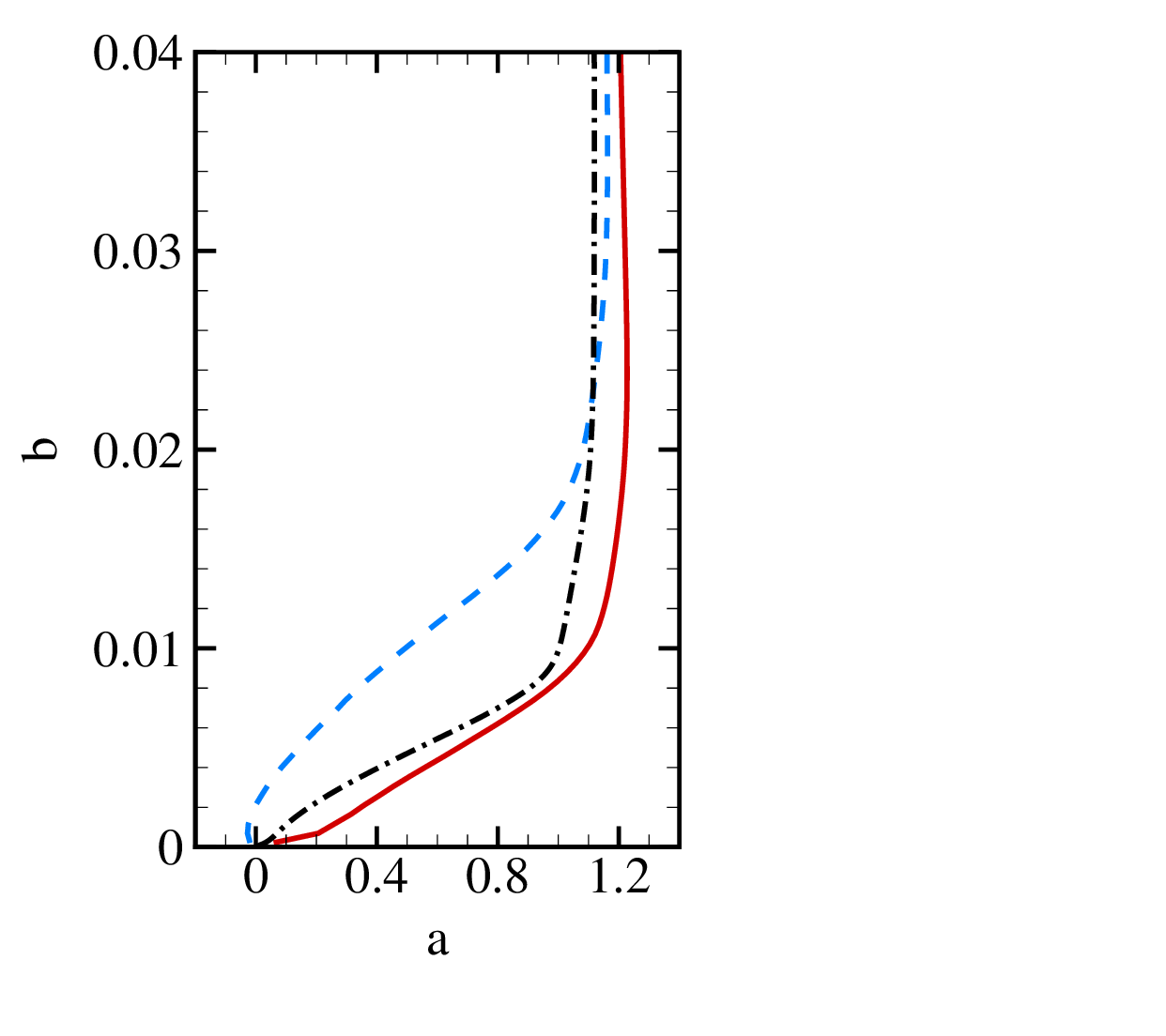}}}
\sidesubfloat[]{
{\psfrag{a}[][]{{$\overline{u}_1/U_\infty$}}
\psfrag{b}[][]{{$x_2/L$}}
\includegraphics[width=0.22\textwidth,trim={1.7cm 0.5cm 7.7cm 0.6cm},clip]{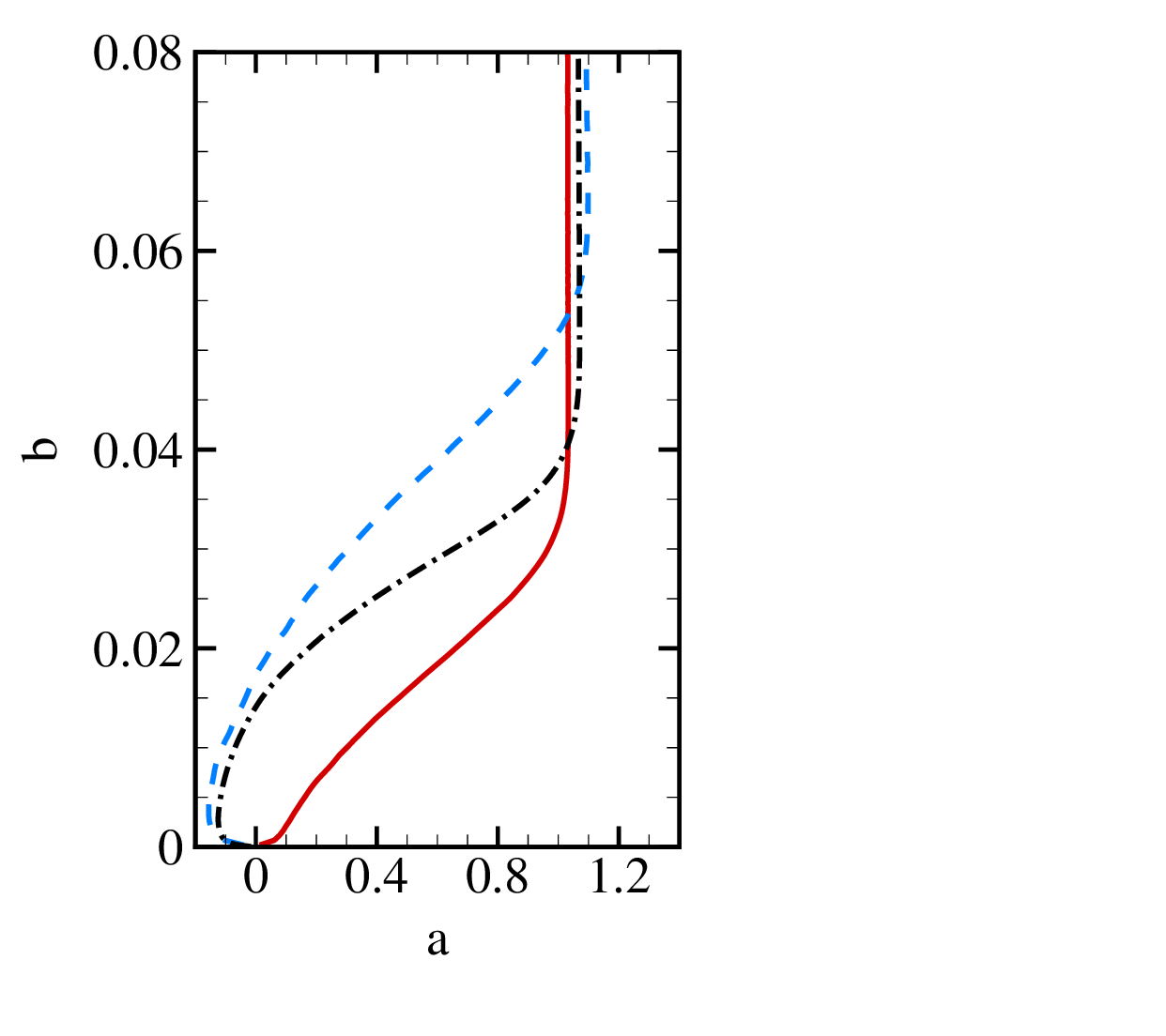}}}
\caption{{The profiles of mean streamwise velocity at $x/L = -0.7$ (a), $x/L = 0$ (b), $x/L = 0.1$ (c) and $x/L = 0.2$ (d) for the SM, MSM, and the reference DNS \citep{uzun2022high}.}}
\label{mean_vel_profile_SGS}
\end{figure}

A comparison of boundary layer profiles from the medium-mesh simulations with the DNS results of \citet{uzun2022high} is shown in figure~\ref{mean_vel_profile_SGS}, where the mean streamwise velocity $\overline{u}_1$ is plotted at {four} streamwise stations in the computational domain. The results capture the flow acceleration on the windward side of the bump, followed by deceleration and boundary-layer thickening or separation on the leeward side. Upstream of the bump peak, the two SGS model simulations agree reasonably well with each other and with the DNS. Downstream of the peak, however, clear differences emerge: the boundary layer thickens more rapidly in the MSM simulation, leading to the formation of a separation bubble, while the boundary layer in the SM simulation remains attached throughout the domain.

\begin{figure}
\centering
{\psfrag{a}[][]{{$x/L$}}
\psfrag{b}[][]{{$\overline{u}_1/U_\infty$}}
\includegraphics[width=.65\textwidth,trim={0.1cm 6.5cm 0.1cm 0.1cm},clip]{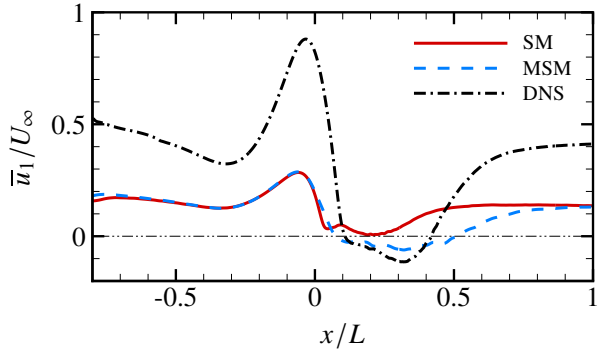}}
\caption{Mean streamwise velocity at the first off-wall cell center from the medium-mesh simulations with the SM and MSM, and the reference DNS results \citep{uzun2022high} at the same wall-normal location. $\overline{u}_1 = 0$ is indicated by the horizontal line.}
\label{mean_vel_1st_SGS}
\end{figure}

Since all simulations impose a wall-shear stress matched to the local mean wall-shear stress from the reference DNS \citep{uzun2022high}, the length of the predicted separation bubble is estimated using the mean streamwise velocity at the first off-wall cell center, as shown in figure~\ref{mean_vel_1st_SGS}. Upstream of the bump peak, the mean velocity at the first off-wall cell center agrees well across all simulations. Downstream of the peak, the flow in the SM simulation approaches separation but remains attached, whereas the MSM simulation clearly exhibits a separation bubble. It should be noted that the wall-shear stress is prescribed rather than derived from a wall-law constraint as in conventional wall-modeling approaches. In addition, {since the SGS model coefficients are not tuned for this configuration and no near-wall damping is applied}, the inner-layer velocity profiles are not expected to match the DNS exactly even in the upstream region, despite the correct mean wall-shear stress being imposed.

\begin{figure}
\centering
{\psfrag{a}[][]{{$\Delta_{\text{c}}/L{\times10^3}$}}
\psfrag{b}[][]{{$L_\text{s}/L$}}
\includegraphics[width=.65\textwidth,trim={0.1cm 7.5cm 0.1cm 0.1cm},clip]{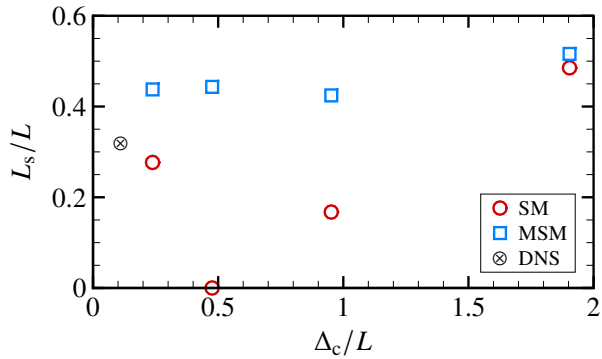}}
\caption{Mean separation bubble length on the leeward side of the bump from the simulations using the SM and MSM for different mesh resolutions and the reference DNS \citep{uzun2022high}. Symbols represent data point for each case.}
\label{separation_bubble_size}
\end{figure}

To quantitatively assess the effects of mesh resolution and SGS model on the predicted mean separation bubble size, figure~\ref{separation_bubble_size} shows the mean horizontal length ($L_\text{s}/L$) as a function of the characteristic mesh resolution ($\Delta_\text{c}/L$). The DNS value of approximately $0.32L$ is included for reference \citep{uzun2022high}. For the SM simulations, convergence with mesh refinement is non-monotonic, producing a spurious reduction of the separation bubble upon refinement, a behaviour reported in previous studies with various isotropic SGS models \citep{whitmore2021large, agrawal2022non, zhou2024sensitivity} and attributed to limitations in the SGS model. In contrast, the MSM predicts a larger separation bubble but yields consistent results across mesh resolutions, suggesting that including an anisotropic SGS stress term provides a beneficial effect. Similar consistency has been observed with the mixed model \citep{bardina1983improved, sarghini1999scale}, as discussed in {Appendix~\ref{appA}} and {in \citet{zhou2024sensitivity}}. Since the current MSM has not been optimized and employs fixed coefficients, further improvement may be achieved through dynamic coefficients or other optimization strategies. As the mesh is refined to fine-mesh resolution, results from all simulations converge toward the DNS, with the dependence on SGS model significantly reduced at near-WRLES resolution.

The results highlight the sensitivity of the predicted mean velocity field in WMLES to both the SGS model and mesh resolution, consistent with observations from previous studies \citep{rezaeiravesh2019systematic, lozano2019error, whitmore2020requirements, iyer2022wall, zhou2024sensitivity}. 
Two key questions follow: (i) why does the mean velocity on the leeward side differ qualitatively between isotropic and anisotropic SGS models at medium resolution, and (ii) why do anisotropic models yield more consistent separation bubble predictions across mesh resolutions? These questions are addressed in the following sections.

\subsection{Identification of the critical region for SGS anisotropy effect}
\label{sec:identification}

\begin{figure}
\centering
{\includegraphics[width=.85\textwidth,trim={0.1cm 0.1cm 0.1cm 0.1cm},clip]{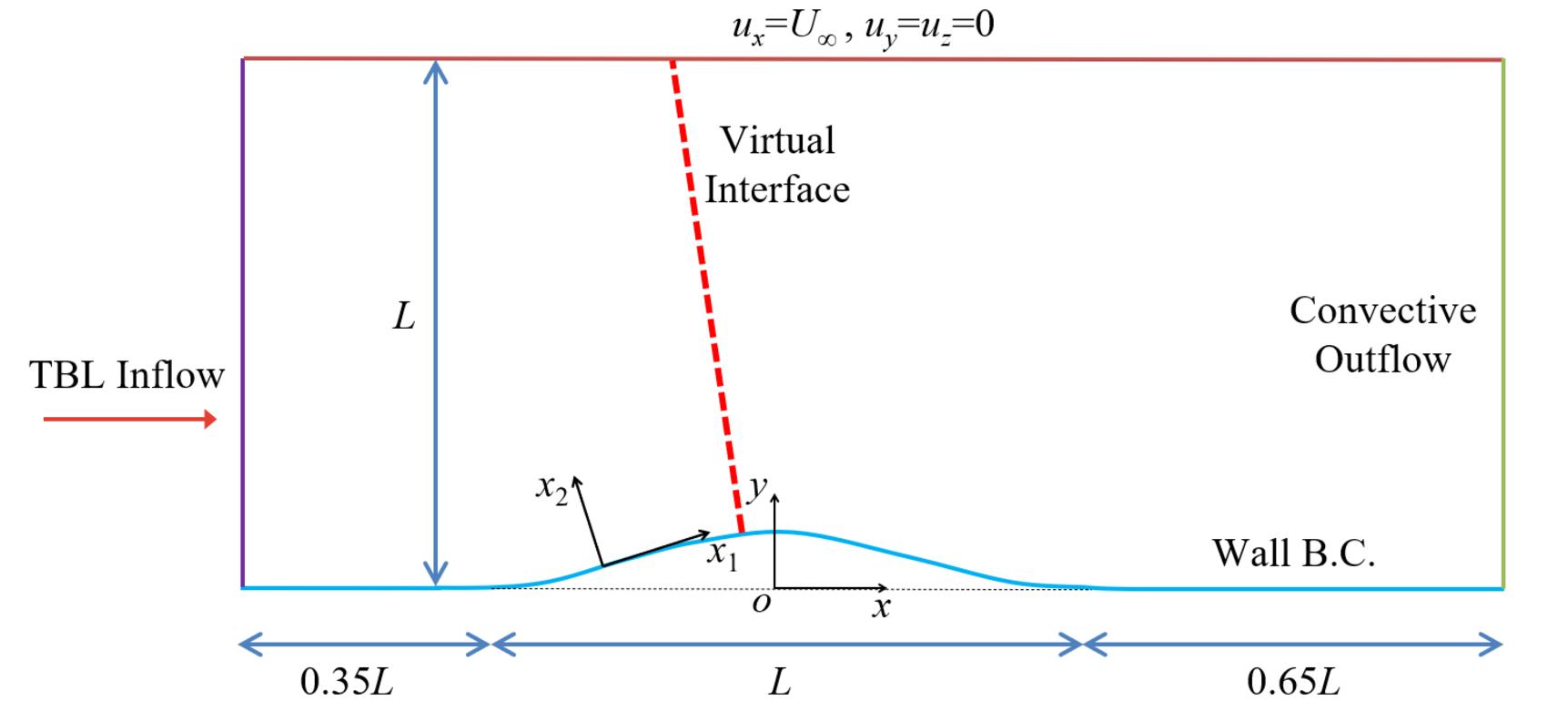}}
\caption{Virtual-interface setup dividing the domain into upstream and downstream regions using different SGS models for the flow over a Gaussian bump.}
\label{experiment_setup}
\end{figure}

To identify where the anisotropy effects are most critical, a numerical experiment is designed in which the computational domain is divided into upstream and downstream sections, as illustrated in figure~\ref{experiment_setup}, using the medium mesh. A different SGS model is assigned to each section, with a logistic function providing a smooth transition at the virtual interface. All simulations use the same boundary conditions as before. 

The SM and MSM are selectively assigned to the two domain sections. Five interface locations are considered near the bump peak, which prior studies have identified as important for downstream separation \citep{uzun2022high, prakash2024streamline, xu2024wall}. The interface is aligned with the local surface normal, and its $x$-coordinate at the bump surface is defined as $x_0$. Details are listed in {t}able~\ref{tab:table2}. In group SM-MSM, the upstream section uses the SM and the downstream section uses the MSM; in group MSM-SM, the assignment is reversed. The SGS stress is given by
\begin{equation}\label{Eq:interface}
\tau^\text{sgs}_{ij} = \tau^\text{iso}_{ij} + g \cdot \tau^\text{ani}_{ij} = \tau^\text{SM}_{ij} + g \cdot \tau^\text{ani}_{ij}\hspace{3pt}.
\end{equation}
Here, $g$ is a logistic function used to achieve a smooth transition from one SGS model to another in the computational domain, and its form for the two groups of simulations is given by
\begin{equation}
g = \left\{
\begin{split}
&\frac{g_0}{1+e^{k\times d \times \varphi}} \hspace{25pt} \text{(SM-MSM)}\\
&\frac{g_0}{1+e^{-k\times d \times \varphi}} \hspace{20pt} \text{(MSM-SM)}
\end{split}
\right. \hspace{7pt},
\end{equation}
with $g_0=1$, $d$ the spatial distance from a point to the interface at $x_0$, $k=5000$ giving an effective interface width of $5\times10^{-3}L$ (ten cells in the medium mesh). Tests with interface widths ranging from $1\times 10^{-3}L$ to $1\times 10^{-2}L$ showed no significant influence on the resulting mean velocity fields. The parameter $\varphi=1$ upstream and $-1$ downstream of the interface centerline. Statistics are collected over three flow-through times after discarding two for transient removal.

\begin{table}
  \begin{center}
\def~{\hphantom{0}}
\noindent\rule{\textwidth}{0.4pt}
\small
\begin{tabular}{lcc}
  \\ Group index \hspace{2pt}  & Virtual interface location $x_0/L$ \hspace{2pt}  & SGS models in the upstream / downstream \\[3pt]
  \multirow{4}{*}{SM-MSM} 
    & \multirow{8}{*}{
        \begin{tabular}{c}
          $-0.3$ \\[1.0ex]
          $-0.2$ \\[1.0ex]
          $-0.1$ \\[1.0ex]
          $0.0$ \\[1.0ex]
          $0.05$ \\
        \end{tabular}
      } 
    & \multirow{4}{*}{SM / MSM} \\
    & & \\
    & & \\
    & & \\
  \multirow{4}{*}{MSM-SM} 
    & & \multirow{4}{*}{MSM / SM} \\
    & & \\
    & & \\
    & & \\[1pt]
\end{tabular}
\noindent\rule{\textwidth}{0.4pt}  
\caption{List of parameters for the virtual interface setup.}
\label{tab:table2}
\end{center}
\end{table}

\begin{figure}
\centering
\sidesubfloat[]{
{\psfrag{a}[][]{{$x_0/L$}}
\psfrag{b}[][]{{$L_{\text{s}}/L$}}
\includegraphics[width=.46\textwidth,trim={0.5cm 4.5cm 1.8cm 1.5cm},clip]{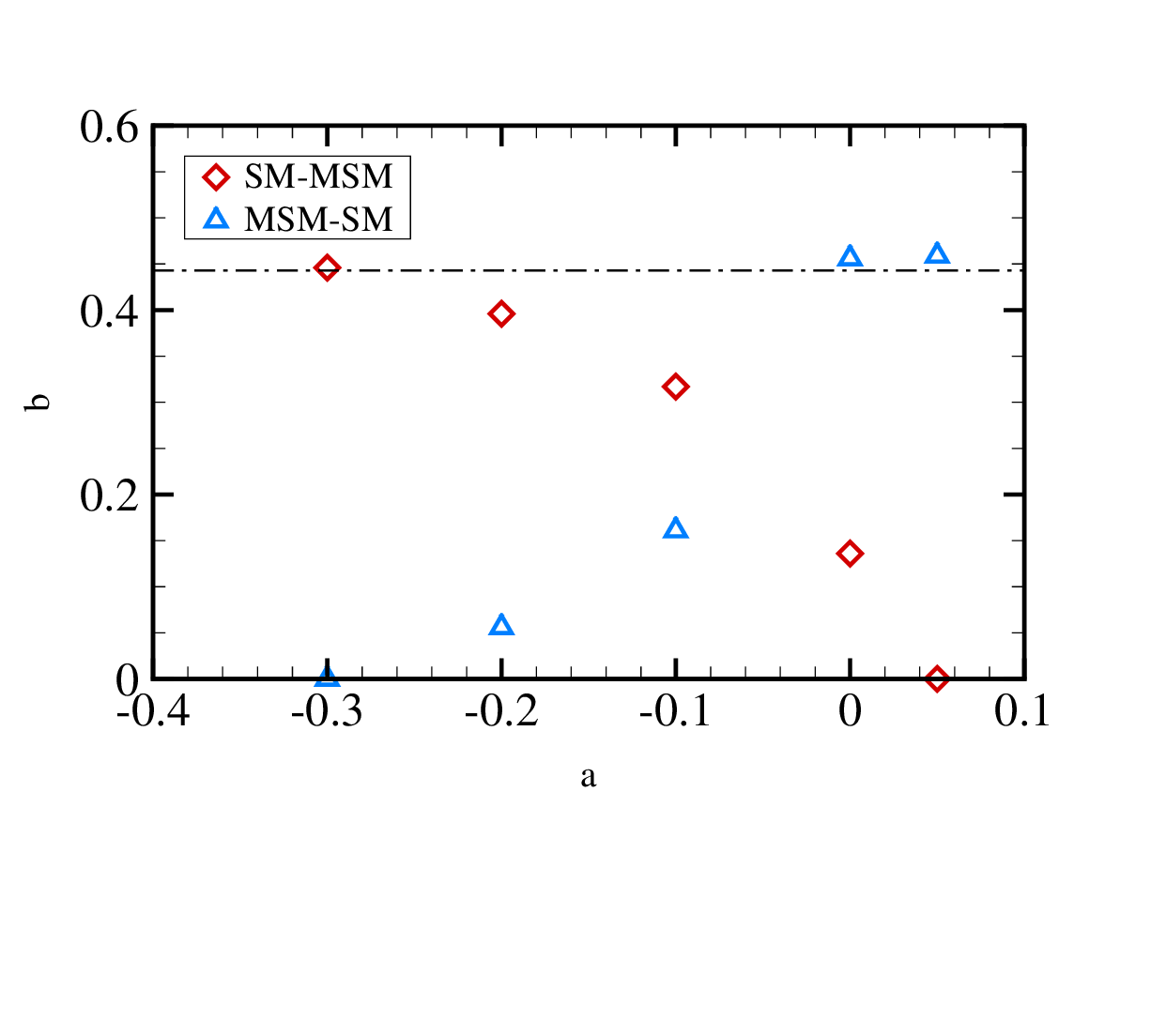}}
}
\sidesubfloat[]{
{\psfrag{a}[][]{{$x_0/L$}}
\psfrag{b}[][]{{$x_{\text{s}}/L$}}
\includegraphics[width=.46\textwidth,trim={0.5cm 4.5cm 1.8cm 1.5cm},clip]{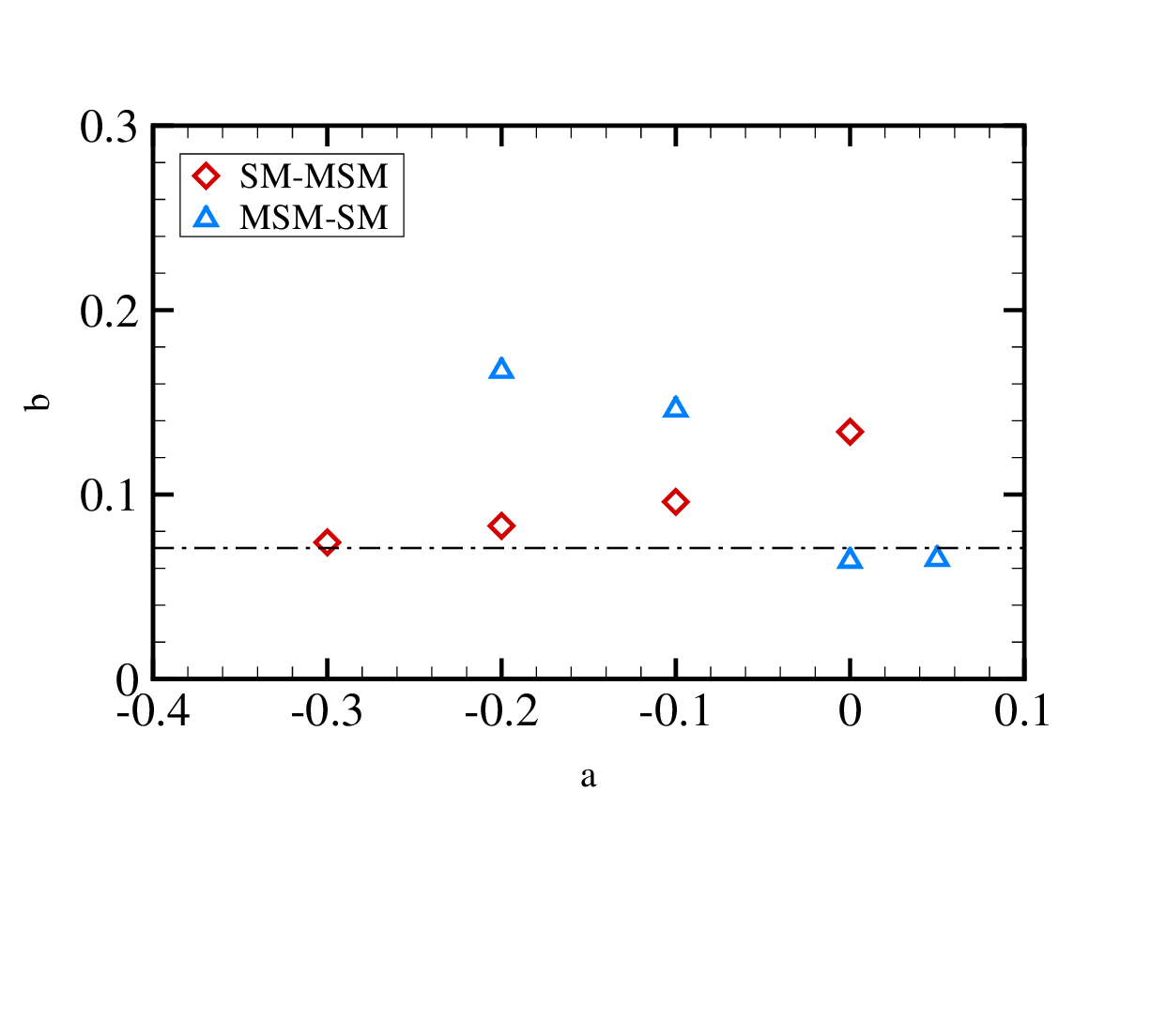}}
}
\caption{Mean separation bubble length on the leeward side of the bump (a) and the location of mean separation point (b) from medium-mesh simulations using the SM-MSM and MSM-SM with different virtual interface locations. Horizontal lines indicate the results from the medium-mesh simulation with MSM.}
\label{separation_point_twoSGS}
\end{figure}

Figure~\ref{separation_point_twoSGS} shows the predicted mean separation bubble size ($L_\text{s}/L$) and separation point location ($x_\text{s}/L$). When the interface is at the most upstream location ($x_0/L=-0.3$), predictions closely match those from the simulation using the downstream SGS model over the entire domain. As the interface shifts downstream, results increasingly reflect the upstream model. For example, in group SM-MSM with the interface at $x_0/L=-0.3$, the separation bubble matches that of the full-domain MSM simulation ($L_\text{s}\approx0.44L$). As the interface moves downstream, the bubble shrinks progressively, disappearing when the interface reaches $x_0/L=0.05$. An opposite trend is observed for group MSM-SM. The most pronounced variations occur as the interface moves from $x_0/L=-0.2$ to the bump peak, within the strong FPG region; beyond the peak, further downstream shifts have diminishing influence.

\begin{figure}
\centering

\sidesubfloat[]{
{\psfrag{a}[][]{{$\overline{u}_1/U_\infty$}}
\psfrag{b}[][]{{$x_2/L$}}
\includegraphics[width=0.28\textwidth,trim={1.7cm 0.6cm 7.6cm 0.5cm},clip]{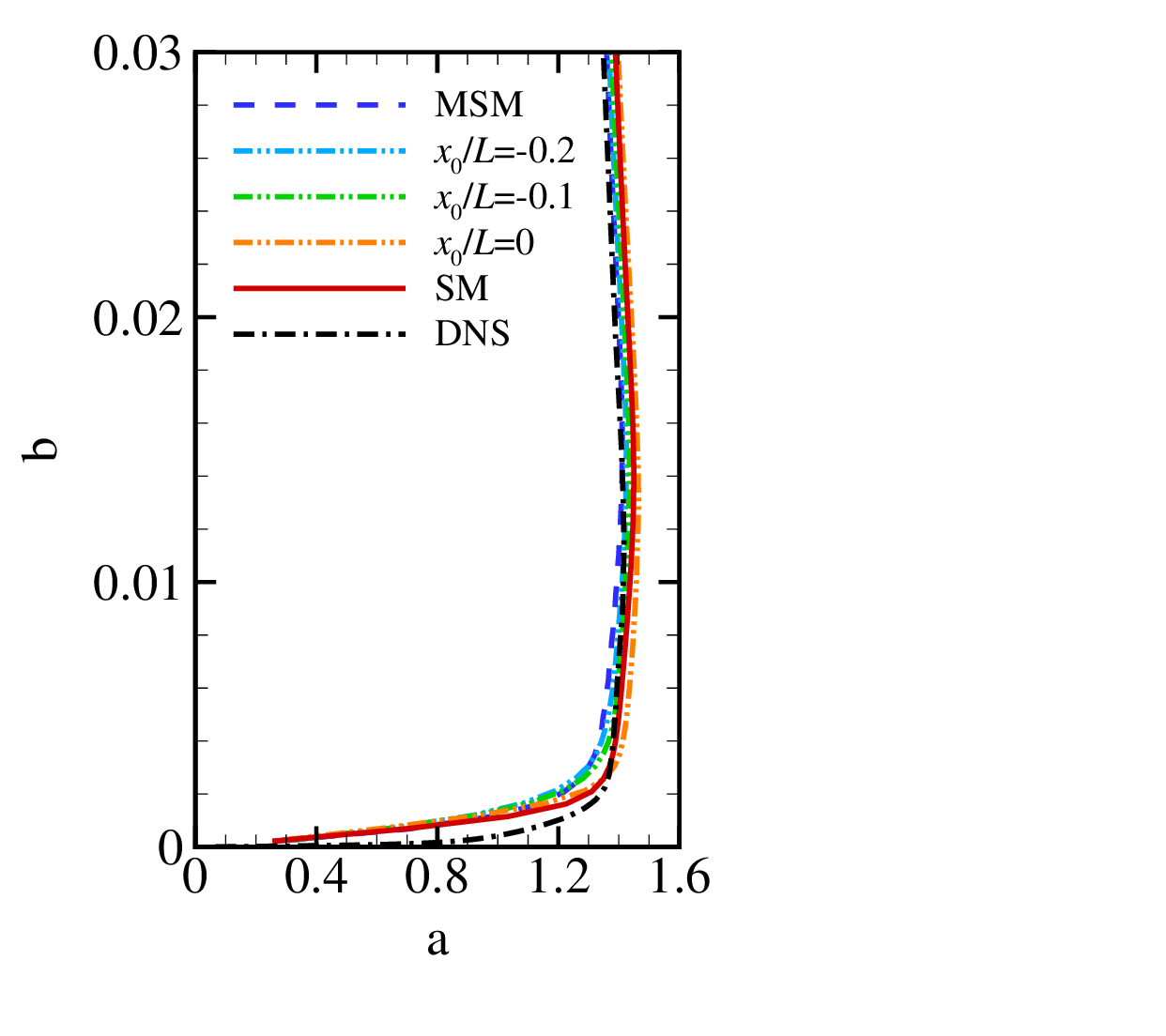}}}
\sidesubfloat[]{
{\psfrag{a}[][]{{$\overline{u'_1u'_2}/U^2_\infty$}}
\psfrag{b}[][]{{$x_2/L$}}
\includegraphics[width=0.28\textwidth,trim={1.8cm 0.6cm 7.6cm 0.5cm},clip]{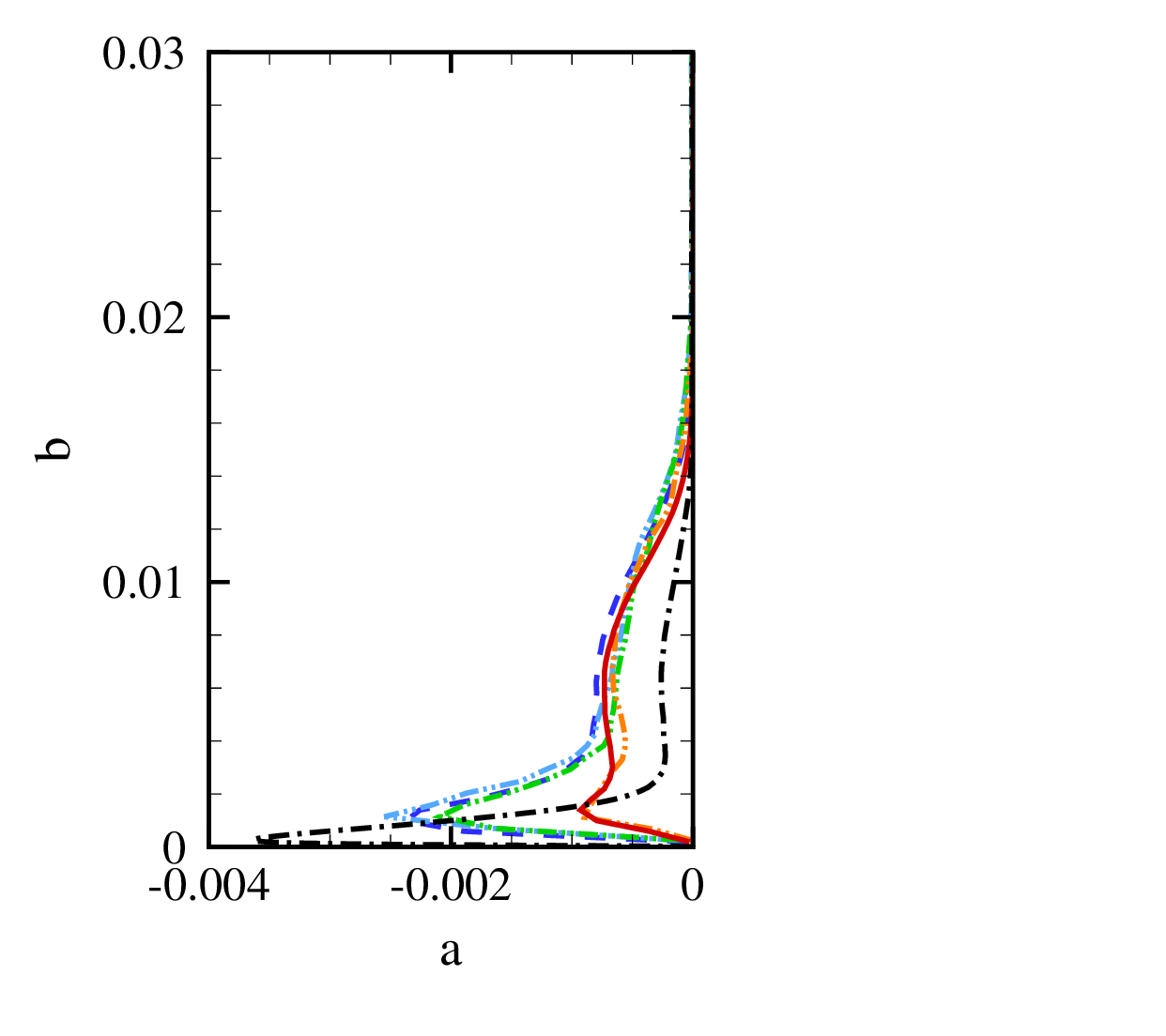}}}
\sidesubfloat[]{
{\psfrag{a}[][]{{$\overline{u'_2u'_2}/U^2_\infty$}}
\psfrag{b}[][]{{$x_2/L$}}
\includegraphics[width=0.28\textwidth,trim={1.8cm 0.6cm 7.6cm 0.5cm},clip]{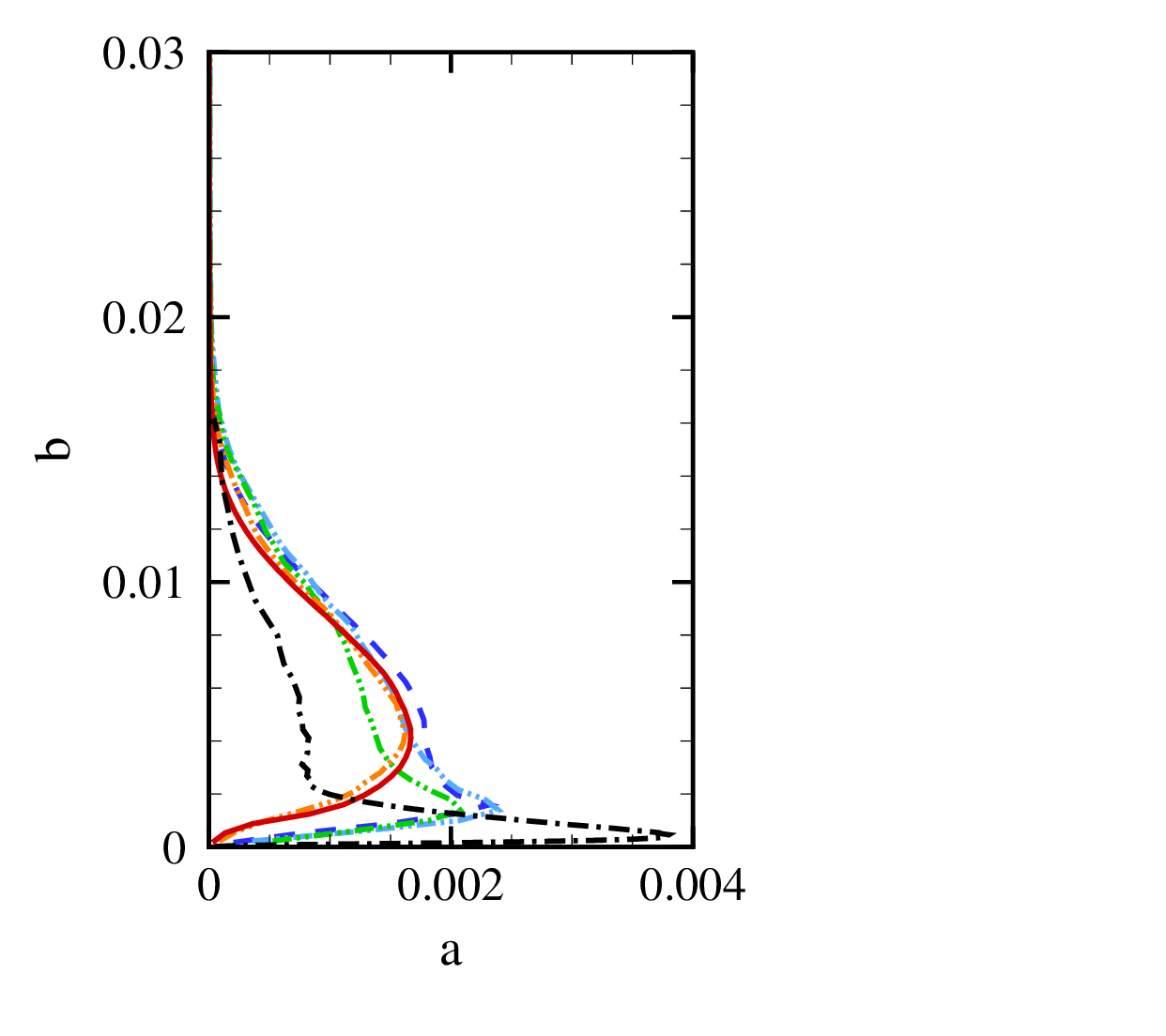}}}
\caption{Mean streamwise velocity $\overline{u}_1$ (a), Reynolds shear stress $\overline{u'_1u'_2}$ (b), and Reynolds normal stress $\overline{u'_2u'_2}$ (c) profiles at the bump peak ($x/L=0$) for SM-MSM with the virtual interface located at $x_0/L = -0.2$, $-0.1$, and $0$. DNS \citep{uzun2022high}, SM and MSM results are shown for reference.}
\label{stats_SM-MSM}
\end{figure}

\begin{figure}
\captionsetup[subfigure]{oneside,margin={0.6cm,0cm}}
\centering

\sidesubfloat[]{
{\psfrag{a}[][]{{$\overline{u}_1/U_\infty$}}
\psfrag{b}[][]{{$x_2/L$}}
\includegraphics[width=0.28\textwidth,trim={1.7cm 0.6cm 7.6cm 0.5cm},clip]{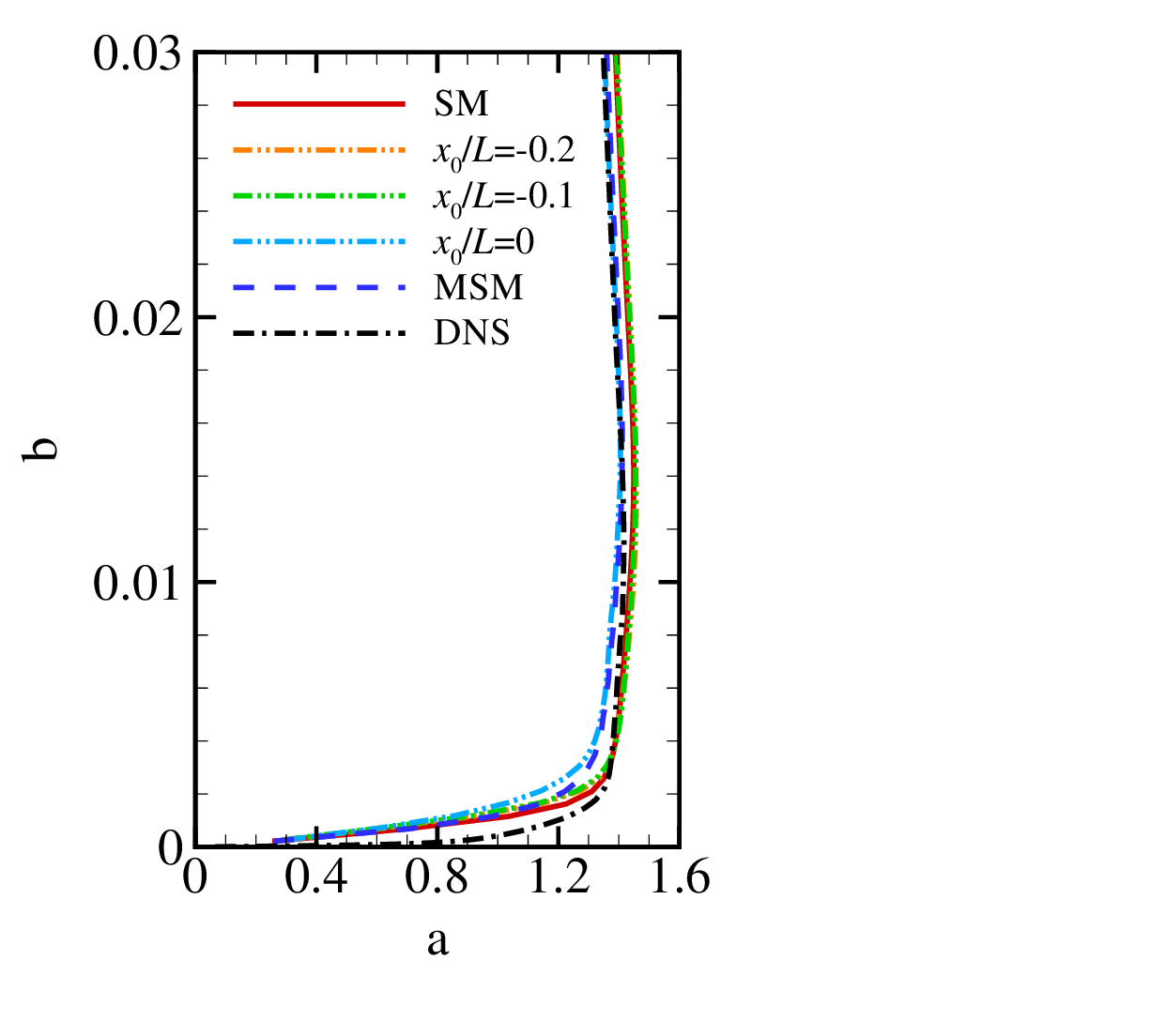}}}
\sidesubfloat[]{
{\psfrag{a}[][]{{$\overline{u'_1u'_2}/U^2_\infty$}}
\psfrag{b}[][]{{$x_2/L$}}
\includegraphics[width=0.28\textwidth,trim={1.8cm 0.6cm 7.6cm 0.5cm},clip]{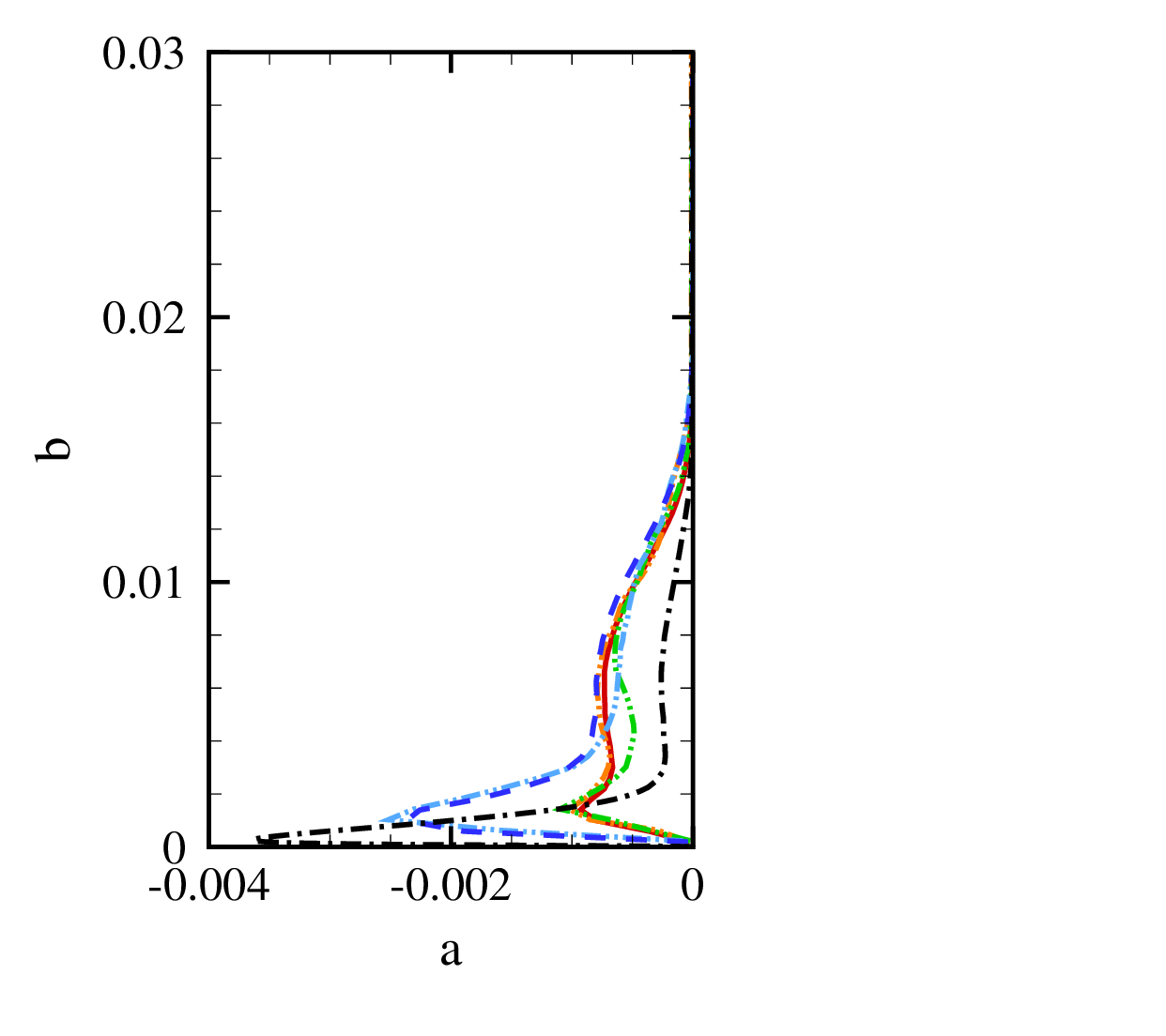}}}
\sidesubfloat[]{
{\psfrag{a}[][]{{$\overline{u'_2u'_2}/U^2_\infty$}}
\psfrag{b}[][]{{$x_2/L$}}
\includegraphics[width=0.28\textwidth,trim={1.8cm 0.6cm 7.6cm 0.5cm},clip]{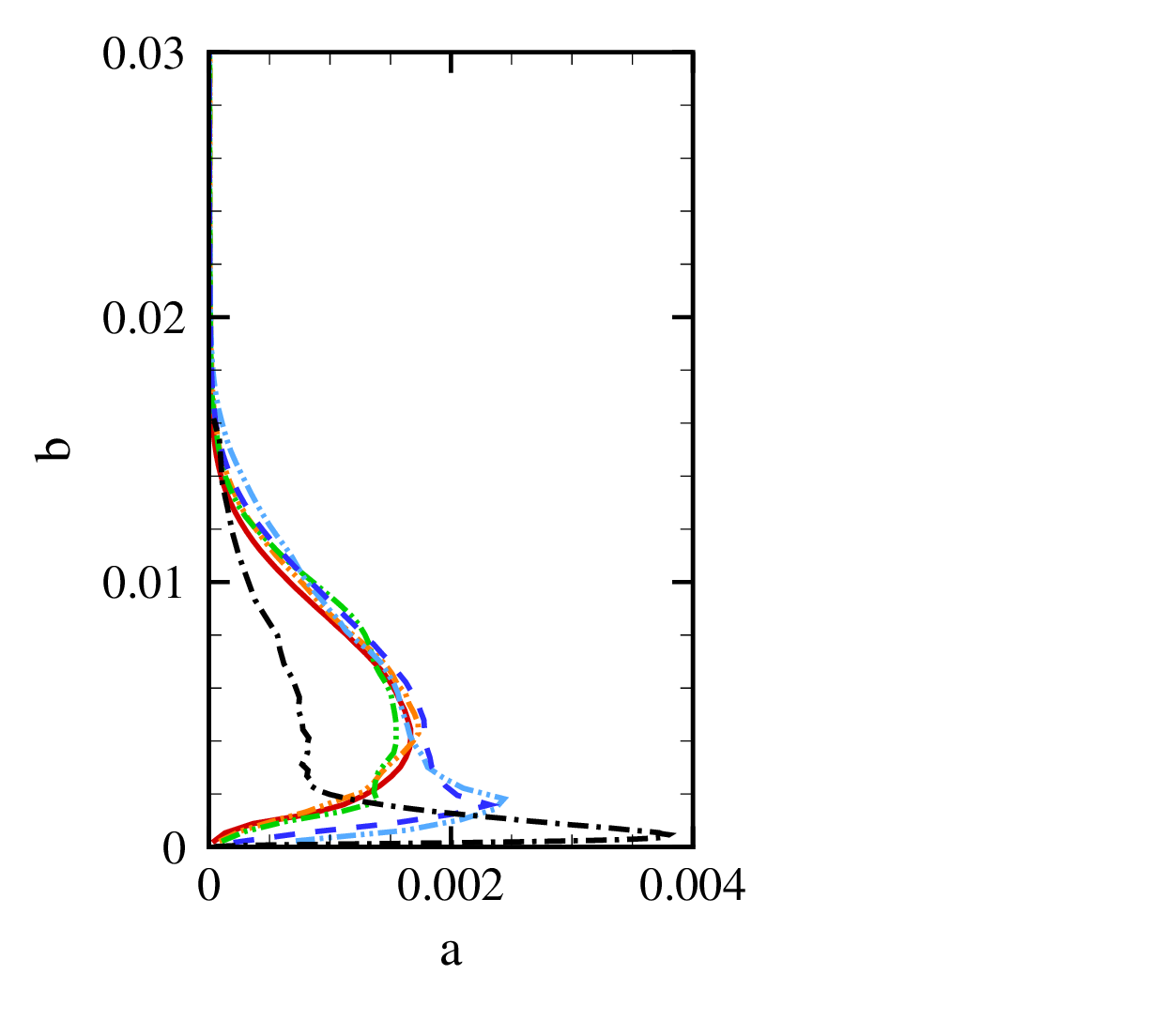}}}
\caption{Mean streamwise velocity $\overline{u_1}$ (a), Reynolds shear stress $\overline{u'_1u'_2}$ (b), and Reynolds normal stress $\overline{u'_2u'_2}$ (c) profiles at the bump peak ($x/L=0$) for MSM–SM with the virtual interface located at $x_0/L = -0.2$, $-0.1$, and $0$. DNS \citep{uzun2022high}, SM and MSM results are shown for reference.}
\label{stats_MSM-SM}
\end{figure}

To examine how the TBL evolves with interface location within the critical FPG region ($x\in[-0.2,0]$), velocity statistics at the bump peak ($x/L=0$) are analysed. In addition to the mean streamwise velocity $\overline{u}_1$, the Reynolds shear stress $\overline{u'_1u'_2}$ and wall-normal Reynolds normal stress $\overline{u'_2u'_2}$ are examined, as these play important roles in shaping the near-wall mean flow, as discussed in \Cref{sec:mean_budgets}--\Cref{sec:Reynolds_budgets}. Results are shown in figures~\ref{stats_SM-MSM} and \ref{stats_MSM-SM}. As the interface shifts downstream, results gradually approach those of the upstream model applied over the entire domain. The DNS profiles exhibit pronounced internal peaks in the Reynolds stresses at this location, which \citet{uzun2022high} identify as crucial for the downstream TBL evolution and onset of separation. 

From the comparisons among the simulations, it is found that applying the MSM within the critical FPG region improves the capture of these internal peaks. {Since the resolved Reynolds stress is governed by a transport equation with mean convection, these stress distributions are carried downstream and directly influence the momentum balance near the separation point, ultimately driving the differences in leeward-side separation.} This mechanism will be examined in detail in \Cref{sec:mean_budgets} and \Cref{sec:Reynolds_budgets}. Overall, these results highlight that anisotropic SGS stress on the windward side is critical: the SGS model used in the strong FPG region governs the wall-normal Reynolds stress distributions and, through them, the downstream separation behaviour. 

\section{Budget analyses} 
\label{sec:budget}

In this section, we analyse the mean streamwise momentum and pressure equations slightly upstream of the mean separation point to isolate the effects of individual budget terms, demonstrating that the Reynolds stresses (particularly $\overline{u_1'u_2'}$ and $\overline{u_2'u_2'}$) have a significant impact on the mean flow field. We then analyse the Reynolds stress transport equations to understand how their distributions within the TBL are influenced by anisotropic SGS stress. Only the medium-mesh simulations are considered in \Cref{sec:mean_budgets}--\ref{sec:Reynolds_budgets}, as this resolution shows the most pronounced differences in separation prediction between the two SGS models (see figure~\ref{separation_bubble_size}). The effect of mesh resolution is then examined in \Cref{sec:mesh_effect}.

\subsection{Mean streamwise momentum {equation}} \label{sec:mean_budgets}

Considering the spanwise homogeneity of the statistically stationary flow, the streamwise mean momentum equation {\citep{Pope_2000, abe2019notable}} is
\begin{equation}
    \overline{u}_1 \frac{\p \overline{u}_1}{\p x_1} + \overline{u}_2 \frac{\p \overline{u}_1}{\p x_2} = P_g + V_{11}+ V_{12} + T_{11} + T_{12}+ R_{11} + R_{12} \hspace{3pt},
    \label{streamwise_momentum}
\end{equation}
where the seven terms on the right-hand side of the equation are
\begin{equation}
    P_g = -\frac{1}{\rho} \frac{\p \overline{p}}{\p x_1}\hspace{2pt}, \hspace{2em}
    V_{11} = \frac{\p}{\p x_1}\left(2\nu \overline{S}_{11}\right), \hspace{2em} V_{12} = \frac{\p}{\p x_2}\left(2\nu \overline{S}_{12}\right),
    \label{Mobud_eq_begin}
\end{equation}
\begin{equation}
    T_{11} = \frac{\p}{\p x_1}\left( -\overline{\tau_{11}^{\text{sgs}}}\right), \hspace{2em} T_{12} = \frac{\p}{\p x_2}\left(-\overline{\tau_{12}^{\text{sgs}}}\right),
\end{equation}
\begin{equation}
    R_{11} = \frac{\p}{\p x_1}\left(-\overline{u'_1 u'_1}\right), \hspace{2em} R_{12} = \frac{\p}{\p x_2}\left(- \overline{u'_1 u'_2}\right),
    \label{Mobud_eq_end}
\end{equation}
corresponding to mean pressure gradient ($P_g$), viscous ($V_{11}$, $V_{12}$), SGS stress ($T_{11}$, $T_{12}$), and Reynolds stress ($R_{11}$, $R_{12}$) contributions, respectively. The magnitudes of these terms influence the distribution of mean momentum and, consequently, the mean velocity field. Curvature effects are neglected as their influence upstream of the mean separation point ($x/L<0.1$) is negligible, as shown in {Appendix~\ref{appA_new}}.

\begin{figure}
\centering
\sidesubfloat[]{
{\psfrag{a}[][]{{}}
\psfrag{b}[][]{{$x_2/L$}}
\psfrag{c}[][]{{$P_g$}}
\psfrag{d}[][]{{$V_{11}$}}
\psfrag{e}[][]{{$V_{12}$}}
\psfrag{f}[][]{{$T_{11}$}}
\psfrag{g}[][]{{$T_{12}$}}
\psfrag{h}[][]{{$R_{11}$}}
\psfrag{i}[][]{{$R_{12}$}}
\psfrag{j}[][]{{$\overline{u}_k \frac{\p \overline{u}_1}{\p x_k}$}}
\includegraphics[width=.45\textwidth,trim={1cm 1.5cm 0.1cm 0.1cm},clip]{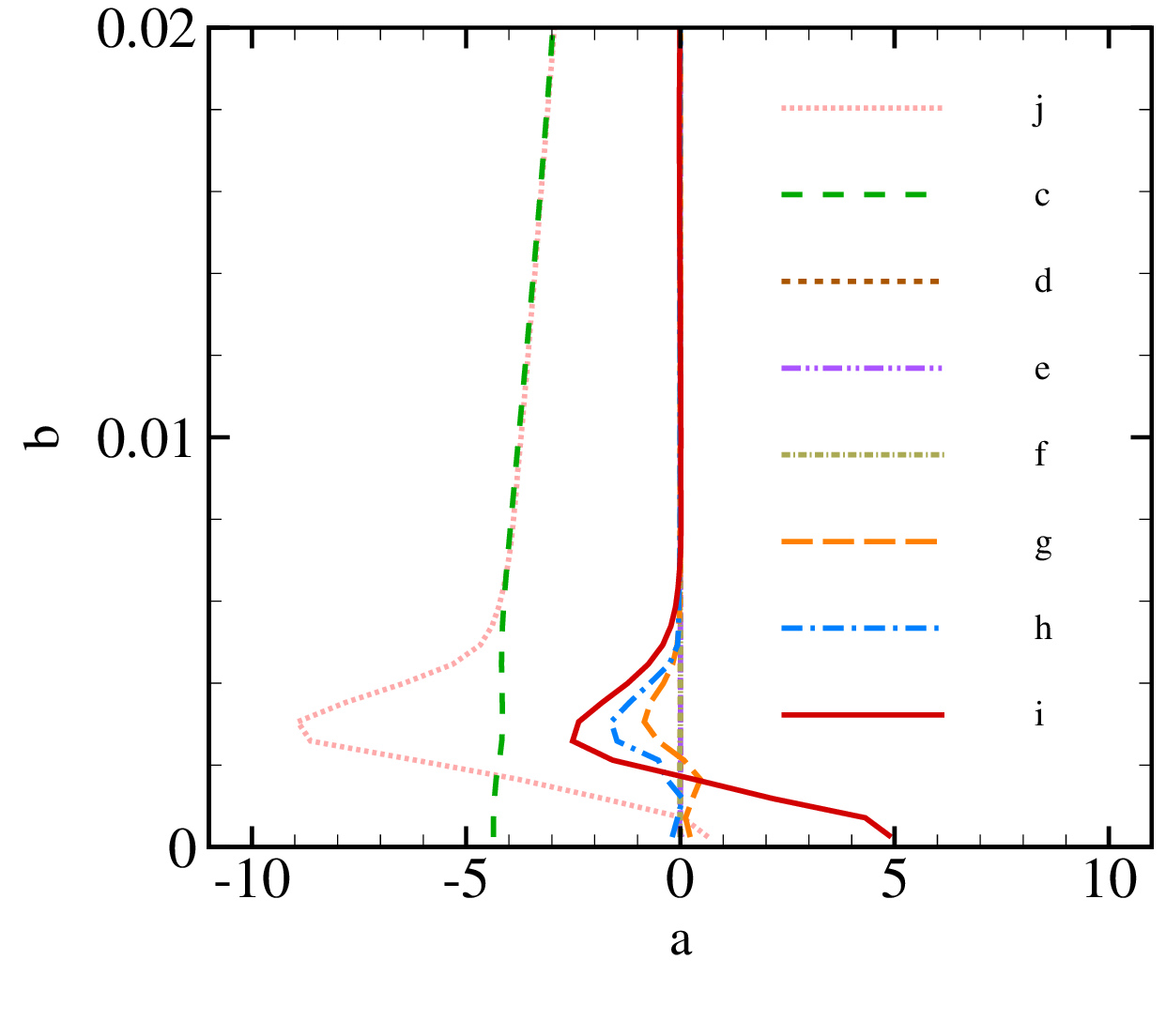}}
}
\sidesubfloat[]{
{\psfrag{a}[][]{{}}
\psfrag{b}[][]{{$x_2/L$}}
\includegraphics[width=.45\textwidth,trim={1cm 1.5cm 0.1cm 0.1cm},clip]{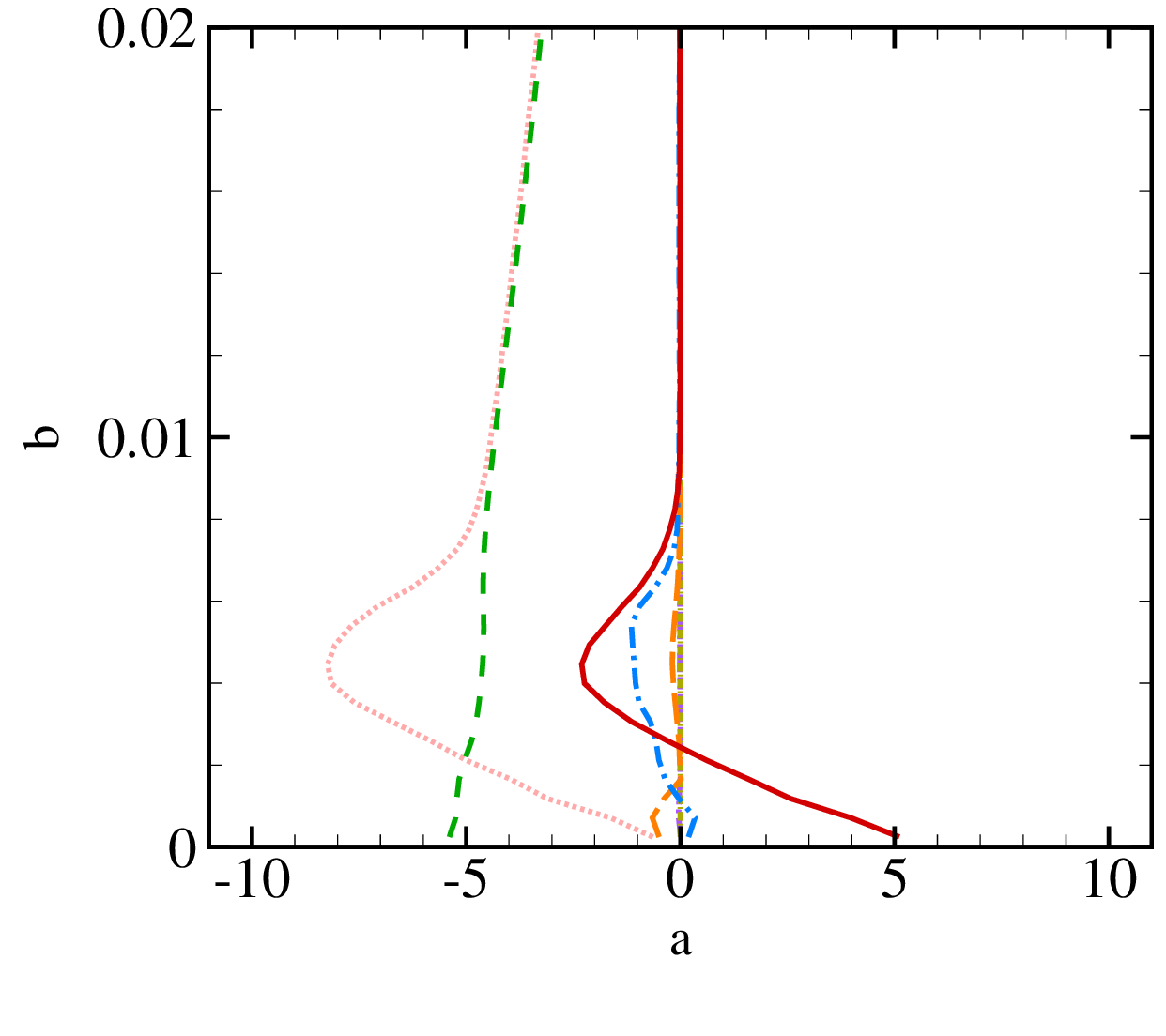}}
}
\caption{Mean streamwise momentum budget terms at $x/L=0.05$ from medium-mesh simulations with the SM (a) and MSM (b). All terms are nondimensionalized using $U_\infty$, $L$ and $\rho$. {The term $\overline{u}_k \frac{\p \overline{u}_1}{\p x_k}$ denotes the net mean convection on the left-hand side of the mean momentum equation, and the other} line notations correspond to equations~\eqref{Mobud_eq_begin}–\eqref{Mobud_eq_end}.}
\label{Mobudget_medium_SGS}
\end{figure}

Figure~\ref{Mobudget_medium_SGS} shows the mean streamwise momentum budget at $x/L=0.05$, approximately one boundary-layer thickness {($0.02L$)} upstream of the separation point in the MSM simulation. The same location is used for the SM case for consistency. {In both simulations, the budget shows that mean convection $\overline{u}_k \frac{\p \overline{u}_1}{\p x_k}$, the mean pressure gradient $P_g$, and the Reynolds shear-stress gradient $R_{12}$ are the leading-order terms governing the near-wall momentum balance. All three terms differ quantitatively between the SM and MSM simulations. The convection term represents the transport of mean momentum by the established mean flow, and its quantitative difference between the two simulations reflects the differences in the upstream mean velocity field that have already been shaped by the SGS model. In contrast, $P_g$ and $R_{12}$ are forcing terms that directly act on the local momentum balance, and the upstream Reynolds stresses that contribute to both $P_g$ and $R_{12}$ are, as shown in \Cref{sec:identification}, the quantities through which the SGS model in the upstream FPG region influences the downstream flow field. The subsequent discussion therefore focuses on $P_g$ and $R_{12}$ as the terms that most directly connect the SGS model to the separation onset. The residual (total sum of all budget terms) is negligible relative to the leading-order terms throughout the boundary layer in both simulations and is not shown for brevity.}

In both cases, $P_g$ is strongly negative across the boundary layer, but the MSM produces a noticeably larger magnitude, indicating stronger mean flow deceleration. The distributions of $R_{12}$ also differ critically: while the peak magnitude is similar between the two simulations, the MSM exhibits a significantly broader region of negative $R_{12}$, implying that momentum is extracted over a thicker portion of the boundary layer and redistributed toward the near-wall region. Consequently, less streamwise momentum remains available farther from the wall to resist the APG, making the near-wall flow more susceptible to separation. In contrast, the SM confines the momentum deficit to a thinner layer, helping the flow remain attached.

\subsection{Mean pressure equation}
\label{sec:pressure_budgets}
The Poisson equation for the mean pressure {\citep{bradshaw1981note, adrian1982comment, Pope_2000}} is
\begin{equation}
\begin{split}
        -\frac{1}{\rho}\nabla^2 \overline{p} 
        &=\frac{\partial \overline{u}_i}{\partial x_j} \frac{\partial
\overline{u}_j}{\partial x_i} + \frac{\partial^2 \overline{u'_i
u'_j}}{\partial x_i\partial x_j}{+\frac{\partial^2 \overline{\tau^{sgs}_{ij}}}{\partial x_i\,\partial x_j}}\\
        &=U_{11}+U_{12}+U_{22}+W_{11}+W_{12}+W_{22}+{G_{11}+G_{12}+G_{22}} \hspace{3pt},
    \end{split}
\end{equation}
where the {nine} terms on the right-hand side of the equation are
\begin{equation}
    U_{11} = \Big(\frac{\partial \overline{u}_1}{\partial x_1}\Big)^2,
    \hspace{2em} U_{12} = 2\Big(\frac{\partial \overline{u}_1}{\partial x_2} \frac{\partial \overline{u}_2}{\partial x_1}\Big)\hspace{3pt},
    \hspace{2em} U_{22} = \Big(\frac{\partial
\overline{u}_2}{\partial x_2}\Big)^2,
\label{Prebud_eq_begin}
\end{equation}
\begin{equation}
    W_{11} = \frac{\partial^2 \overline{u'_1
u'_1}}{\partial x_1^2}\hspace{3pt},
    \hspace{2em} W_{12} = 2\frac{\partial^2 \overline{u'_1
u'_2}}{\partial x_1 \partial x_2}\hspace{3pt},
    \hspace{2em} W_{22} = \frac{\partial^2 \overline{u'_2
u'_2}}{\partial x_2^2}\hspace{3pt},
    \label{Prebud_eq_middle}
\end{equation}
\begin{equation}{
    G_{11} = \frac{\partial^2 \overline{\tau^{sgs}_{11}}}{\partial x_1^2}\hspace{3pt},
    \hspace{2em} G_{12} = 2\,\frac{\partial^2 \overline{\tau^{sgs}_{12}}}{\partial x_1\,\partial x_2}\hspace{3pt},
    \hspace{2em} G_{22} = \frac{\partial^2 \overline{\tau^{sgs}_{22}}}{\partial x_2^2}\hspace{3pt},
    \label{Prebud_eq_end}
}\end{equation}
{representing contributions to the mean pressure from the mean velocity field ($U_{ij}$), Reynolds stress ($W_{ij}$), and mean SGS stress ($G_{ij}$), respectively}.

\begin{figure}
\centering
\sidesubfloat[]{
{\psfrag{a}[][]{{}}
\psfrag{b}[][]{{$x_2/L$}}
\psfrag{c}[][]{{$U_{11}$}}
\psfrag{d}[][]{{$U_{12}$}}
\psfrag{e}[][]{{$U_{22}$}}
\psfrag{f}[][]{{$W_{11}$}}
\psfrag{g}[][]{{$W_{12}$}}
\psfrag{h}[][]{{$W_{22}$}}
\psfrag{i}[][]{{$G_{11}$}}
\psfrag{j}[][]{{$G_{12}$}}
\psfrag{k}[][]{{$G_{22}$}}
\includegraphics[width=.45\textwidth,trim={1cm 1.5cm 0.1cm 0.8cm},clip]{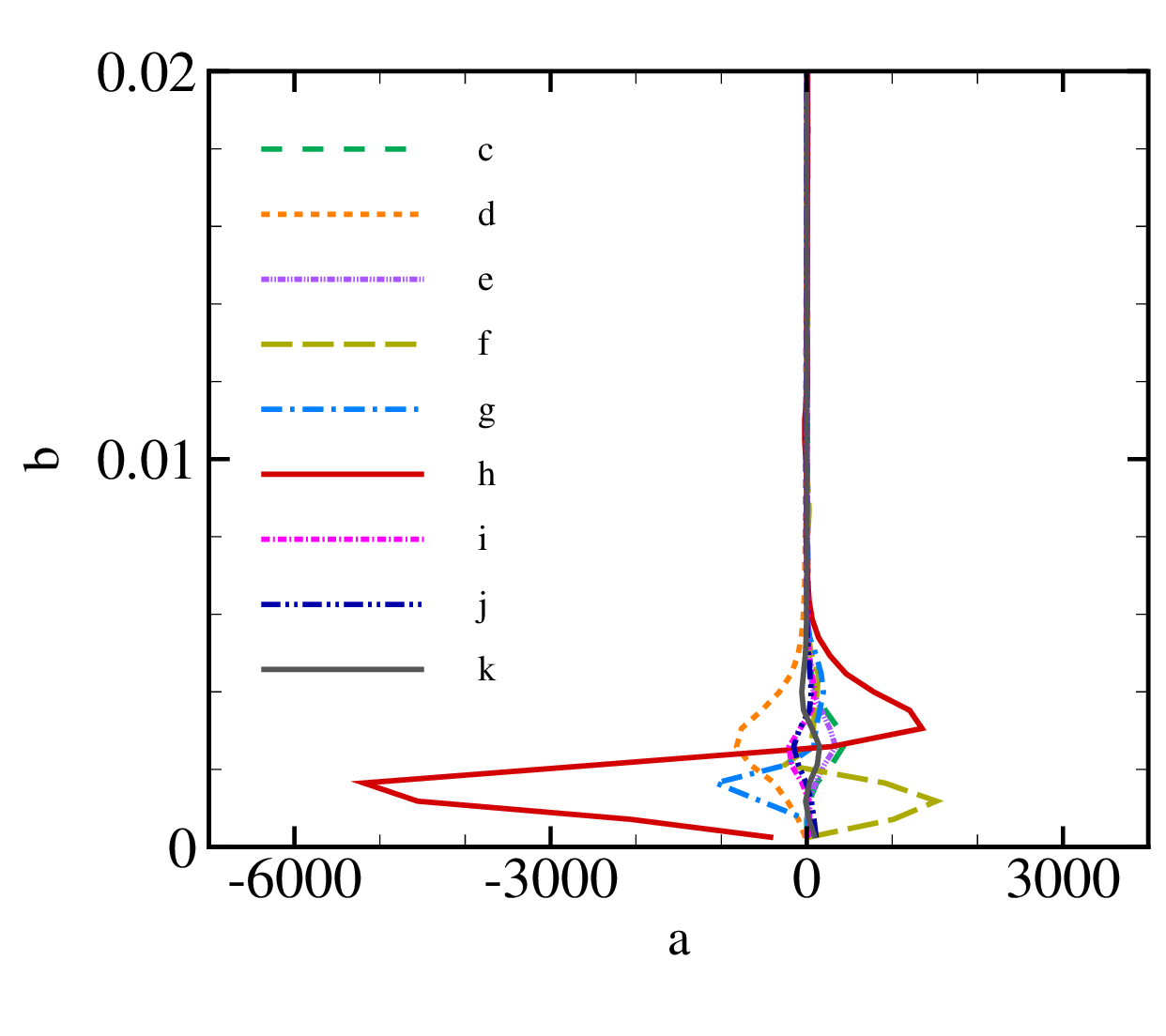}}
}
\sidesubfloat[]{
{\psfrag{a}[][]{{}}
\psfrag{b}[][]{{$x_2/L$}}
\includegraphics[width=.45\textwidth,trim={1cm 1.5cm 0.1cm 0.8cm},clip]{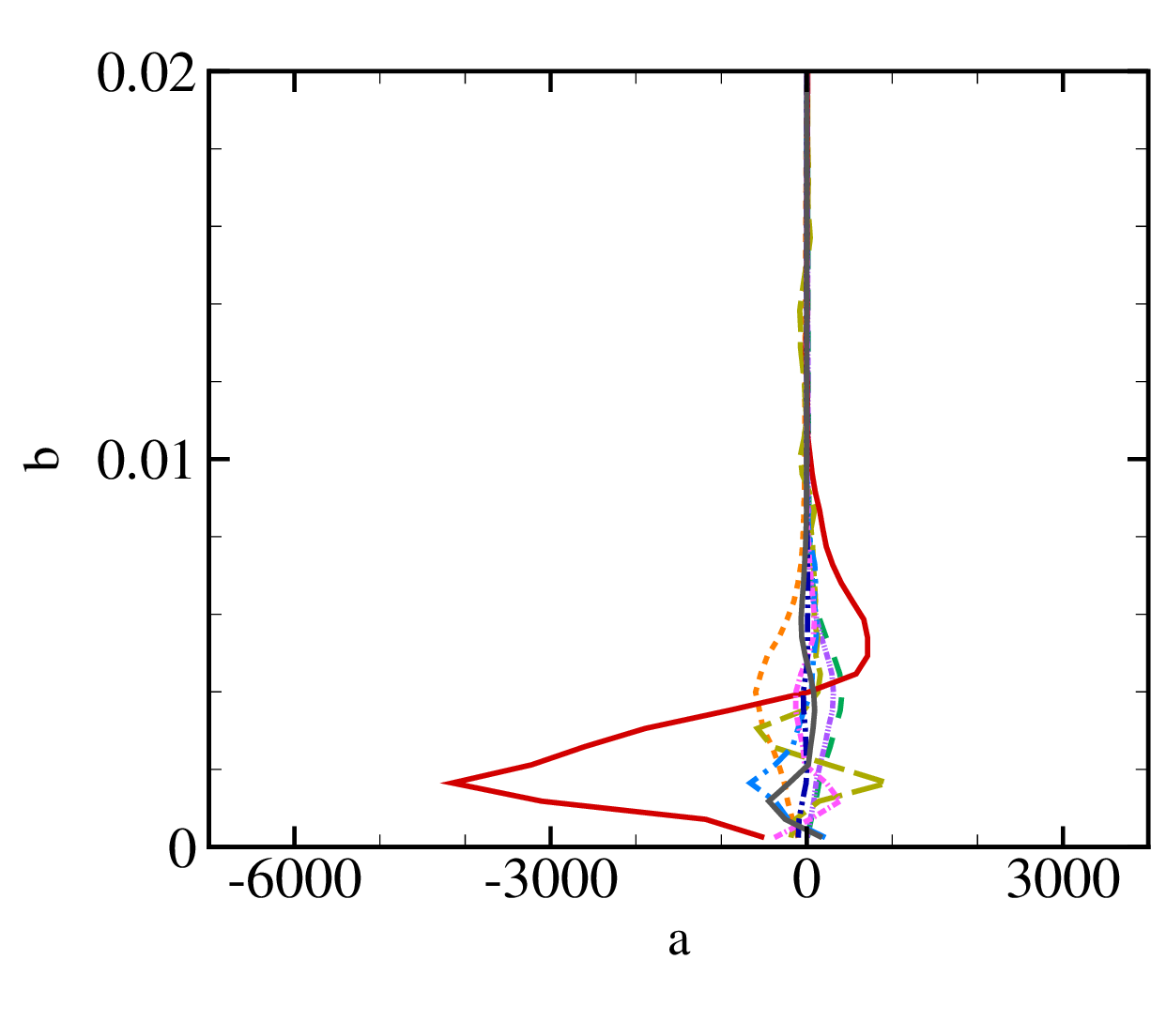}}
}
\caption{Mean pressure budget terms at $x/L=0.05$ from medium-mesh simulations with the SM (a) and MSM (b). All terms are nondimensionalized using $U_\infty$, $L$ and $\rho$. {The line notations correspond to equations~\eqref{Prebud_eq_begin}--\eqref{Prebud_eq_end}.}}
\label{Prebudget_medium_SGS}
\end{figure}

Figure~\ref{Prebudget_medium_SGS} shows profiles of all {nine} terms at $x/L=0.05$. The results are qualitatively similar for both SGS models. Within the TBL, the dominant term is $W_{22}$, associated with the wall-normal Reynolds normal stress $\overline{u'_2u'_2}$, indicating that wall-normal turbulent fluctuations are the primary contributors to the local mean pressure distribution.

Taken together with the momentum budget analysis, these results highlight a consistent picture: in the medium-mesh simulations, the Reynolds stresses, not the mean SGS stresses, govern both the mean velocity and mean pressure fields immediately upstream of the separation point. Although the two SGS models produce different downstream separation behaviour, those differences arise primarily from how each model shapes the Reynolds-stress distributions within the upstream FPG region. These stresses are then carried downstream by the mean flow and influence the momentum balance near the separation point. The mean SGS stresses, by comparison, make only a minor contribution at this location.

\subsection{Reynolds stress transport equation} \label{sec:Reynolds_budgets}

\begin{figure}
\centering

\sidesubfloat[]{
{\psfrag{a}[][]{{$\overline{u'_1u'_2}/U^2_\infty$}}
\psfrag{b}[][]{{$x_2/L$}}
\includegraphics[width=0.28\textwidth,trim={1.8cm 0.6cm 7.6cm 0.5cm},clip]{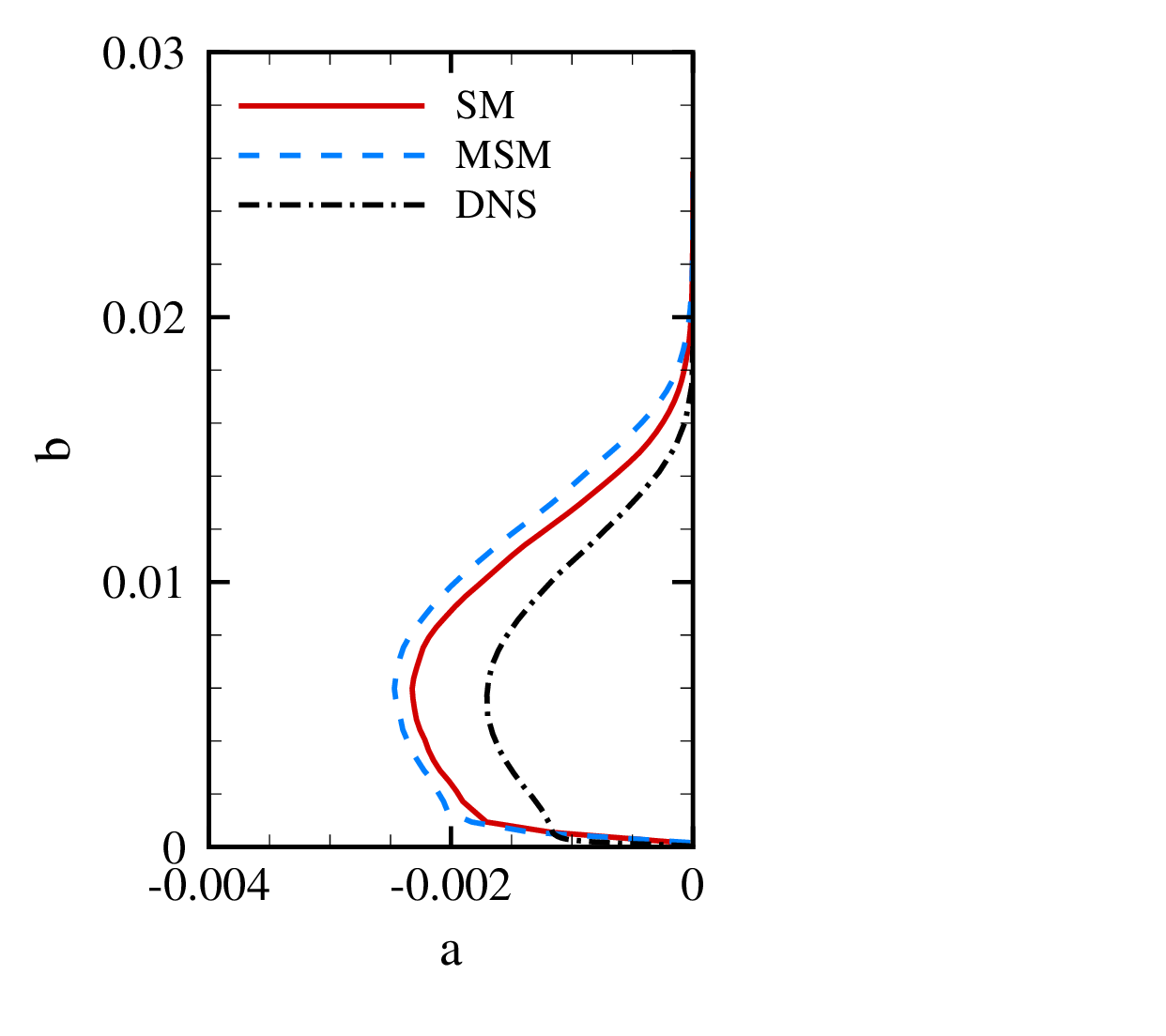}}}
\sidesubfloat[]{
{\psfrag{a}[][]{{$\overline{u'_1u'_2}/U^2_\infty$}}
\psfrag{b}[][]{{$x_2/L$}}
\includegraphics[width=0.28\textwidth,trim={1.8cm 0.6cm 7.6cm 0.5cm},clip]{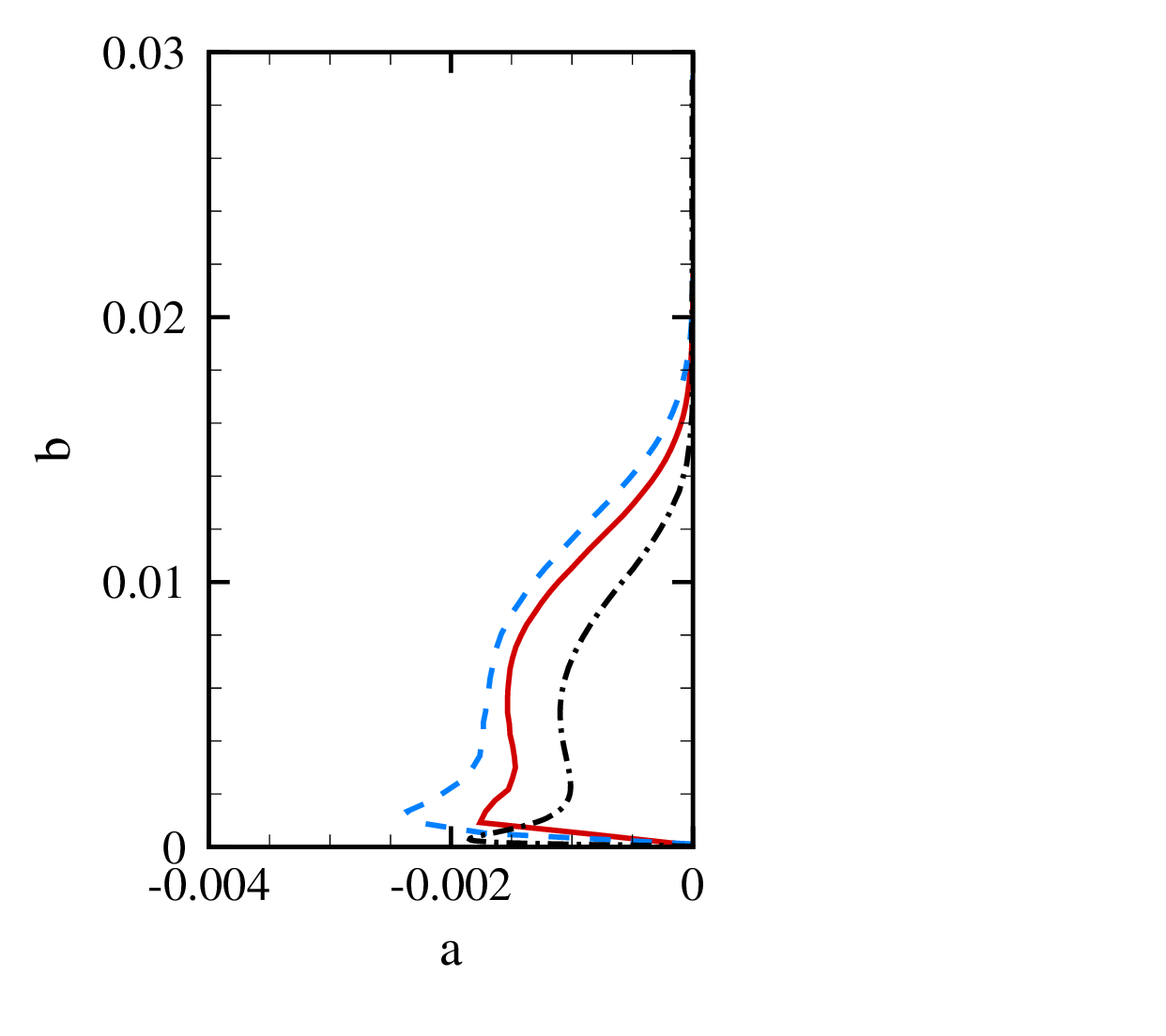}}}
\sidesubfloat[]{
{\psfrag{a}[][]{{$\overline{u'_1u'_2}/U^2_\infty$}}
\psfrag{b}[][]{{$x_2/L$}}
\includegraphics[width=0.28\textwidth,trim={1.8cm 0.6cm 7.6cm 0.5cm},clip]{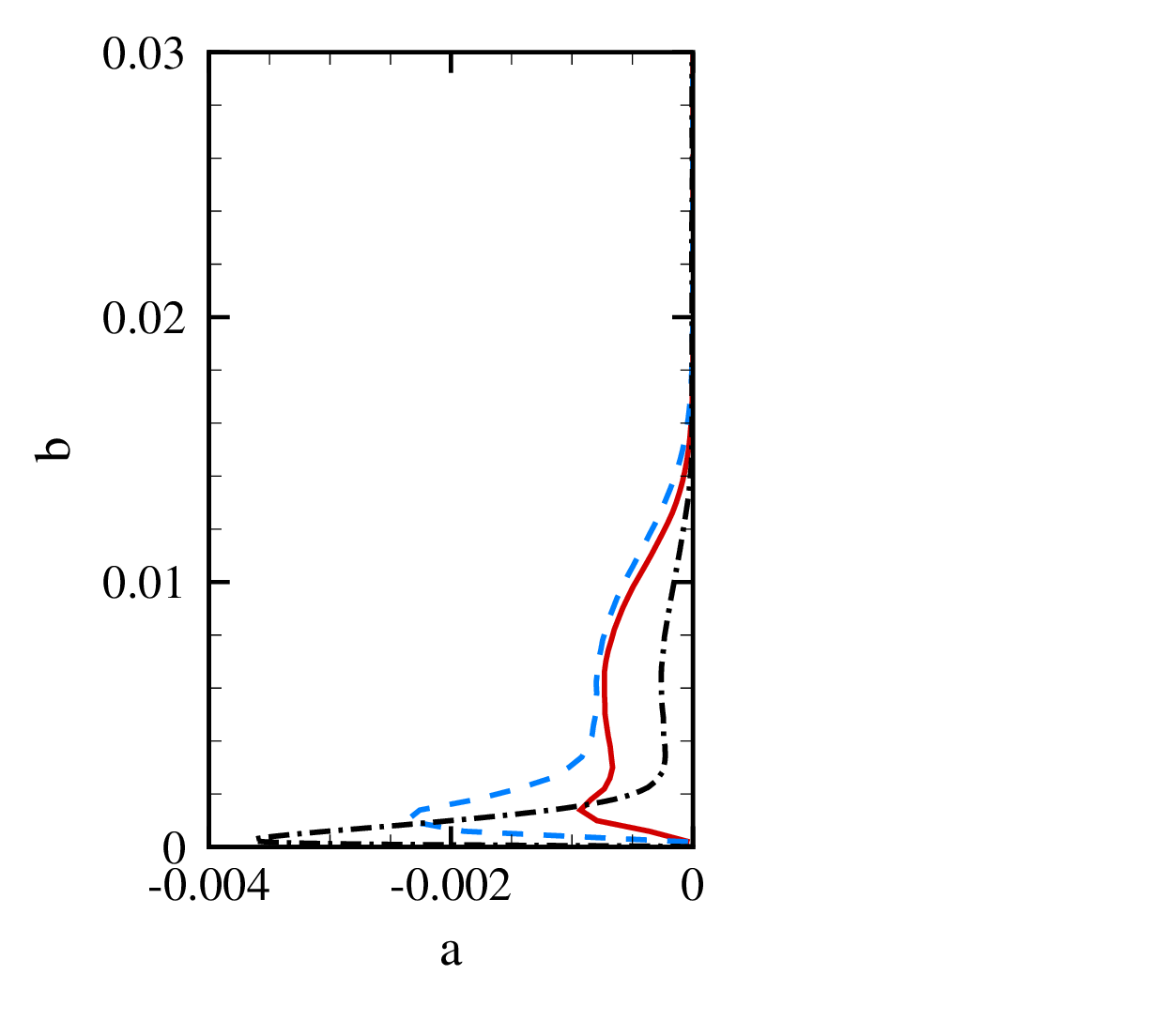}}}

\sidesubfloat[]{
{\psfrag{a}[][]{{$\overline{u'_2u'_2}/U^2_\infty$}}
\psfrag{b}[][]{{$x_2/L$}}
\includegraphics[width=0.28\textwidth,trim={1.8cm 0.6cm 7.6cm 0.5cm},clip]{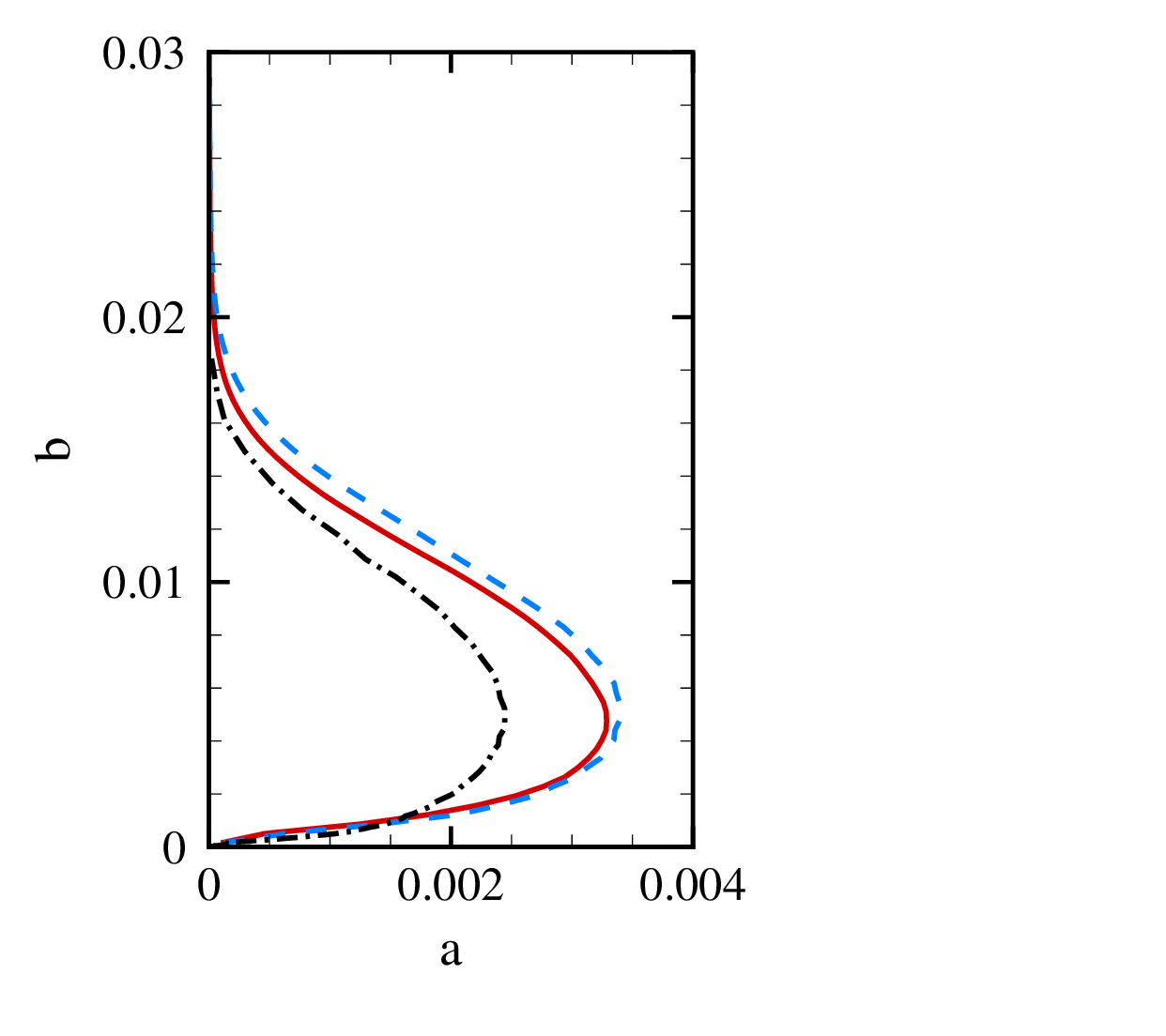}}}
\sidesubfloat[]{
{\psfrag{a}[][]{{$\overline{u'_2u'_2}/U^2_\infty$}}
\psfrag{b}[][]{{$x_2/L$}}
\includegraphics[width=0.28\textwidth,trim={1.8cm 0.6cm 7.6cm 0.5cm},clip]{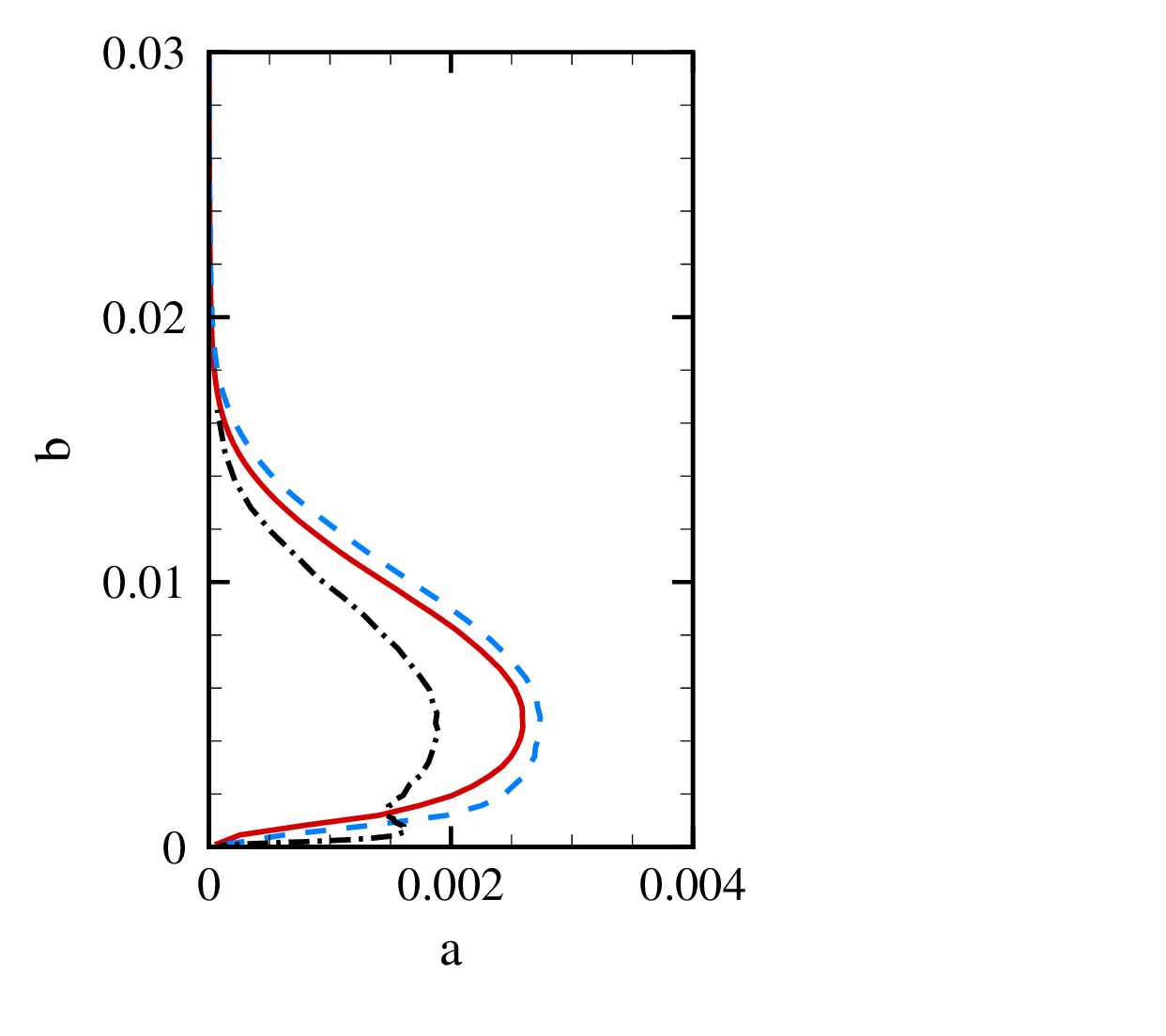}}}
\sidesubfloat[]{
{\psfrag{a}[][]{{$\overline{u'_2u'_2}/U^2_\infty$}}
\psfrag{b}[][]{{$x_2/L$}}
\includegraphics[width=0.28\textwidth,trim={1.8cm 0.6cm 7.6cm 0.5cm},clip]{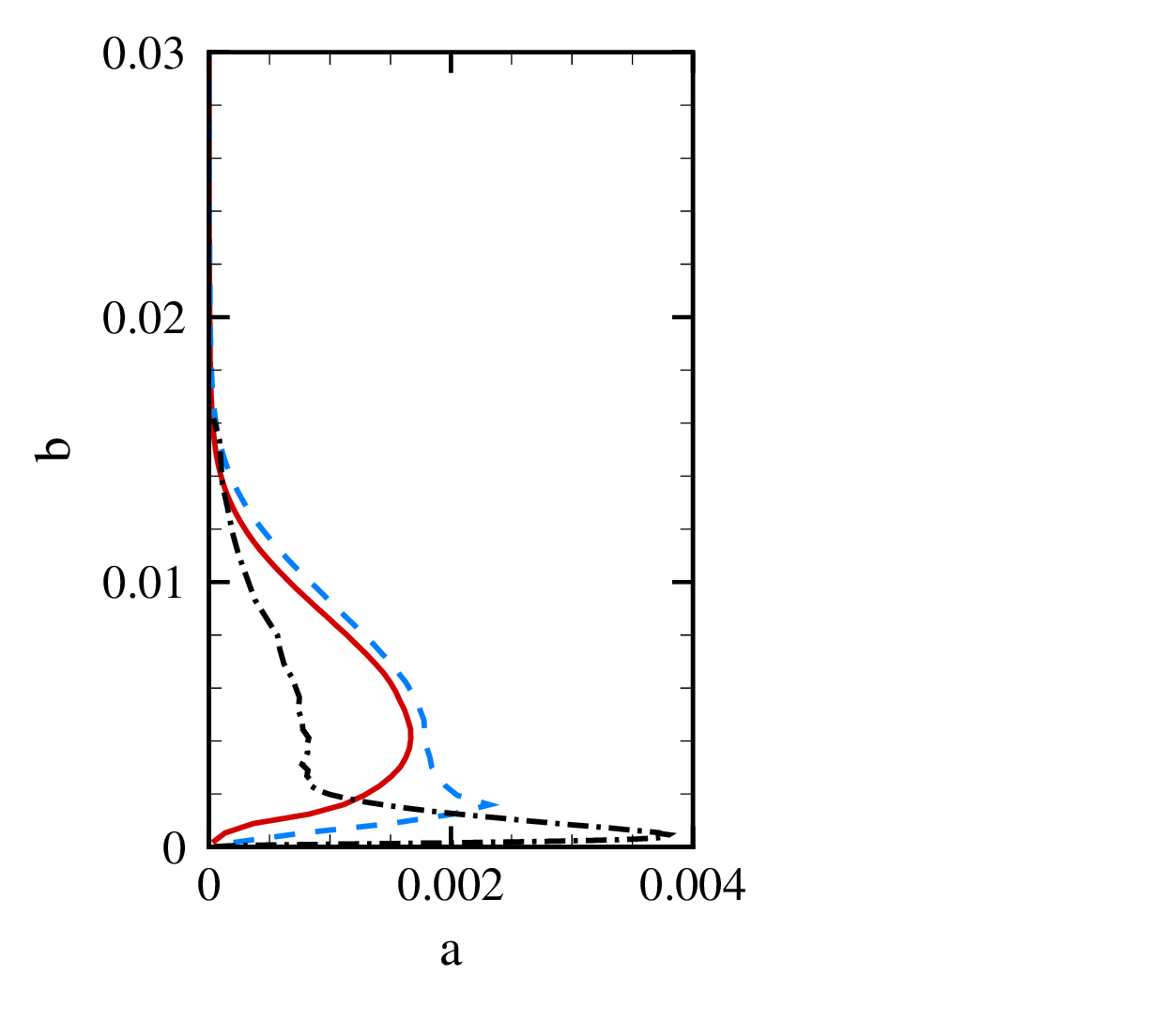}}}
\caption{Reynolds shear stress $\overline{u'_1u'_2}$ (a-c) and Reynolds normal stress $\overline{u'_2u'_2}$ (d-f) profiles at $x/L = -0.2$ (a,d), $x/L = -0.1$ (b,e), and $x/L = 0$ (c,f) for SM, MSM and reference DNS \citep{uzun2022high}.}
\label{R12R22_bump}
\end{figure}

Based on the mean momentum and pressure budget analyses, we examine the Reynolds shear stress $\overline{u'_1u'_2}$ and Reynolds normal stress $\overline{u'_2u'_2}$ at three streamwise locations within the FPG region upstream of the bump peak. Figure~\ref{R12R22_bump} shows wall-normal profiles from the two medium-mesh simulations along with DNS results \citep{uzun2022high}.

The DNS data show that the magnitudes of both stresses gradually increase in the near-wall region and exhibit distinct internal peaks within the FPG region. Both medium-mesh simulations deviate noticeably from the DNS, overpredicting Reynolds stresses in the outer layer. {This is partly attributable to the use of constant, non-optimized model coefficients and the inability of the coarse mesh to resolve near-wall turbulent structures, which can lead to an imbalance in the resolved turbulence, a behaviour widely reported in WMLES \citep{bae2018, rezaeiravesh2019systematic, lozano2019error}}. The most prominent distinction between the two simulations is that the MSM reproduces clear internal peaks of $\overline{u'_1u'_2}$ and $\overline{u'_2u'_2}$ at $x_2/L \approx 1.3\times10^{-3}$, slightly above the DNS location, while the SM does not capture these features. This upward shift is partly due to the coarse mesh resolution.

As shown in \Cref{sec:mean_budgets}, the wall-normal gradient of $\overline{u'_1u'_2}$ shapes mean momentum transport while that of $\overline{u'_2u'_2}$ governs the mean pressure distribution. Accurately capturing the wall-normal variation of these stresses is therefore essential for predicting downstream separation. The improved prediction of Reynolds stress profiles in the MSM, particularly the internal stress peaks under FPG conditions, demonstrates the benefit of incorporating anisotropic SGS stress. As discussed by \citet{uzun2022high} and consistent with the analysis in \Cref{sec:identification}, these internal peaks evolve downstream and strongly influence the mean flow and separation onset. Taken together, the results indicate that the improved separation prediction in the medium-mesh simulation with MSM is linked to its ability to better reproduce Reynolds-stress distributions, particularly the internal peak in the FPG region.

To understand how the SGS model influences these Reynolds-stress distributions, we analyse the Reynolds stress transport equation \citep{meneveau1994statistics, abe2019notable, inagaki2023analysis},
\begin{equation}
 \frac{\partial  \overline{ u_i'u_j'} }{\partial  t} +
\overline{u}_k\frac{\partial  \overline{ u_i'u_j'}
}{\partial  x_k} = P_{ij}-\varepsilon_{ij}+\phi_{ij}+\xi_{ij} +
\frac{\partial }{\partial
x_k}\Big(\zeta_{ijk}+D_{ijk}+T_{ijk}+J_{ijk}\Big) \hspace{3pt},
\label{Re_term_all}
\end{equation}
where
\begin{equation}
    P_{ij} = -\overline{ u_i' u_k' } \frac{\partial
\overline{u}_j}{\partial  x_k} - \overline{ u_j'
u_k'}\frac{\partial  \overline{u}_i}{\partial  x_k}\hspace{3pt}, 
\label{Re_term_first}
\end{equation}

\begin{equation}
    \varepsilon_{ij} = 2\nu \overline{ \Big( S_{ik}' \frac{\partial
u_j'}{\partial  x_k}+ S_{jk}' \frac{\partial  u_i'}{\partial
x_k}\Big)} = 2\nu \overline{\Big( \frac{\partial  u_i'}{\partial  x_k}
\frac{\partial  u_j'}{\partial  x_k}\Big )}\hspace{3pt}, 
\end{equation}

\begin{equation}
\phi_{ij} = \overline{\frac{p'}{\rho} \cdot \left(\frac{\partial
u_i'}{\partial  x_j} + \frac{\partial  u_j'}{\partial  x_i}\right)
} = 2\overline{\frac{p'}{\rho} S_{ij}'}\hspace{3pt},
\end{equation}

\begin{equation}
    \xi_{ij} = \overline{ (\tau^{\text{sgs}}_{ik})' \frac{\partial  u_j'}{\partial x_k} + (\tau^{\text{sgs}}_{jk})' \frac{\partial  u_i'}{\partial  x_k}}\hspace{3pt}, 
\label{EQ:SGS_diss}    
\end{equation}

\begin{equation}
    \frac{\partial }{\partial  x_k}\zeta_{ijk} = -\frac{\partial
}{\partial  x_k}\left[\overline{ (\tau^{\text{sgs}}_{ik})' u_j'}  + \overline{
(\tau^{\text{sgs}}_{jk})' u_i'} \right]\hspace{3pt},
\label{EQ:SGS_diff}
\end{equation}

\begin{equation}
 \frac{\partial }{\partial  x_k} D_{ijk} = \nu \frac{\partial ^2
\overline{ u_i' u_j'} }{\partial  x_k^2} = 2\nu \frac{\partial
}{\partial  x_k}\left(\overline{ S_{ik}'u_j'}  + \overline{
S_{jk}'u_i'} \right)\hspace{3pt}, 
\end{equation}

\begin{equation}
    \frac{\partial }{\partial  x_k} T_{ijk} = -\frac{\partial
}{\partial  x_k}\overline{ u_i' u_j' u_k'}\hspace{3pt},
\end{equation}

\begin{equation}
    \frac{\partial }{\partial  x_k} J_{ijk} = -\frac{\partial
}{\partial  x_k}\left(\overline{ p'u_i'} \delta_{jk} + \overline{p'u_j'} \delta_{ik}\right) \hspace{3pt},
\label{Re_term_final}
\end{equation}
corresponding to production, viscous dissipation, pressure strain, SGS dissipation, SGS diffusion, viscous diffusion, turbulent diffusion, and pressure diffusion, respectively.

The SGS-related terms in equations~\eqref{EQ:SGS_diss} and \eqref{EQ:SGS_diff} directly represent the contributions from the SGS model. {Although they are generally smaller in magnitude than leading-order terms such as production and pressure strain (see Appendix~\ref{appC_add}), they remain physically meaningful}. The SGS dissipation represents local transfer of resolved-scale energy to unresolved scales, while SGS diffusion represents spatial redistribution. {Crucially, differences in these terms between the two models modify the evolution of the resolved Reynolds stresses, which in turn drives changes in production, pressure strain, and other terms through the coupled governing equations. The differences observed in the remaining budget terms between the SM and MSM can therefore be interpreted as a response of the resolved flow to different SGS forcing. The following analysis focuses on these SGS terms.}

\begin{figure}
\centering
\sidesubfloat[]{
{\psfrag{a}[][]{{$\xi_{12}, \frac{\partial }{\partial  x_k}\zeta_{12k}$}}
\psfrag{b}[][]{{$x_2/L$}}
\includegraphics[width=.28\textwidth,trim={1cm 1.3cm 7.5cm 0.5cm},clip]{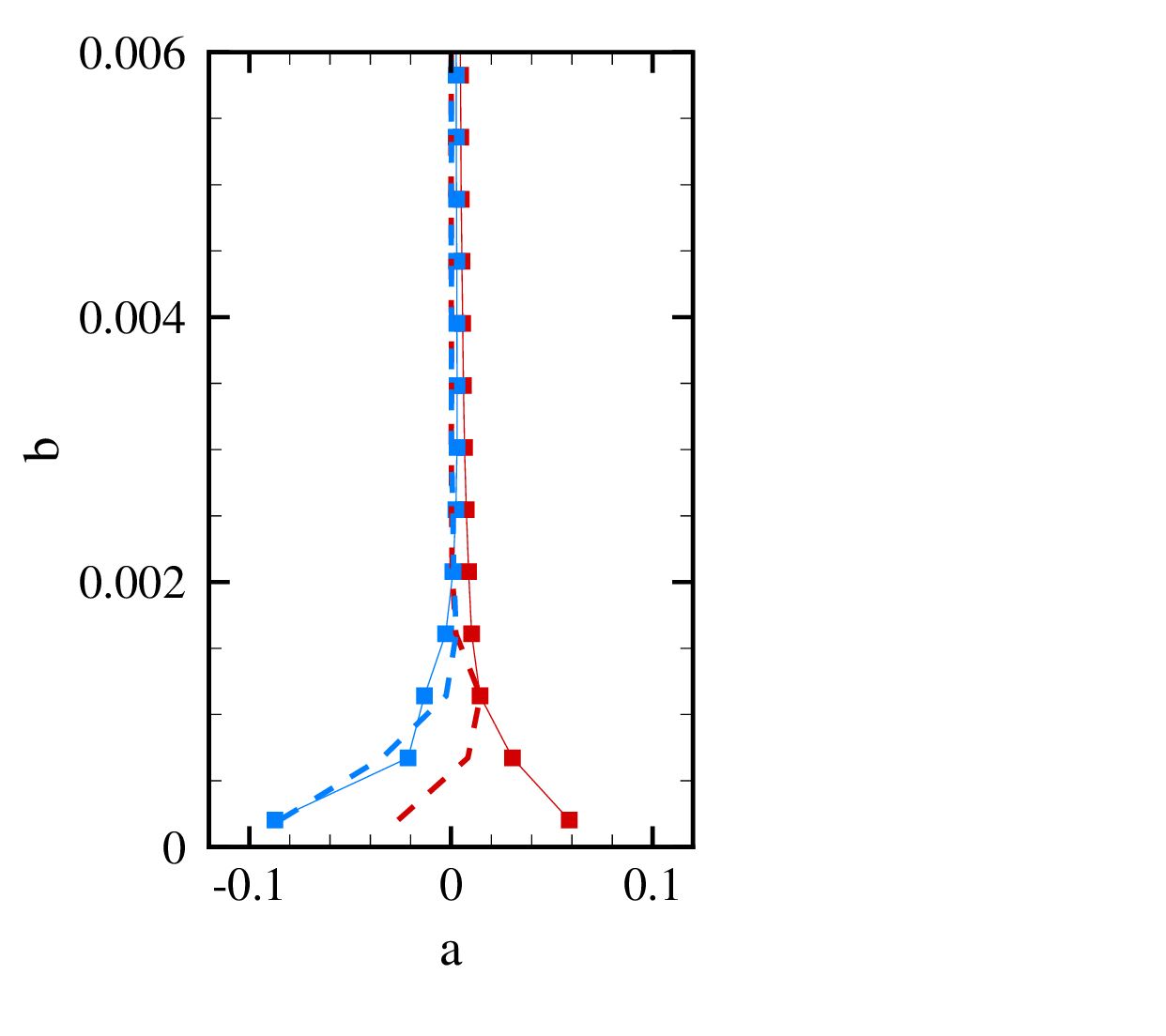}}
}
\sidesubfloat[]{
{\psfrag{a}[][]{{$\xi_{12}, \frac{\partial }{\partial  x_k}\zeta_{12k}$}}
\psfrag{b}[][]{{$x_2/L$}}
\includegraphics[width=.28\textwidth,trim={1cm 1.3cm 7.5cm 0.5cm},clip]{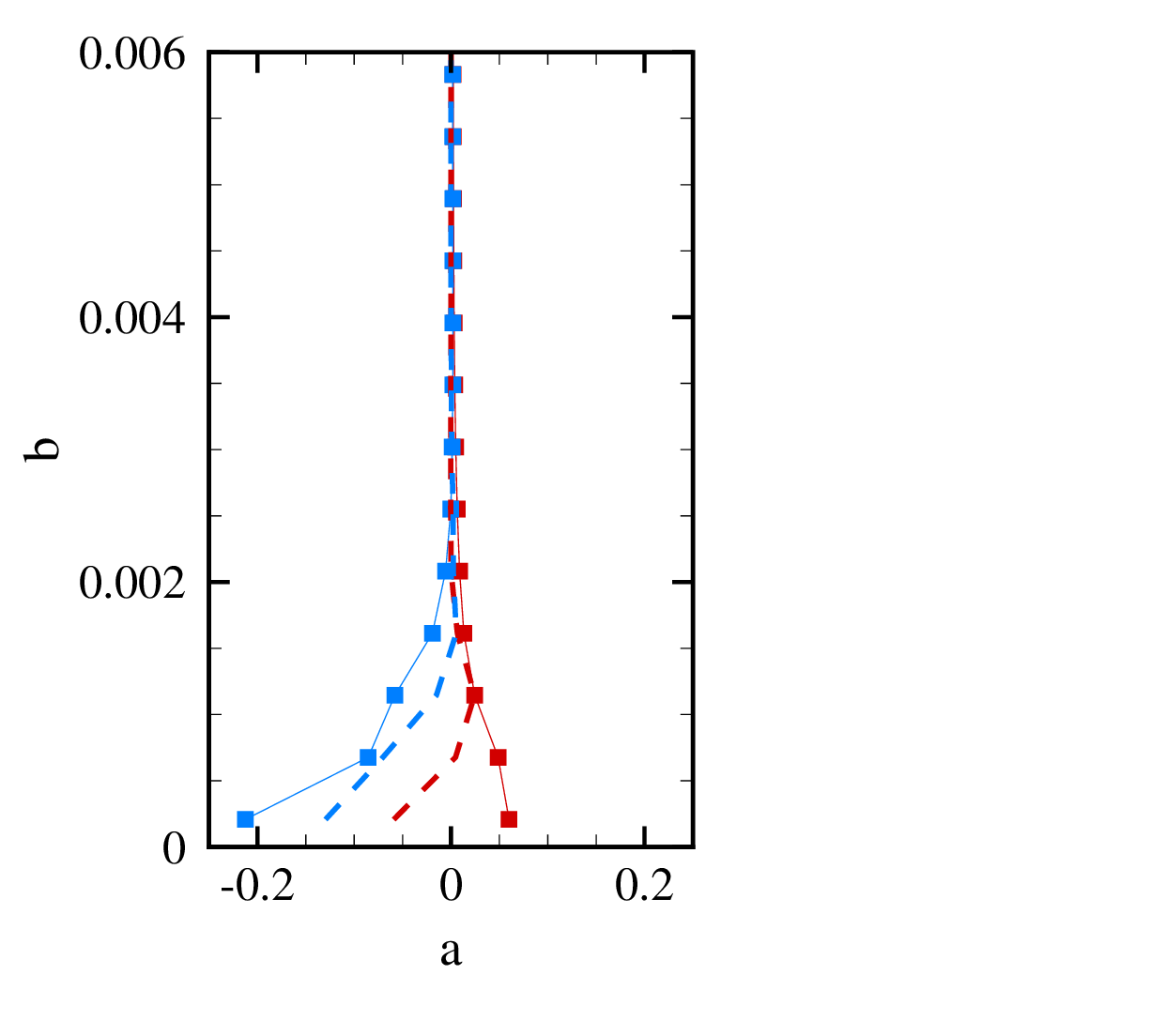}}
}
\sidesubfloat[]{
{\psfrag{a}[][]{{$\xi_{12}, \frac{\partial }{\partial  x_k}\zeta_{12k}$}}
\psfrag{b}[][]{{$x_2/L$}}
\includegraphics[width=.28\textwidth,trim={1cm 1.3cm 7.5cm 0.5cm},clip]{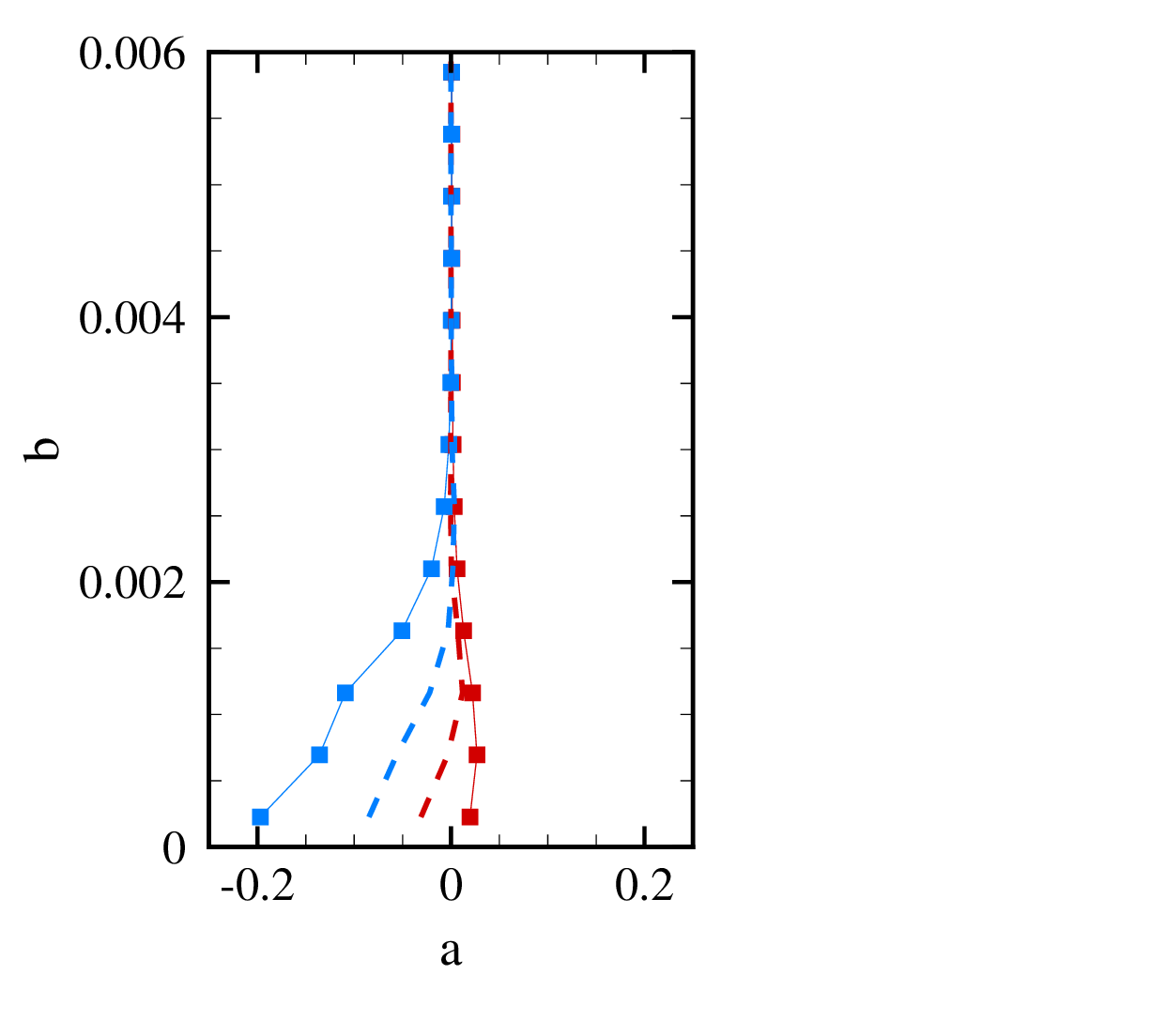}}
}
\vspace{5pt}
\sidesubfloat[]{
{\psfrag{a}[][]{{$\xi_{22}, \frac{\partial }{\partial  x_k}\zeta_{22k}$}}
\psfrag{b}[][]{{$x_2/L$}}
\includegraphics[width=.28\textwidth,trim={1cm 1.3cm 7.5cm 0.5cm},clip]{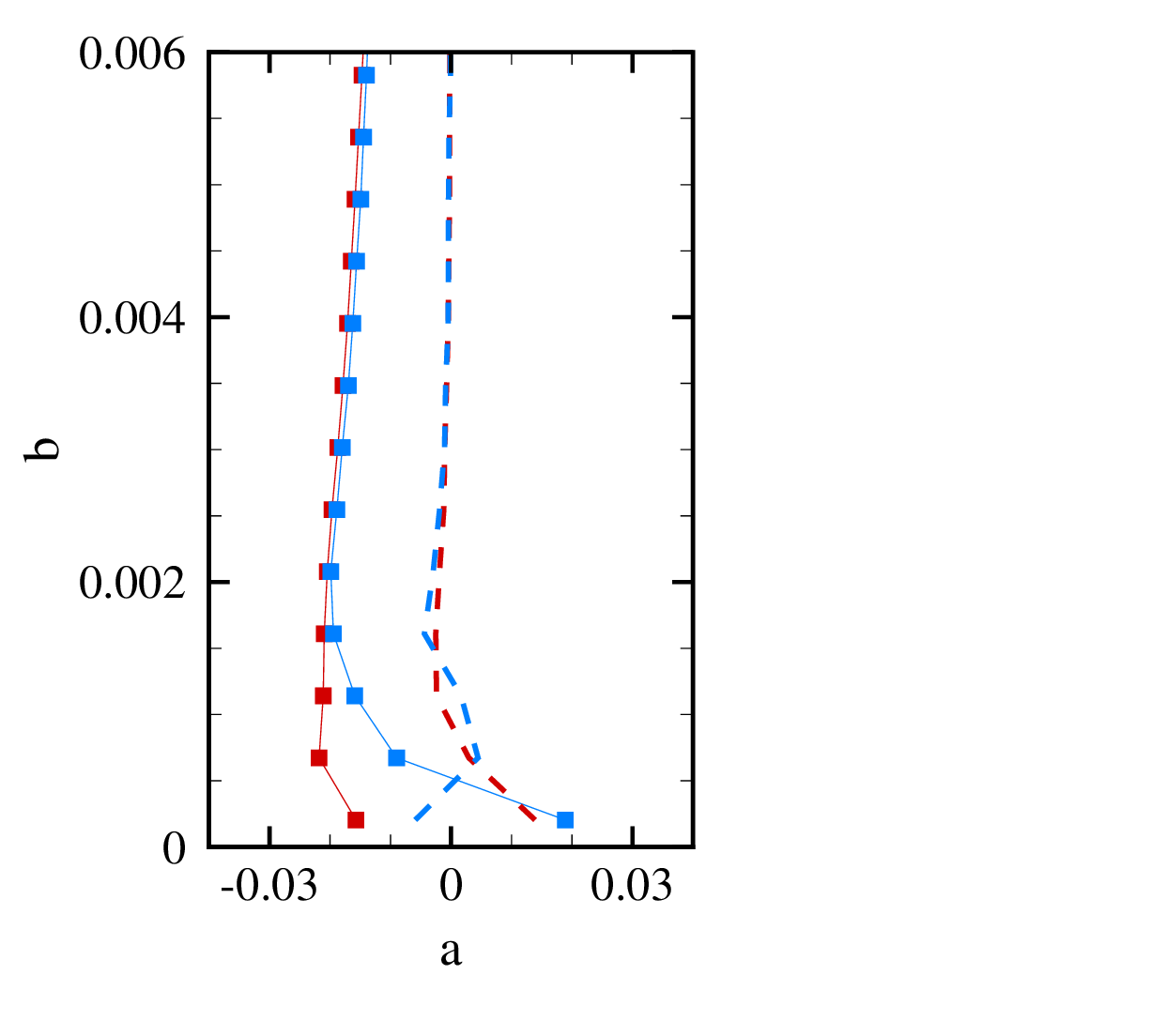}}
}
\sidesubfloat[]{
{\psfrag{a}[][]{{$\xi_{22}, \frac{\partial }{\partial  x_k}\zeta_{22k}$}}
\psfrag{b}[][]{{$x_2/L$}}
\includegraphics[width=.28\textwidth,trim={1cm 1.3cm 7.5cm 0.5cm},clip]{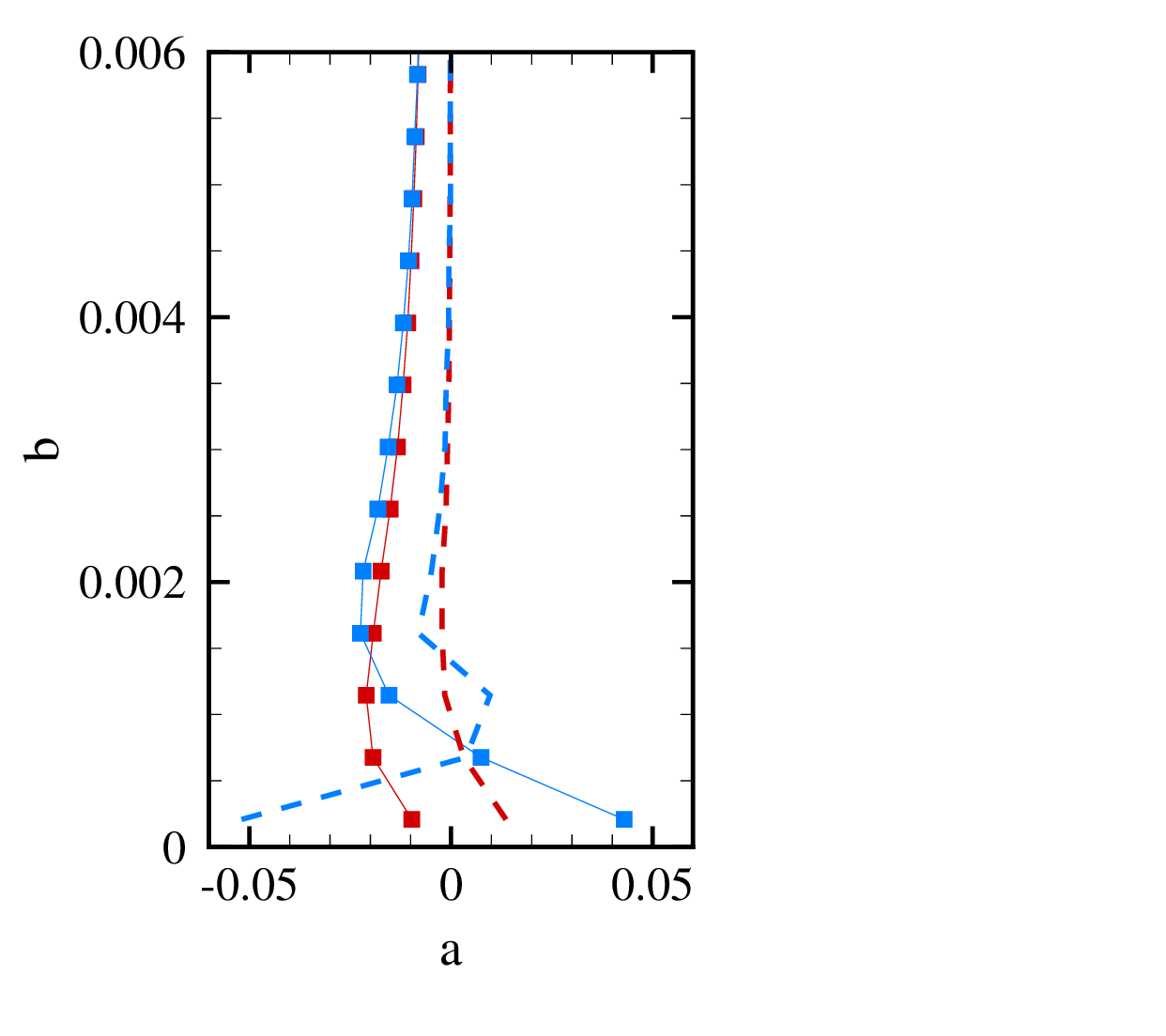}}
}
\sidesubfloat[]{
{\psfrag{a}[][]{{$\xi_{22}, \frac{\partial }{\partial  x_k}\zeta_{22k}$}}
\psfrag{b}[][]{{$x_2/L$}}
\includegraphics[width=.28\textwidth,trim={1cm 1.3cm 7.5cm 0.5cm},clip]{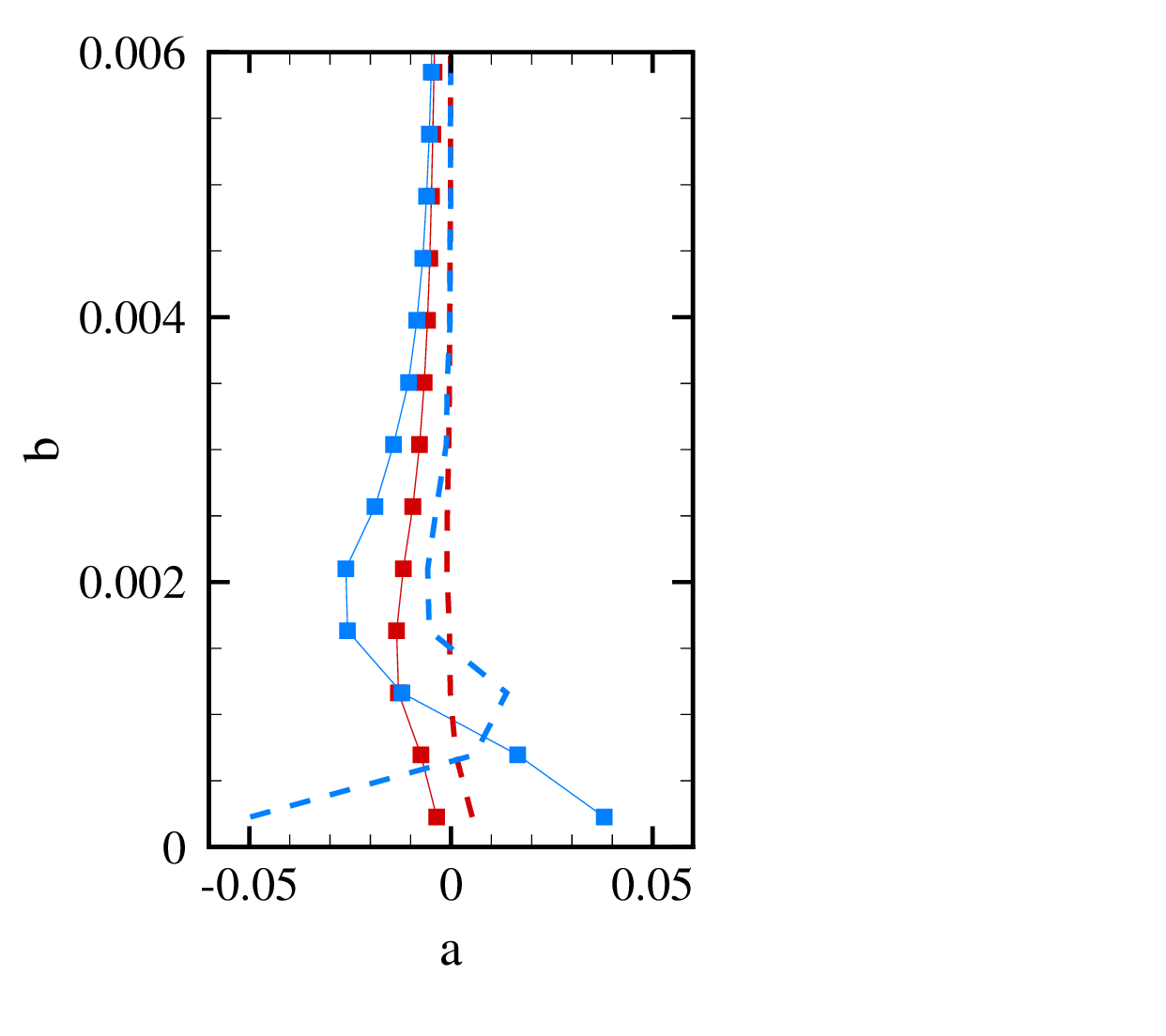}}
}
\caption{SGS dissipation $\xi_{ij}$ (solid square) and diffusion $\frac{\partial }{\partial  x_k}\zeta_{ijk}$ (dashed) for the Reynolds shear stress $\overline{u_1'u_2'}$ (a–c) and the wall-normal Reynolds normal stress $\overline{u_2'u_2'}$ (d–f) at $x/L=-0.2$ (a,d), $x/L=-0.1$ (b,e), and $x/L=0$ (c,f) for SM (red) and MSM (blue) simulations on the medium-mesh. All terms are nondimensionalized using $U_\infty$, $L$ and $\rho$.}
\label{Reynolds_SGSDp_compare}
\end{figure}

Figure~\ref{Reynolds_SGSDp_compare} shows the SGS dissipation and diffusion from the medium-mesh simulations at the three streamwise locations. The SM and MSM produce qualitatively different behaviours for both $\overline{u_1'u_2'}$ and $\overline{u_2'u_2'}$. In the SM, SGS dissipation remains positive for $\overline{u_1'u_2'}$ and negative for $\overline{u_2'u_2'}$, consistently acting as a sink of resolved energy. In contrast, the MSM yields negative SGS dissipation for $\overline{u_1'u_2'}$ in the inner layer and positive values for $\overline{u_2'u_2'}$ very near the wall, indicating net backscatter, a transfer of energy from unresolved to resolved scales that locally enhances the Reynolds stresses. This backscatter is physically significant: on a coarse mesh, the cutoff lies in the energy-containing range where unresolved motions carry substantial turbulent energy and momentum fluxes, and the near-wall cycle involves essential small-scale dynamics that become especially important under pressure gradients. A coarse-mesh LES omitting these near-wall scales loses an important pathway by which energy is both removed from and returned to the larger scales.

The SGS diffusion results reveal further important differences. For $\overline{u_1'u_2'}$, the SM produces negative diffusion near the wall that becomes positive farther away, implying a redistribution of energy from the outer part of the inner layer toward the near-wall region. The MSM yields negative diffusion throughout, continuously moving $\overline{u_1'u_2'}$ toward the wall. For $\overline{u_2'u_2'}$, the SM gives positive near-wall diffusion decreasing with distance, redistributing energy from the outer region toward the wall. The MSM shows negative diffusion very close to the wall, positive values near the internal peak, and negative values farther out, moving $\overline{u_2'u_2'}$ into the internal-peak region from both sides. These SGS diffusion behaviours, combined with the differences in SGS dissipation, help explain why the MSM produces the internal Reynolds-stress peaks in figure~\ref{R12R22_bump} whereas the SM does not.

\begin{figure}
\centering
\sidesubfloat[]{
{\psfrag{a}[][]{{$\xi_{12}, \frac{\partial }{\partial  x_k}\zeta_{12k}$}}
\psfrag{b}[][]{{$x_2/L$}}
\includegraphics[width=.28\textwidth,trim={1cm 1.3cm 7.5cm 0.5cm},clip]{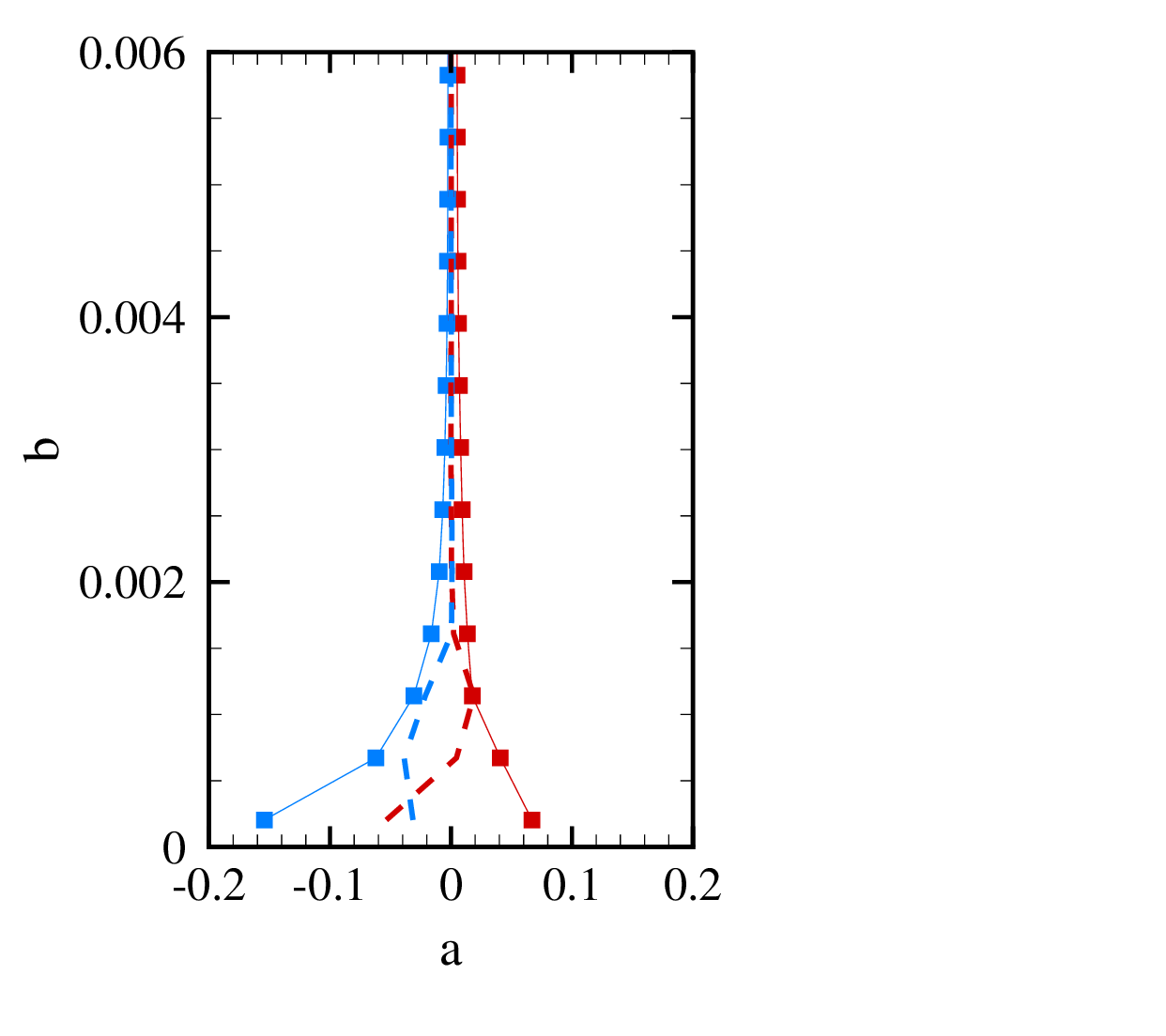}}
}
\sidesubfloat[]{
{\psfrag{a}[][]{{$\xi_{12}, \frac{\partial }{\partial  x_k}\zeta_{12k}$}}
\psfrag{b}[][]{{$x_2/L$}}
\includegraphics[width=.28\textwidth,trim={1cm 1.3cm 7.5cm 0.5cm},clip]{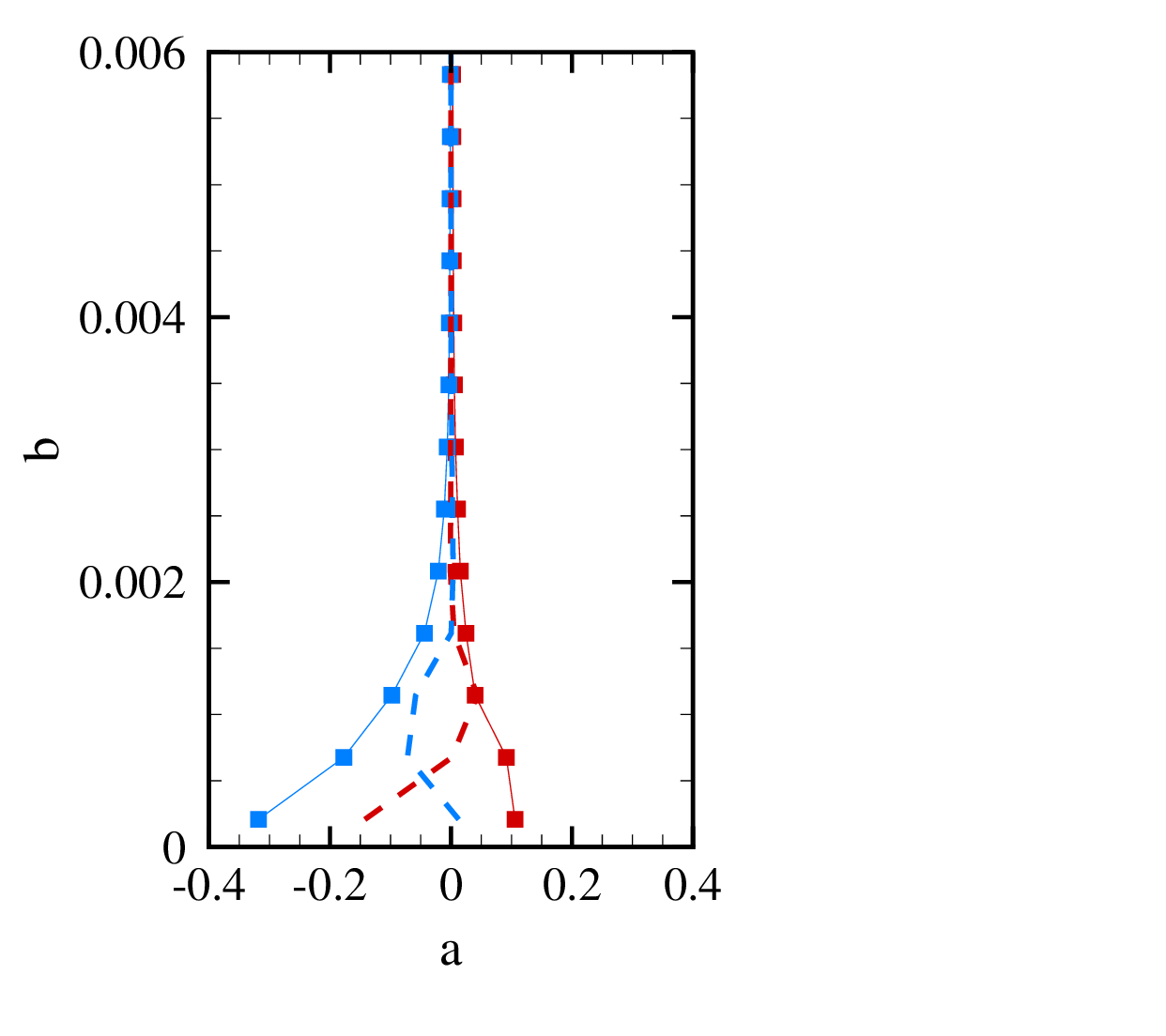}}
}
\sidesubfloat[]{
{\psfrag{a}[][]{{$\xi_{12}, \frac{\partial }{\partial  x_k}\zeta_{12k}$}}
\psfrag{b}[][]{{$x_2/L$}}
\includegraphics[width=.28\textwidth,trim={1cm 1.3cm 7.5cm 0.5cm},clip]{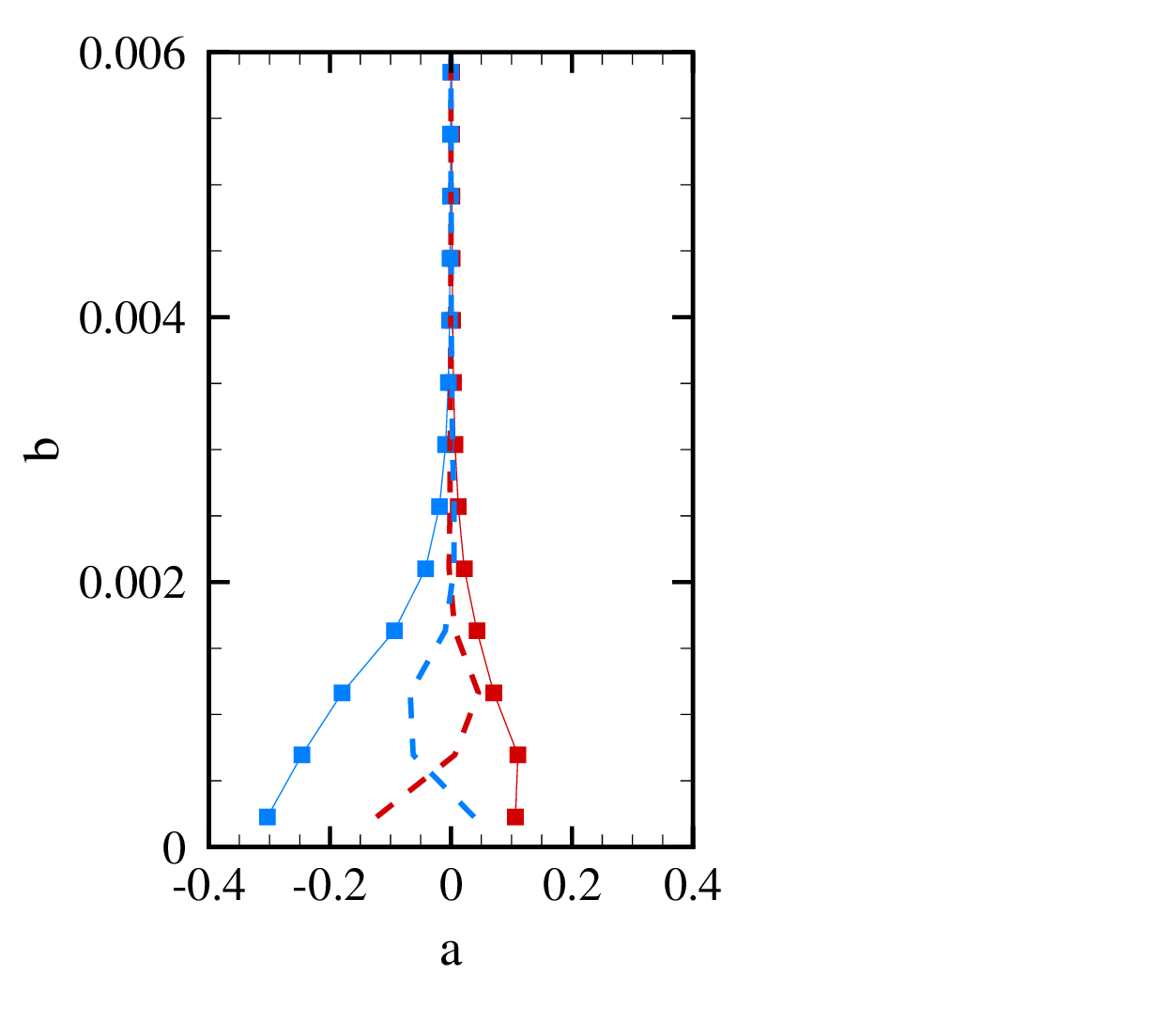}}
}
\vspace{5pt}
\sidesubfloat[]{
{\psfrag{a}[][]{{$\xi_{22}, \frac{\partial }{\partial  x_k}\zeta_{22k}$}}
\psfrag{b}[][]{{$x_2/L$}}
\includegraphics[width=.28\textwidth,trim={1cm 1.3cm 7.5cm 0.5cm},clip]{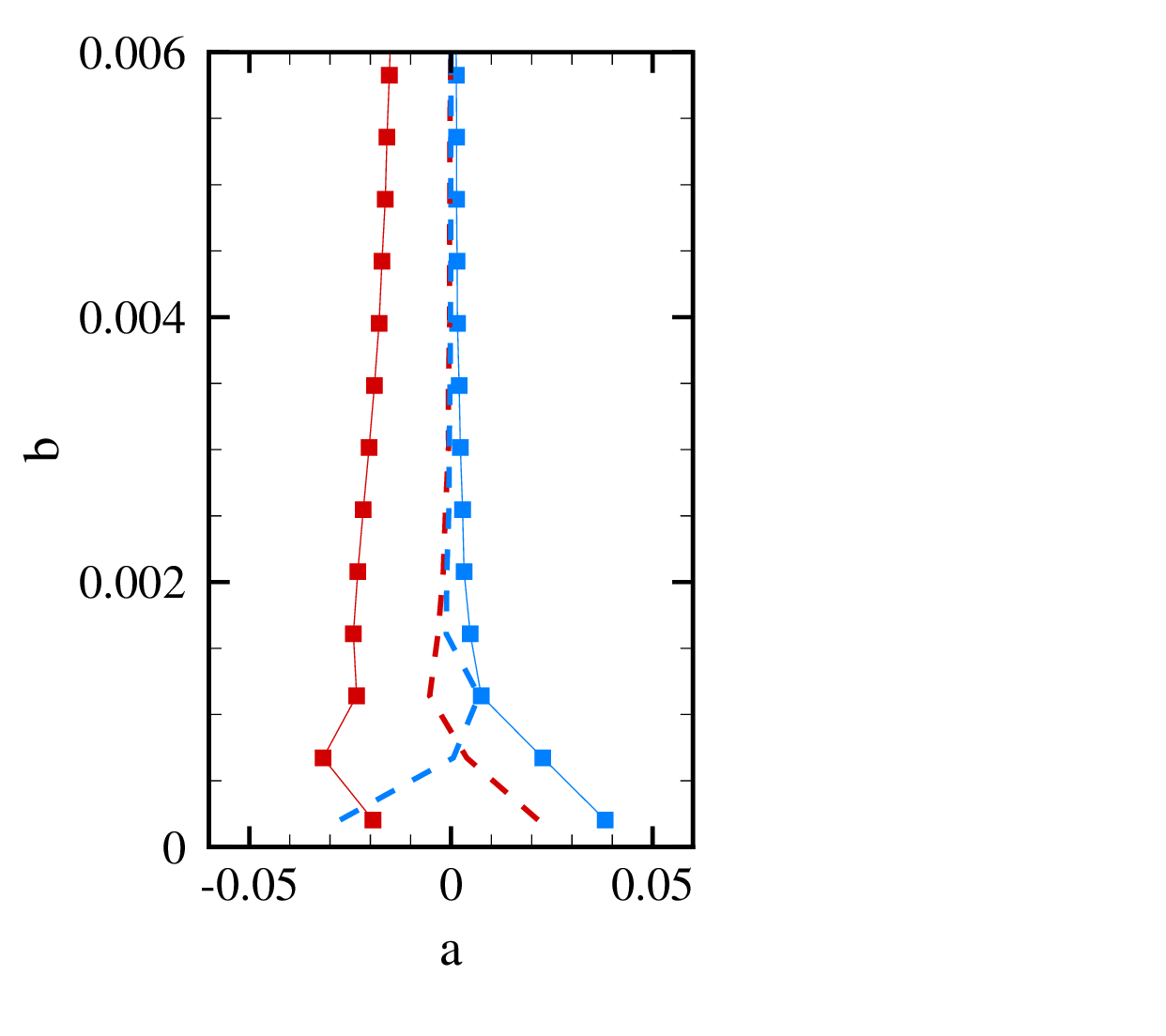}}
}
\sidesubfloat[]{
{\psfrag{a}[][]{{$\xi_{22}, \frac{\partial }{\partial  x_k}\zeta_{22k}$}}
\psfrag{b}[][]{{$x_2/L$}}
\includegraphics[width=.28\textwidth,trim={1cm 1.3cm 7.5cm 0.5cm},clip]{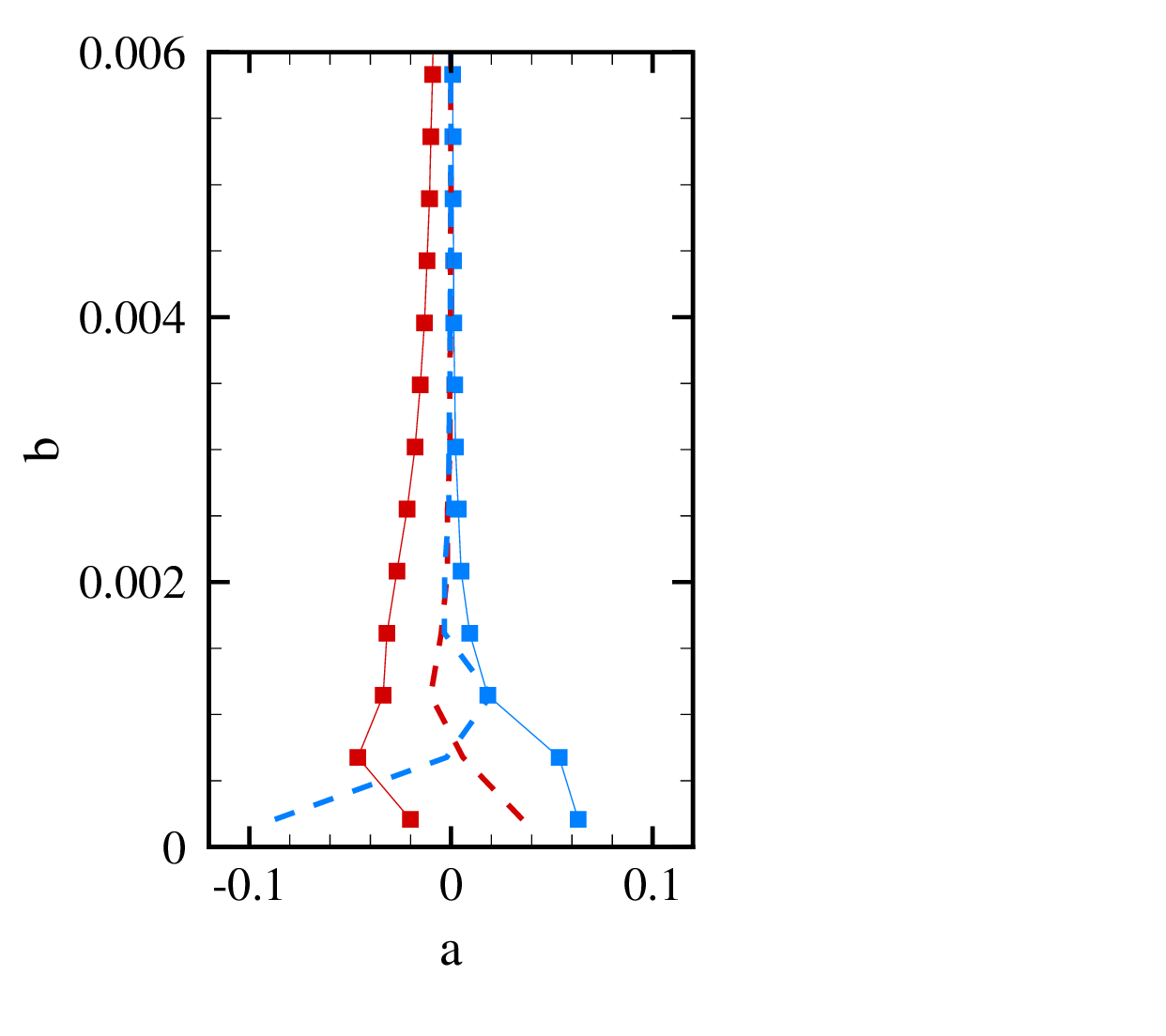}}
}
\sidesubfloat[]{
{\psfrag{a}[][]{{$\xi_{22}, \frac{\partial }{\partial  x_k}\zeta_{22k}$}}
\psfrag{b}[][]{{$x_2/L$}}
\includegraphics[width=.28\textwidth,trim={1cm 1.3cm 7.5cm 0.5cm},clip]{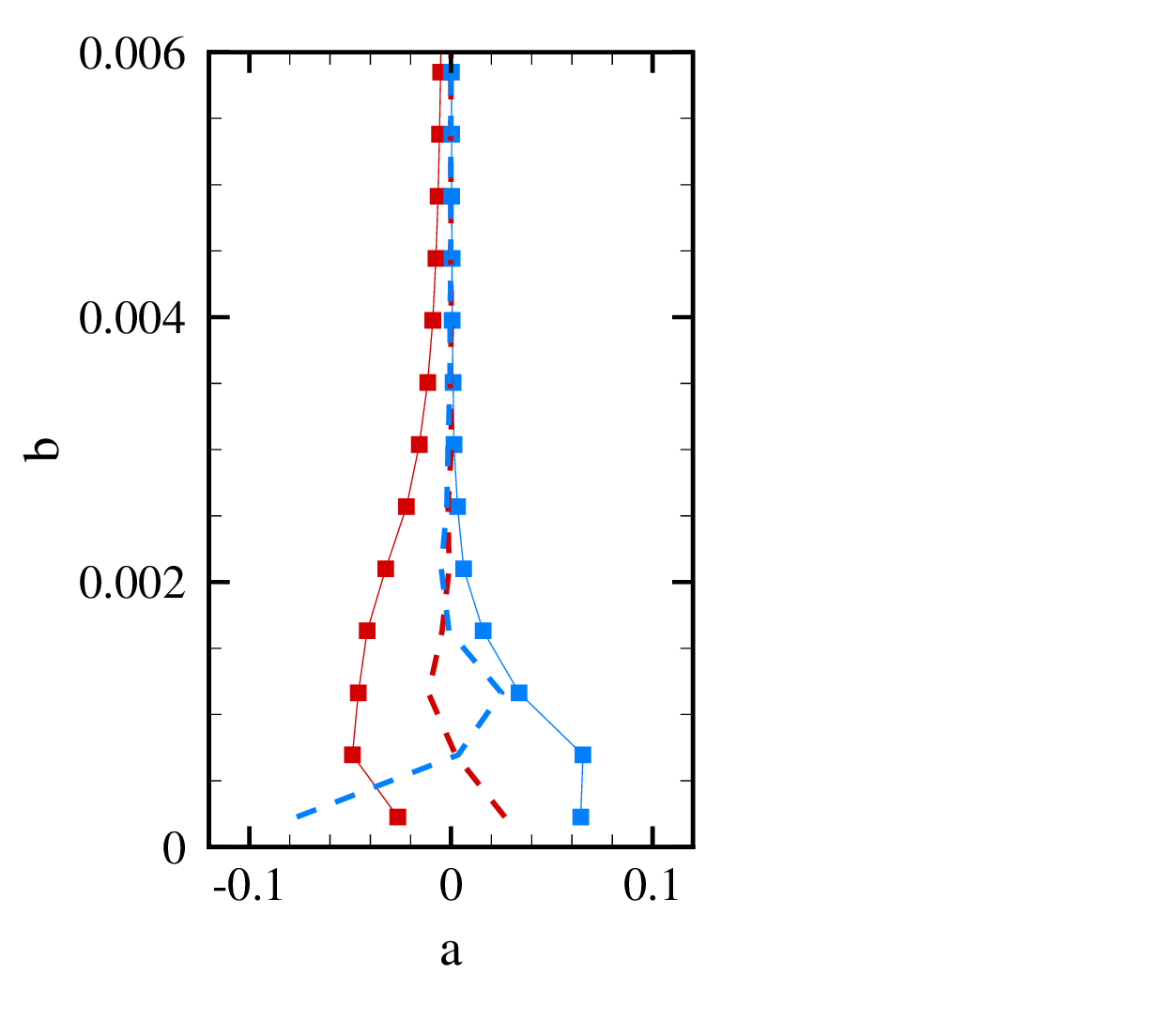}}
}
\caption{SGS dissipation $\xi_{ij}$ (solid square) and diffusion $\frac{\partial }{\partial  x_k}\zeta_{ijk}$ (dashed) from the isotropic (red) and anisotropic (blue) stress components of the MSM for the Reynolds shear stress $\overline{u_1'u_2'}$ (a–c) and the wall-normal Reynolds normal stress $\overline{u_2'u_2'}$ (d–f) at $x/L=-0.2$ (a,d), $x/L=-0.1$ (b,e), and $x/L=0$ (c,f). All terms are nondimensionalized using $U_\infty$, $L$ and $\rho$.}
\label{Reynolds_SGSDp_aniso_compare}
\end{figure}

To clarify the roles of the isotropic and anisotropic components, figure~\ref{Reynolds_SGSDp_aniso_compare} shows their individual contributions to SGS dissipation and diffusion in the MSM. For both $\overline{u_1'u_2'}$ and $\overline{u_2'u_2'}$, the isotropic component behaves similarly to the SM, consistently acting as a dissipative sink with larger magnitude. In contrast, the anisotropic component generates significant local production in the near-wall inner layer, confirming it as the primary source of backscatter in the MSM. For the SGS diffusion of $\overline{u_1'u_2'}$, the isotropic part mirrors the SM pattern (negative near the wall, positive farther out), while the anisotropic part shows the opposite, with negative values near the internal peak at $x_2/L\approx1.3\times10^{-3}$. For $\overline{u_2'u_2'}$, the isotropic part redistributes energy from the outer region toward the wall, while the anisotropic part redistributes energy into the internal-peak location from neighbouring wall-normal positions. Taken together, the anisotropic SGS stress in the MSM is responsible for the redistribution of energy toward the wall-normal locations of the internal Reynolds-stress peaks.

The analyses in this section demonstrate that SGS stress fluctuations strongly influence Reynolds-stress distributions within the critical FPG region. The isotropic SGS stress acts primarily as a dissipative sink for $\overline{u_1'u_2'}$ and $\overline{u_2'u_2'}$, while the anisotropic SGS stress provides significant backscatter and wall-normal redistribution of energy, facilitating the formation of internal Reynolds-stress peaks in the near-wall region of the TBL. This mechanism drives the downstream Reynolds-stress evolution and, in turn, alters the mean flow on the leeward side, particularly the onset of separation. Accurate WMLES predictions therefore require properly representing these complex near-wall SGS dynamics, including SGS stress fluctuations.

\subsection{Influence of mesh resolution} \label{sec:mesh_effect}

We now address the second question from \Cref{sec:flow_stats}: why do anisotropic SGS models yield more consistent separation bubble predictions across mesh resolutions than isotropic models? To this end, we examine mean streamwise momentum budgets at a location immediately upstream of the mean separation point across all mesh resolutions, following the approach of \Cref{sec:mean_budgets}.

\begin{figure}
\centering
{\psfrag{a}[][]{{$\Delta_{\text{c}}/L{\times10^3}$}}
\psfrag{b}[][]{{$x_{\text{s}}/L$}}
\includegraphics[width=.65\textwidth,trim={0.1cm 7.5cm 0.1cm 0.1cm},clip]{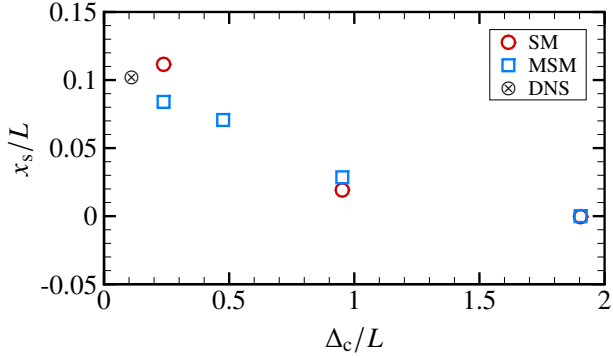}}
\caption{Mean separation point from simulations using the SM and MSM for different mesh resolutions and from the reference DNS \citep{uzun2022high}. No separation is detected in the medium-mesh simulation ($\Delta_{\text{c}}/L\approx4.76\times10^{-4}$) with the SM.}
\label{separation_point}
\end{figure}

The predicted mean separation points are shown in figure~\ref{separation_point}. On the coarsest mesh, both models capture separation with similar bubble sizes and separation locations near the bump peak. As the mesh is refined, the separation point shifts downstream and differences between the two models grow, with the SM losing the separation bubble entirely on the medium mesh. On the fine mesh, both models again predict separation, and their predictions approach the DNS reference \citep{uzun2022high}.

\begin{figure}
\centering
\sidesubfloat[]{
{\psfrag{h}[][]{{$x/L$}}
\psfrag{b}[][]{{$C_p$}}
\includegraphics[width=0.46\textwidth,trim={1.3cm 6cm 1.0cm 0.5cm},clip]{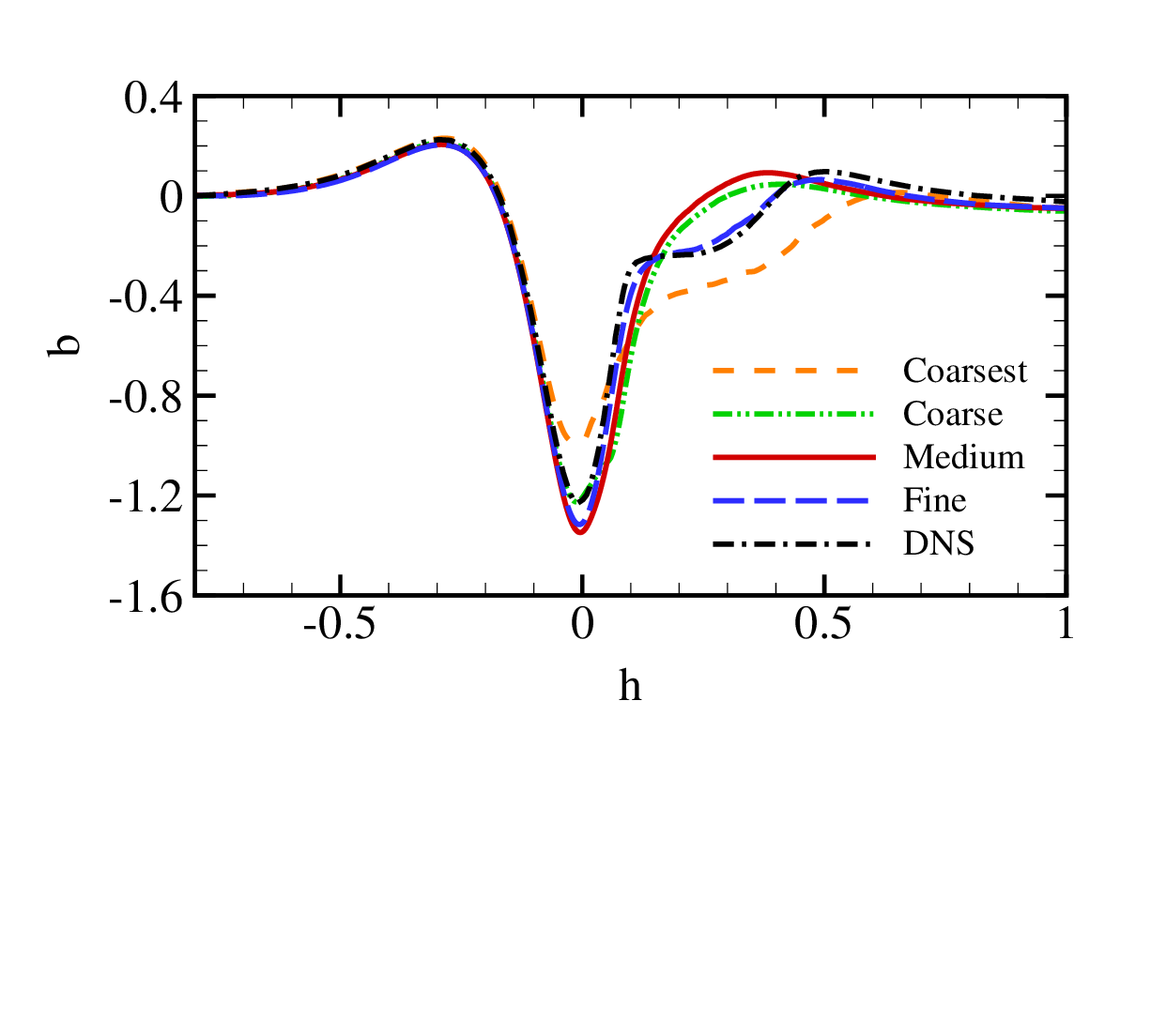}}}
\sidesubfloat[]{
{\psfrag{a}[][]{{$x/L$}}
\psfrag{b}[][]{{$C_p$}}
\includegraphics[width=0.46\textwidth,trim={1.3cm 6cm 1.0cm 0.5cm},clip]{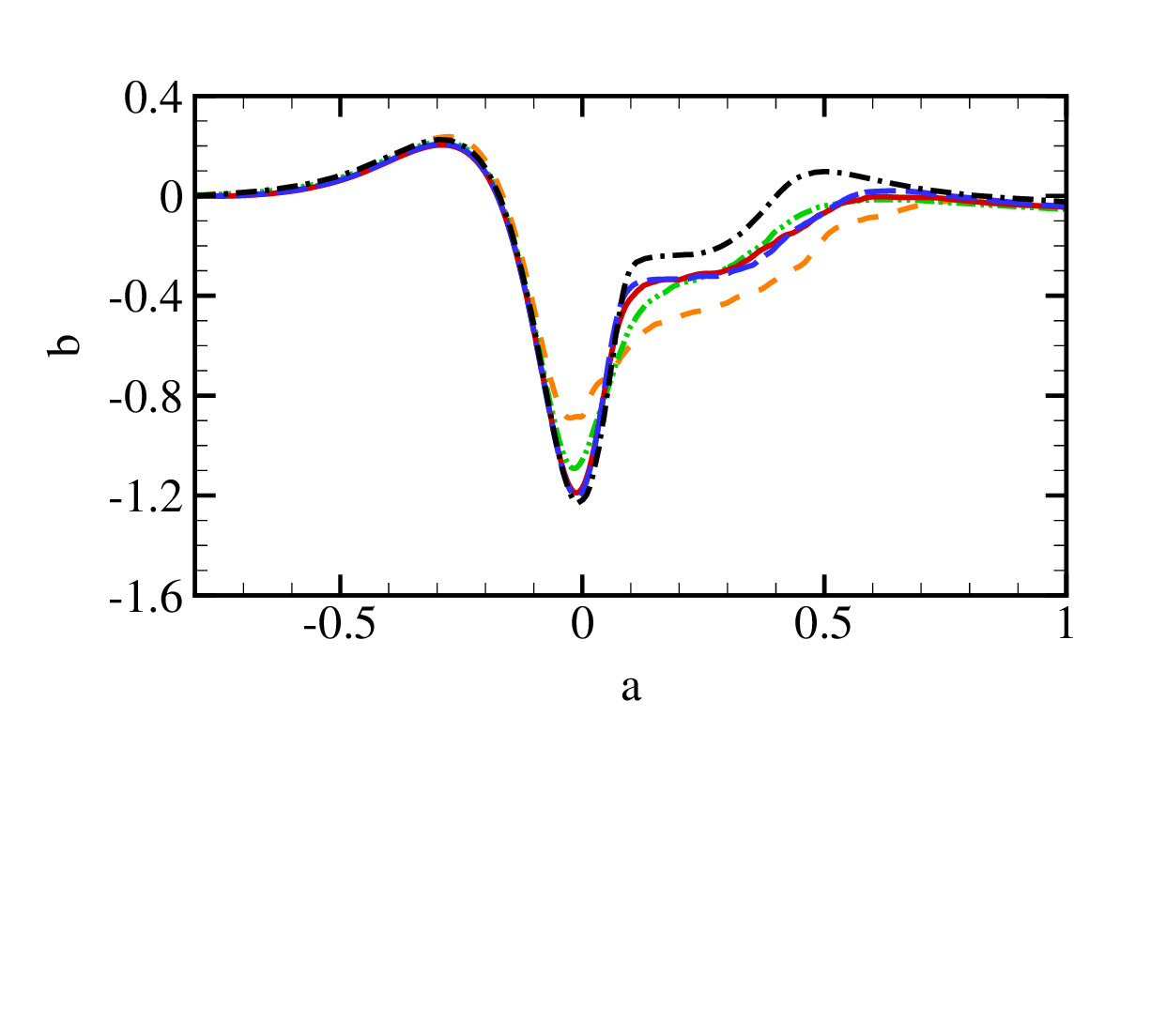}}}
\caption{{Mean pressure coefficient on the bottom surface from the SM (a) and MSM (b) simulations with different computational meshes and from the reference DNS \citep{uzun2022high}.}}
\label{mean_Cp_compare}
\end{figure}

{Figure~\ref{mean_Cp_compare} shows the distributions of the mean pressure coefficient $C_p$ on the bottom surface for all mesh resolutions and both SGS models. Upstream of the bump peak, results agree well across resolutions and models. On the leeward side, the SM simulations are sensitive to mesh resolution: the medium mesh produces no pressure plateau (consistent with the absence of separation), while the coarsest, coarse, and fine meshes all produce some plateau or inflection. In contrast, the MSM consistently produces a separation-indicative pressure plateau across all resolutions. With the fine mesh, both models approach the DNS. These $C_p$ distributions directly determine the streamwise pressure gradient and thus the $P_g$ term in the mean momentum equation.}

\begin{figure}
\centering
\sidesubfloat[]{
{\psfrag{a}[][]{{$\text{}$}}
\psfrag{b}[][]{{$x_2/L$}}
\includegraphics[width=.45\textwidth,trim={1cm 1.5cm 0.1cm 3cm},clip]{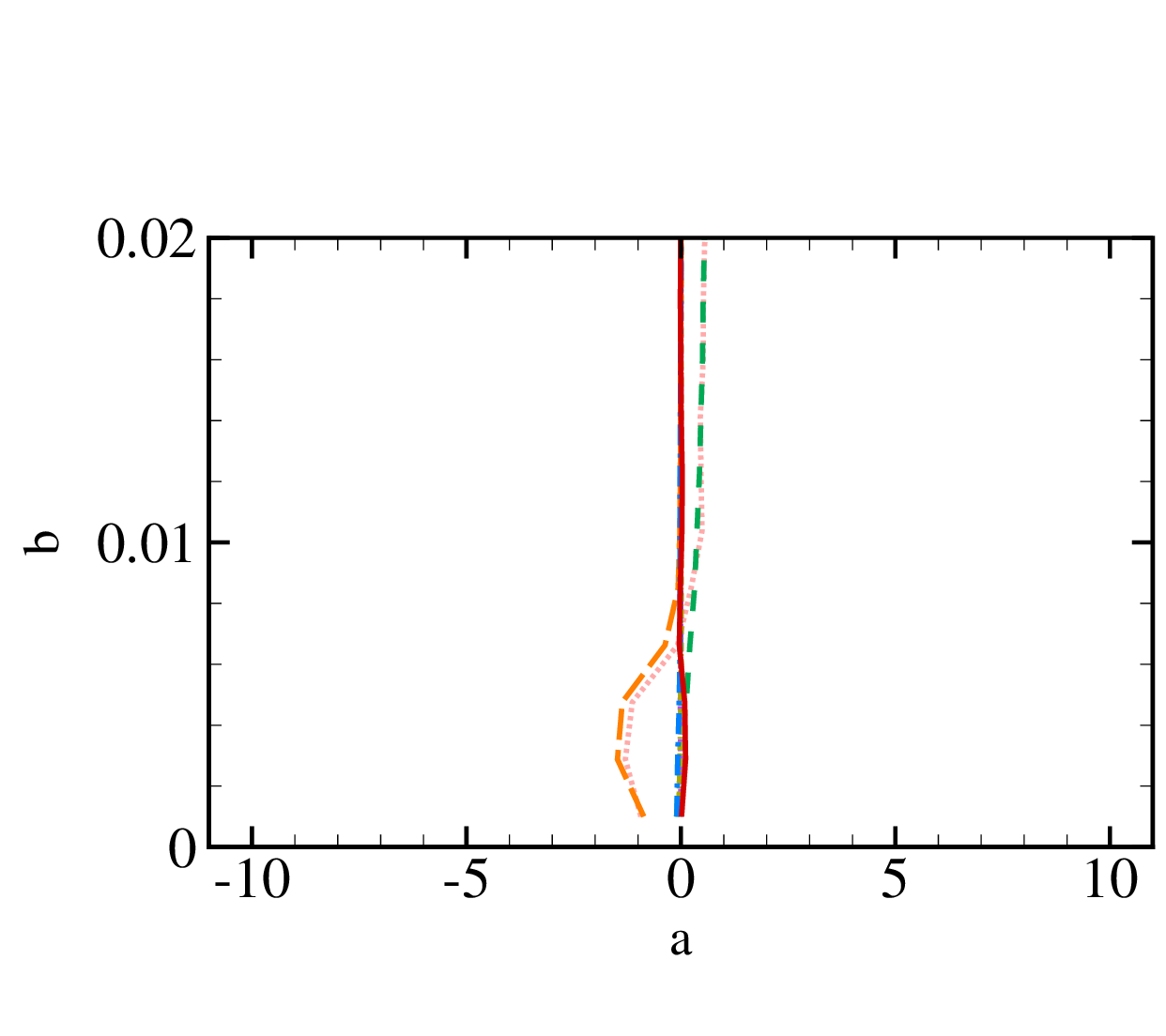}}
}
\sidesubfloat[]{
{\psfrag{a}[][]{{$\text{}$}}
\psfrag{b}[][]{{$x_2/L$}}
\includegraphics[width=.45\textwidth,trim={1cm 1.5cm 0.1cm 3cm},clip]{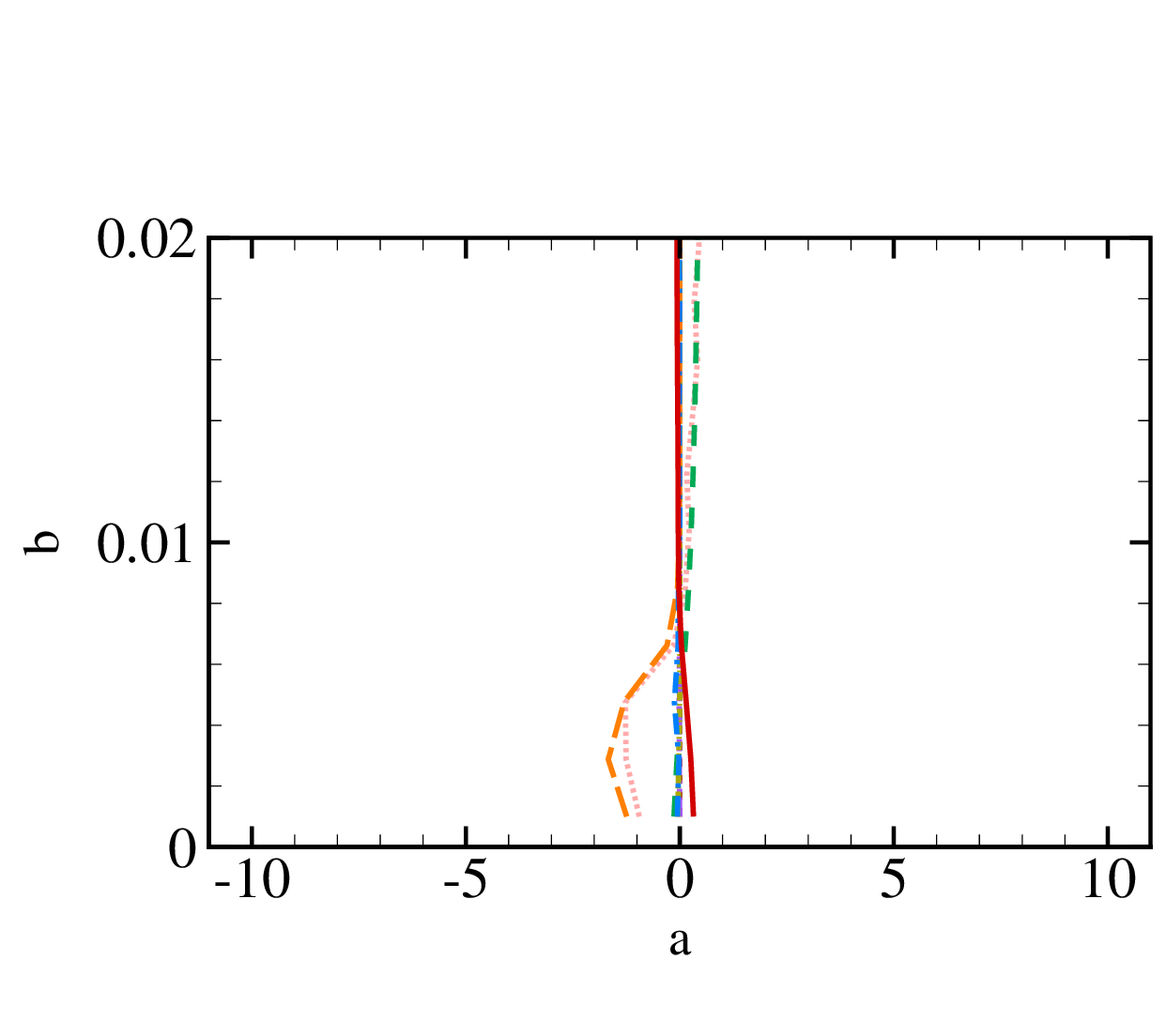}}
}
\vspace{5pt}
\sidesubfloat[]{
{\psfrag{a}[][]{{$\text{}$}}
\psfrag{b}[][]{{$x_2/L$}}
\includegraphics[width=.45\textwidth,trim={1cm 1.5cm 0.1cm 3cm},clip]{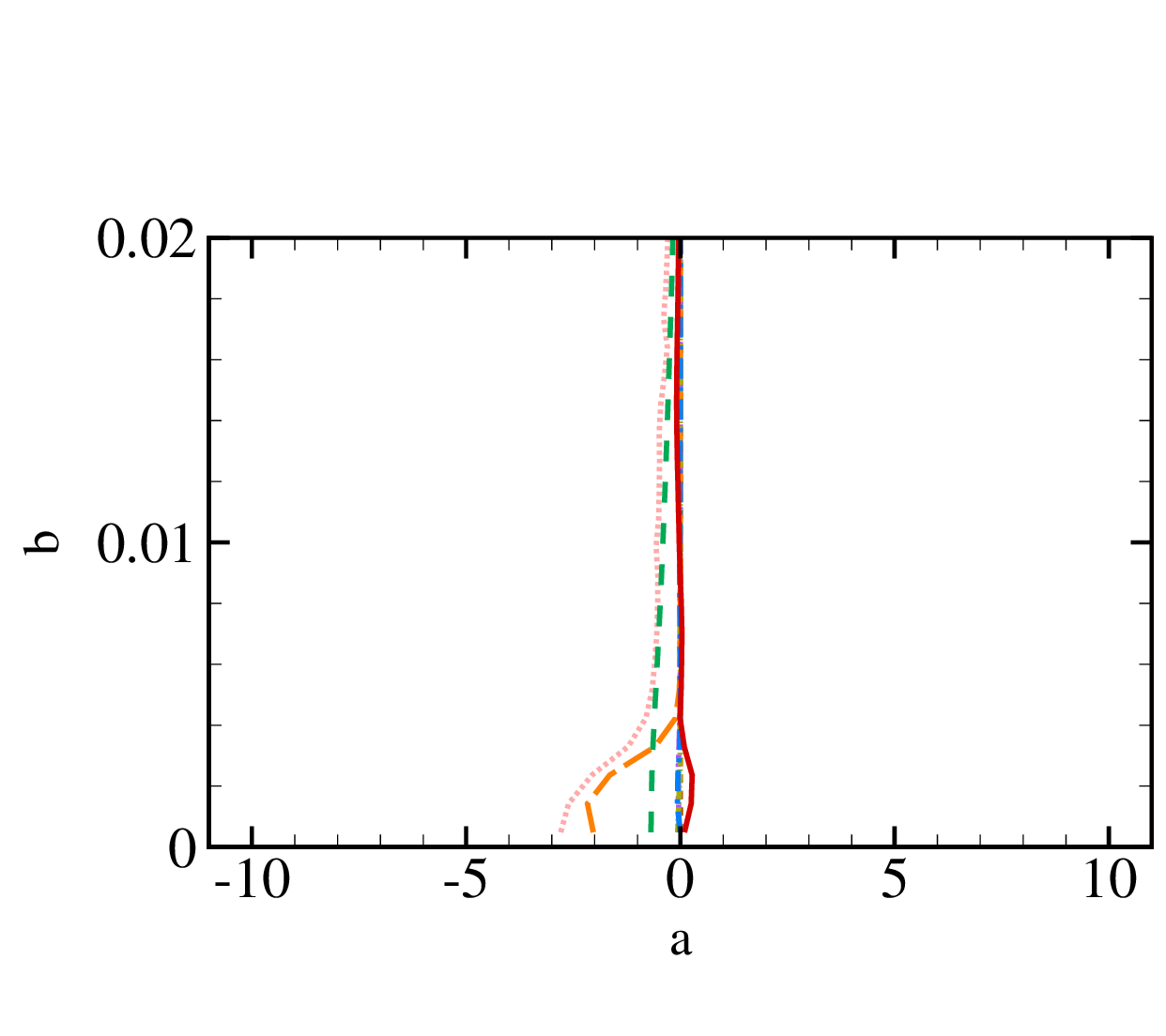}}
}
\sidesubfloat[]{
{\psfrag{a}[][]{{$\text{}$}}
\psfrag{b}[][]{{$x_2/L$}}
\includegraphics[width=.45\textwidth,trim={1cm 1.5cm 0.1cm 3cm},clip]{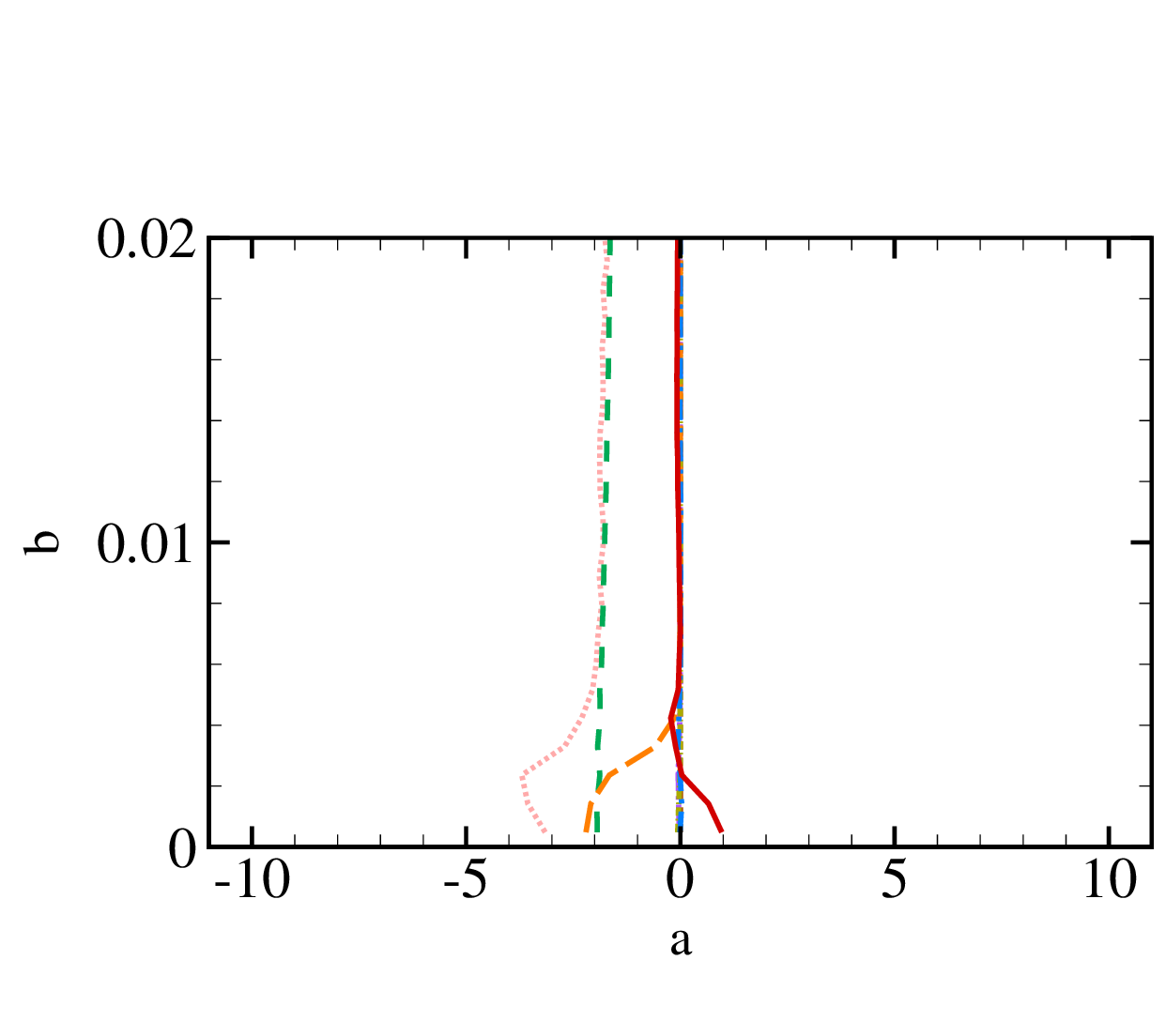}}
}
\vspace{5pt}
\sidesubfloat[]{
{\psfrag{a}[][]{{$\text{}$}}
\psfrag{b}[][]{{$x_2/L$}}
\includegraphics[width=.45\textwidth,trim={1cm 1.5cm 0.1cm 3cm},clip]{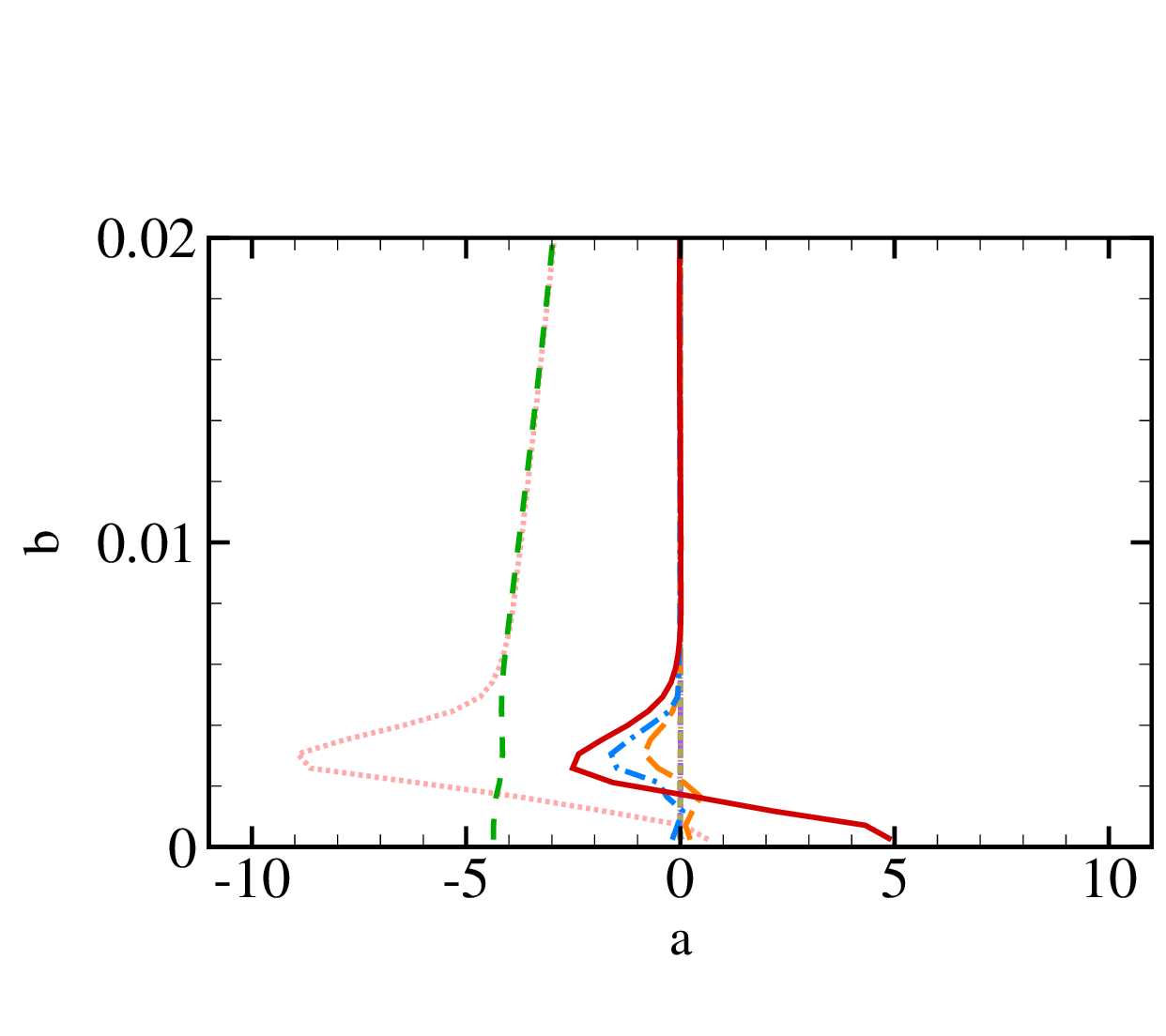}}
}
\sidesubfloat[]{
{\psfrag{a}[][]{{$\text{}$}}
\psfrag{b}[][]{{$x_2/L$}}
\includegraphics[width=.45\textwidth,trim={1cm 1.5cm 0.1cm 3cm},clip]{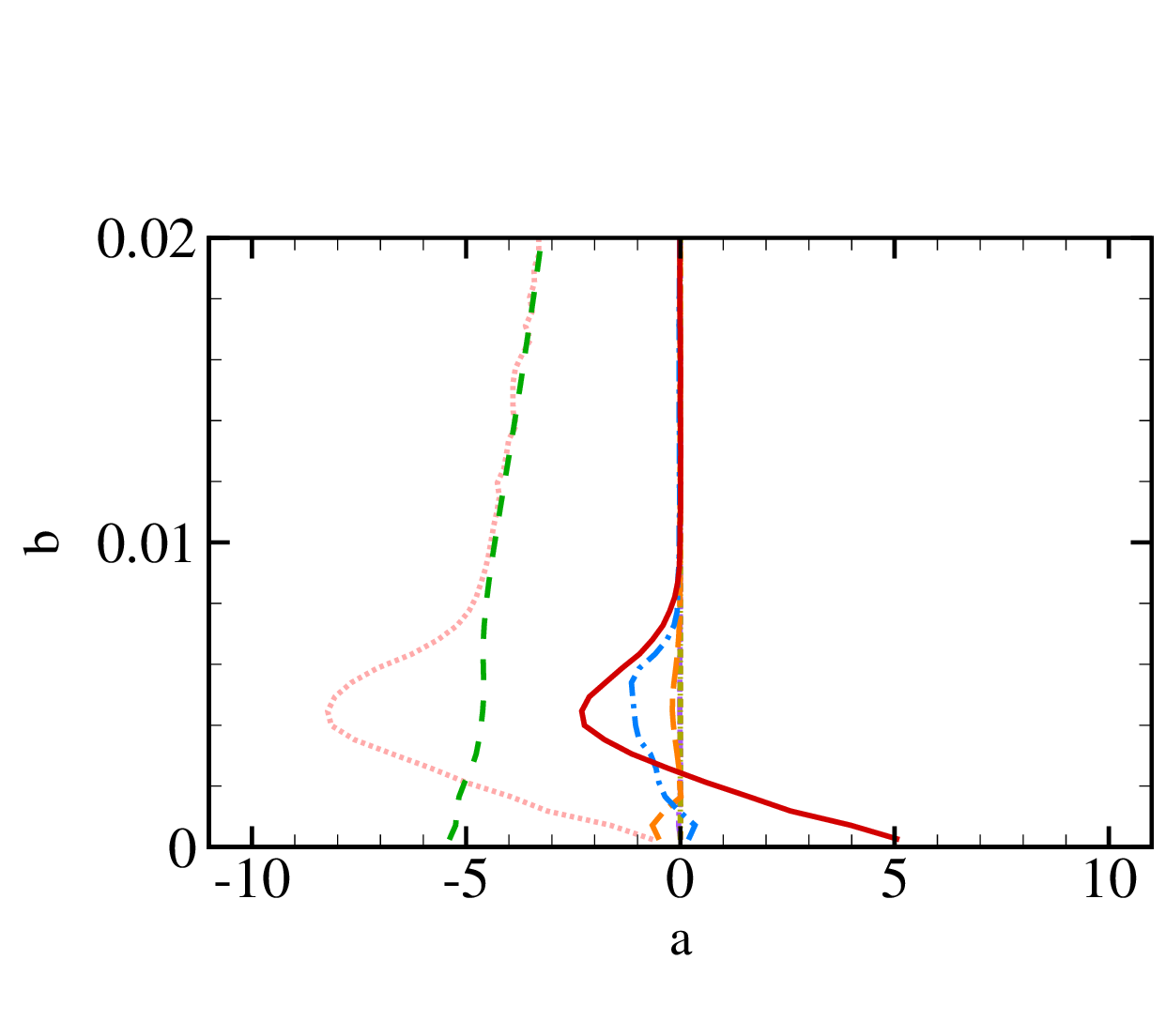}}
}
\vspace{5pt}
\sidesubfloat[]{
{\psfrag{a}[][]{{$\text{}$}}
\psfrag{b}[][]{{$x_2/L$}}
\includegraphics[width=.45\textwidth,trim={1cm 1.5cm 0.1cm 3cm},clip]{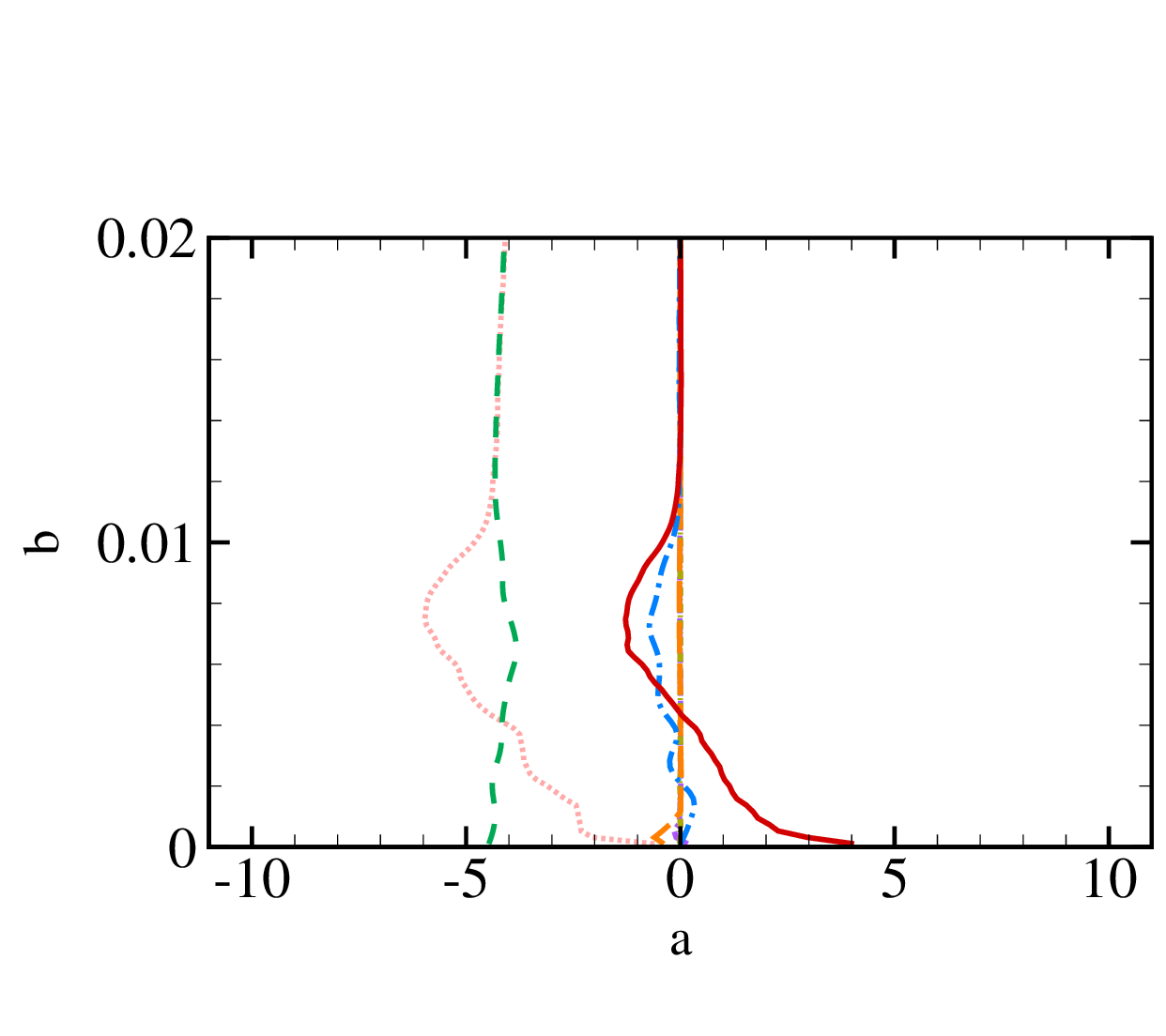}}
}
\sidesubfloat[]{
{\psfrag{a}[][]{{$\text{}$}}
\psfrag{b}[][]{{$x_2/L$}}
\includegraphics[width=.45\textwidth,trim={1cm 1.5cm 0.1cm 3cm},clip]{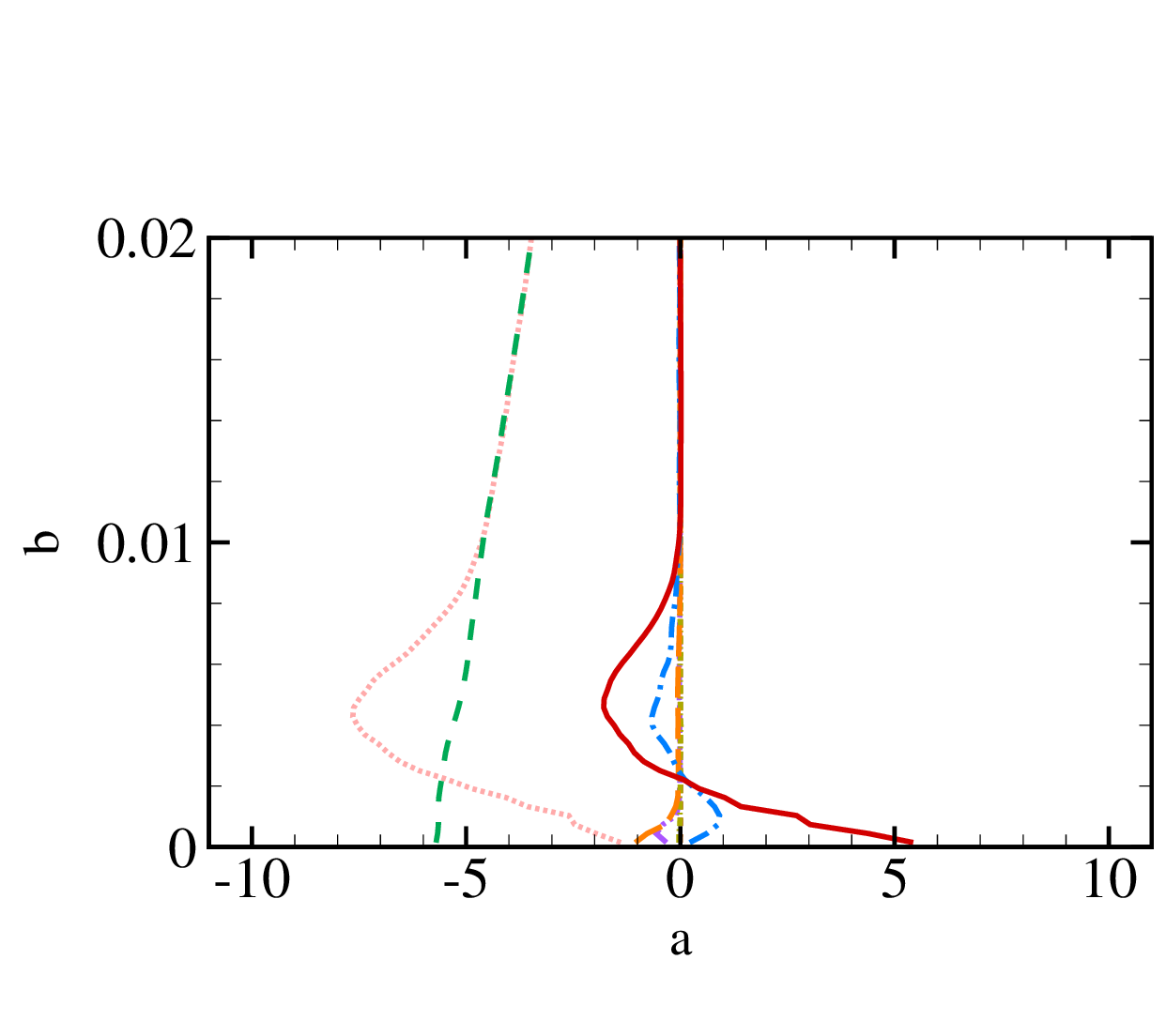}}
}
\caption{{Mean streamwise momentum budget terms at $0.02L$ upstream of the {corresponding} mean separation point along the $x$ direction from cases with SM (a,c,e,g) and MSM (b,d,f,h) for the coarsest mesh (a,b), coarse mesh (c,d), medium mesh (e,f), and fine mesh (g,h). All terms are nondimensionalized using $U_\infty$, $L$ and $\rho$. {The line notations correspond to those in figure~\ref{Mobudget_medium_SGS}}.}}
\label{Momentum_budget_SM_mesh_compare}
\end{figure}

The momentum budget analysis is performed at $0.02L$ upstream of each simulation's mean separation point along the $x$ direction, approximately one boundary-layer thickness upstream of separation, with results shown in {figure~\ref{Momentum_budget_SM_mesh_compare}}. For the medium-mesh simulation with the SM, where no separation bubble forms on the leeward side of the bump, the same location as in the medium-mesh MSM simulation is used for consistency. On the coarsest mesh, both models behave similarly, with the mean SGS shear-stress gradient $T_{12}$ dominating the lower half of the TBL. On the coarse mesh, the separation point shifts downstream and the APG strengthens. While $T_{12}$ remains important, $R_{12}$ begins to play a more notable role in the MSM simulation. On the medium mesh, $T_{12}$ decreases substantially in both cases, and the qualitative difference in separation behaviour emerges as discussed in \Cref{sec:Reynolds_budgets}. On the fine mesh, $R_{12}$ and $P_g$ become the dominant contributions in both models. Notably, the negative region of $R_{12}$ in the SM case extends over a wider wall-normal range than on the medium mesh, extracting momentum over a thicker layer and redistributing it toward the wall. Such redistribution reduces the streamwise momentum available downstream to resist the APG, ultimately enabling separation. The contribution of $R_{11}$, associated with $\overline{u_1'u_1'}$, also becomes important at this resolution.

{The mean convection term itself exhibits a resolution-dependent character that is consistent with the broader picture above. On the coarsest and coarse meshes, where the SGS shear-stress gradient $T_{12}$ dominates the budget, the convection term is modest in magnitude throughout the boundary layer. As the mesh is refined to medium and fine resolutions, the convection term develops substantially richer structure. Specifically, it becomes large in the outer layer, where it grows to balance the pressure gradient, and it develops pronounced structure in the near-wall region as well. This shift parallels the broader transition from an SGS-dominated near-wall momentum balance at coarse resolution to a balance dominated by mean convection, pressure gradient, and Reynolds stress at finer resolutions. As discussed in \Cref{sec:mean_budgets}, the convection term is a kinematic consequence of the established mean flow rather than a direct forcing of the local momentum balance, and the present discussion therefore focuses on the forcing terms $P_g$ and $R_{12}$ that connect the SGS model to the separation onset mechanism. The residual is small relative to the leading-order terms at all four mesh resolutions, confirming the numerical accuracy of the budgets, and is not shown for brevity.}

These results reveal how the relative importance of mean SGS stress and Reynolds stress shifts with mesh resolution. {At coarse resolutions, many flow structures, including the internal Reynolds-stress peaks in the FPG region, remain unresolved,} and the mean SGS shear stress $\overline{\tau^{\text{sgs}}_{12}}$ dominates the momentum balance. Since the anisotropic term's contribution to $\overline{\tau^{\text{sgs}}_{12}}$ in the near-wall region is limited (as shown in \Cref{sec:SGS_properties}), both models behave similarly. At medium resolution, more scales are resolved, Reynolds stresses play a more significant role, and the anisotropic SGS term improves flow prediction through its effect on SGS stress fluctuations, as discussed in \Cref{sec:Reynolds_budgets}. At fine resolution, the SGS model accounts only for the smallest motions, and the influence of both models on mean velocity and Reynolds stress predictions weakens. These findings highlight that robust SGS models for WMLES must accurately represent both mean SGS stress and SGS stress fluctuations across varying mesh resolutions.

\section{Properties of SGS stress} \label{sec:SGS_properties}

In this section, we examine the properties of the SGS stress and the underlying mechanism of the anisotropic SGS stress. The investigation focuses on the critical FPG region upstream of the bump peak, where the anisotropic SGS stress has a strong effect and significantly influences the downstream mean flow separation.

\subsection{Mean SGS stress}

\begin{figure}
\centering
\sidesubfloat[]{
{\psfrag{a}[][]{{$\overline{\tau_{ij}^{\text{sgs}}}/(\rho U_\infty^2)$}}
\psfrag{b}[][]{{$x_2/L$}}
\psfrag{c}[][]{{\scriptsize$\overline{\tau_{11}^{\text{sgs}}}$}}
\psfrag{d}[][]{{\scriptsize$\overline{\tau_{12}^{\text{sgs}}}$}}
\psfrag{e}[][]{{\scriptsize$\overline{\tau_{13}^{\text{sgs}}}$}}
\psfrag{f}[][]{{\scriptsize$\overline{\tau_{22}^{\text{sgs}}}$}}
\psfrag{g}[][]{{\scriptsize$\overline{\tau_{23}^{\text{sgs}}}$}}
\psfrag{h}[][]{{\scriptsize$\overline{\tau_{33}^{\text{sgs}}}$}}
\includegraphics[width=.45\textwidth,trim={1cm 1.0cm 0.1cm 3cm},clip]{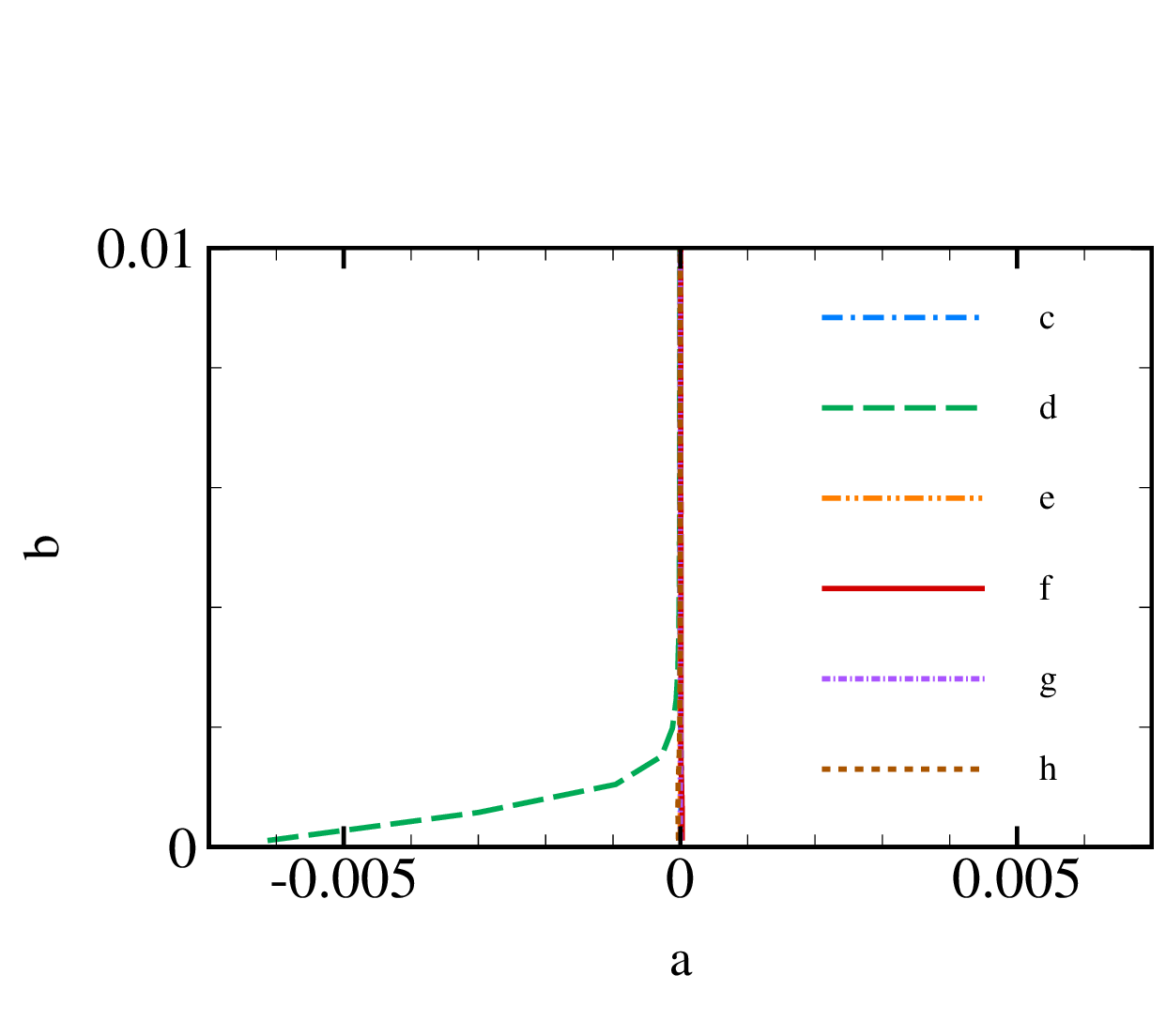}}
}
\sidesubfloat[]{
{\psfrag{a}[][]{{$\overline{\tau_{ij}^{\text{sgs}}}/(\rho U_\infty^2)$}}
\psfrag{b}[][]{{$x_2/L$}}
\includegraphics[width=.45\textwidth,trim={1cm 1.0cm 0.1cm 3cm},clip]{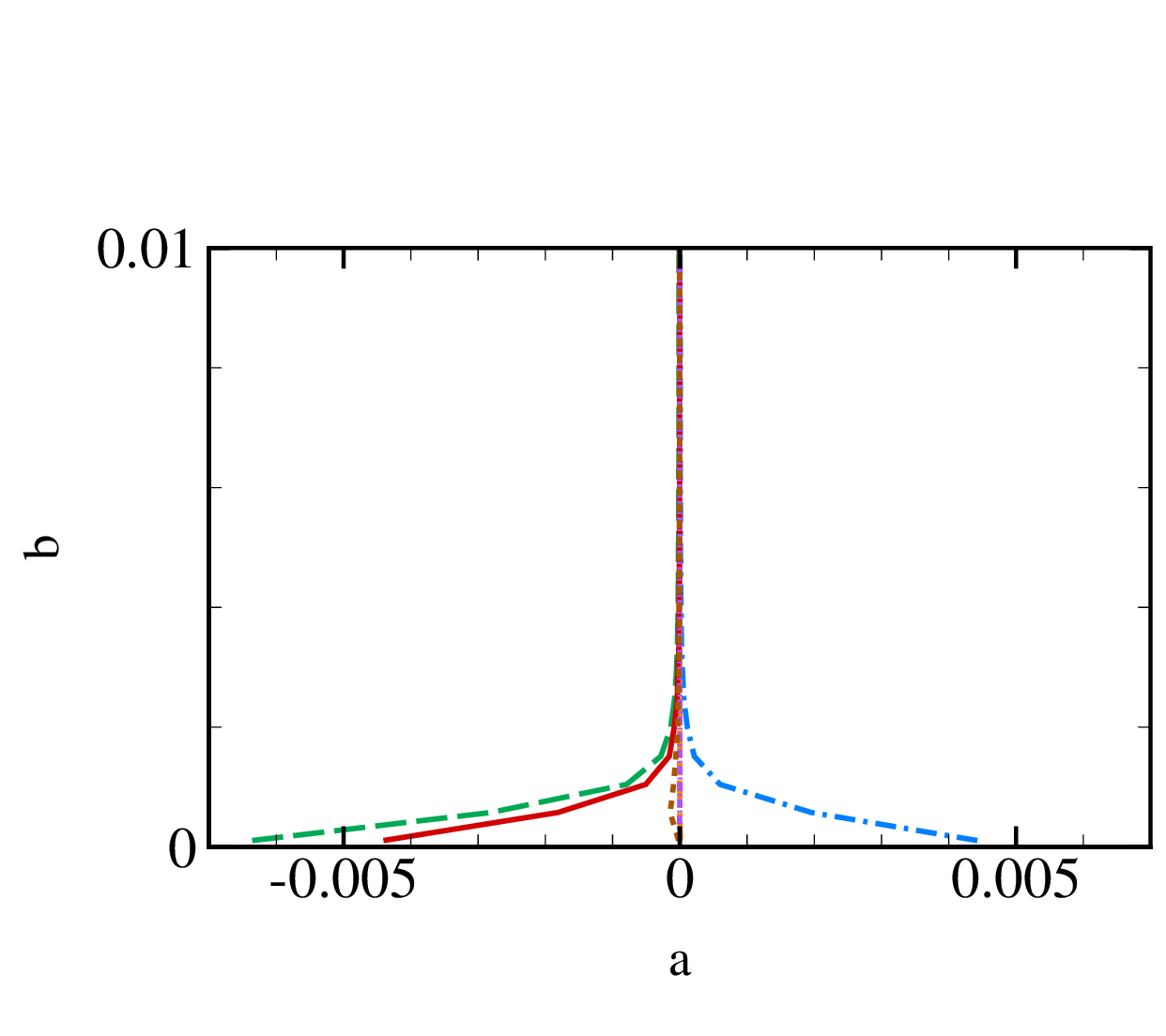}}
}
\caption{Mean SGS stress tensor components $\overline{\tau_{ij}^{\text{sgs}}}$ at $x/L=-0.1$ for medium-mesh simulations with the SM (a) and MSM (b).}
\label{mean_SGS_compare}
\end{figure}

We first examine the mean SGS stress at $x/L=-0.1$, near the centre of the critical FPG region where the TBL remains attached in all simulations. Figure~\ref{mean_SGS_compare} shows the wall-normal distributions of the six independent components of the mean SGS stress from the medium-mesh simulations with the SM and MSM.

In both simulations, the mean SGS stress is significant only very near the wall, with the dominant component being the shear stress $\overline{\tau_{12}^{\text{sgs}}}$, as expected from the large wall-normal gradient of the streamwise velocity. In the SM, the remaining components are negligible relative to this dominant shear stress. In contrast, the MSM produces two additional components of appreciable magnitude, $\overline{\tau_{11}^{\text{sgs}}}$ and $\overline{\tau_{22}^{\text{sgs}}}$, which modify the principal directions of the mean SGS stress tensor and alter the associated momentum transfer. This qualitative behaviour persists throughout the FPG region. Since $\overline{\tau_{11}^{\text{sgs}}}$, $\overline{\tau_{22}^{\text{sgs}}}$, and $\overline{\tau_{12}^{\text{sgs}}}$ are the three main components and the others are negligible, only these are presented in subsequent figures.

{The behaviour of the mean SGS stress at $x/L=-0.1$ is consistent across mesh resolutions. As the mesh is refined, the magnitude of the dominant SGS stress components decreases and the wall-normal extent of significant SGS shear stress narrows. For the SM, $\overline{\tau_{12}^{\text{sgs}}}$ remains dominant at all resolutions, though at fine resolution the normal components $\overline{\tau_{11}^{\text{sgs}}}$ and $\overline{\tau_{22}^{\text{sgs}}}$ become non-negligible near the wall. This indicates that the anisotropic dynamics of near-wall turbulence begin to be resolved, with a shift in the principal directions of the mean SGS stress and strain-rate tensors. The signs of these normal components are consistent with those from the MSM, suggesting that the principal direction of the mean SGS stress in the fine-mesh SM case begins to align with that of the MSM. For the MSM, the contributions of all three main components remain significant near the wall and qualitatively unchanged across resolutions.}

\begin{figure}
\centering
\sidesubfloat[]{
{\psfrag{a}[][]{{$\overline{\tau_{ij}^{\text{iso}}}/(\rho U_\infty^2)$}}
\psfrag{b}[][]{{$x_2/L$}}
\psfrag{c}[][]{{\scriptsize$\overline{\tau_{11}^{\text{iso}}}$}}
\psfrag{d}[][]{{\scriptsize$\overline{\tau_{12}^{\text{iso}}}$}}
\psfrag{f}[][]{{\scriptsize$\overline{\tau_{22}^{\text{iso}}}$}}
\includegraphics[width=.45\textwidth,trim={1cm 1.0cm 0.1cm 3cm},clip]{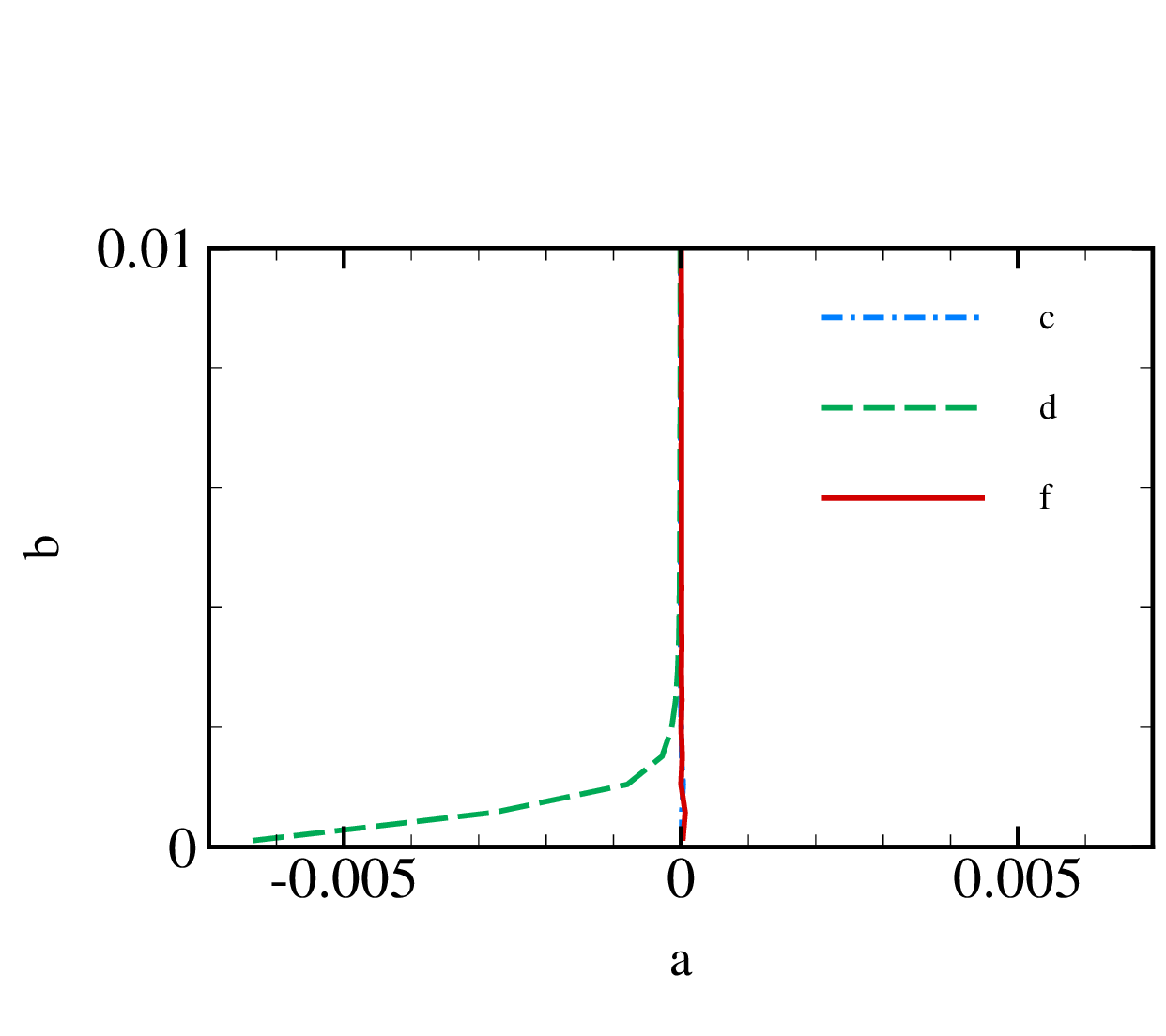}}
}
\sidesubfloat[]{
{\psfrag{a}[][]{{$\overline{\tau_{ij}^{\text{ani}}}/(\rho U_\infty^2)$}}
\psfrag{b}[][]{{$x_2/L$}}
\psfrag{c}[][]{{\scriptsize$\overline{\tau_{11}^{\text{ani}}}$}}
\psfrag{d}[][]{{\scriptsize$\overline{\tau_{12}^{\text{ani}}}$}}
\psfrag{f}[][]{{\scriptsize$\overline{\tau_{22}^{\text{ani}}}$}}
\includegraphics[width=.45\textwidth,trim={1cm 1.0cm 0.1cm 3cm},clip]{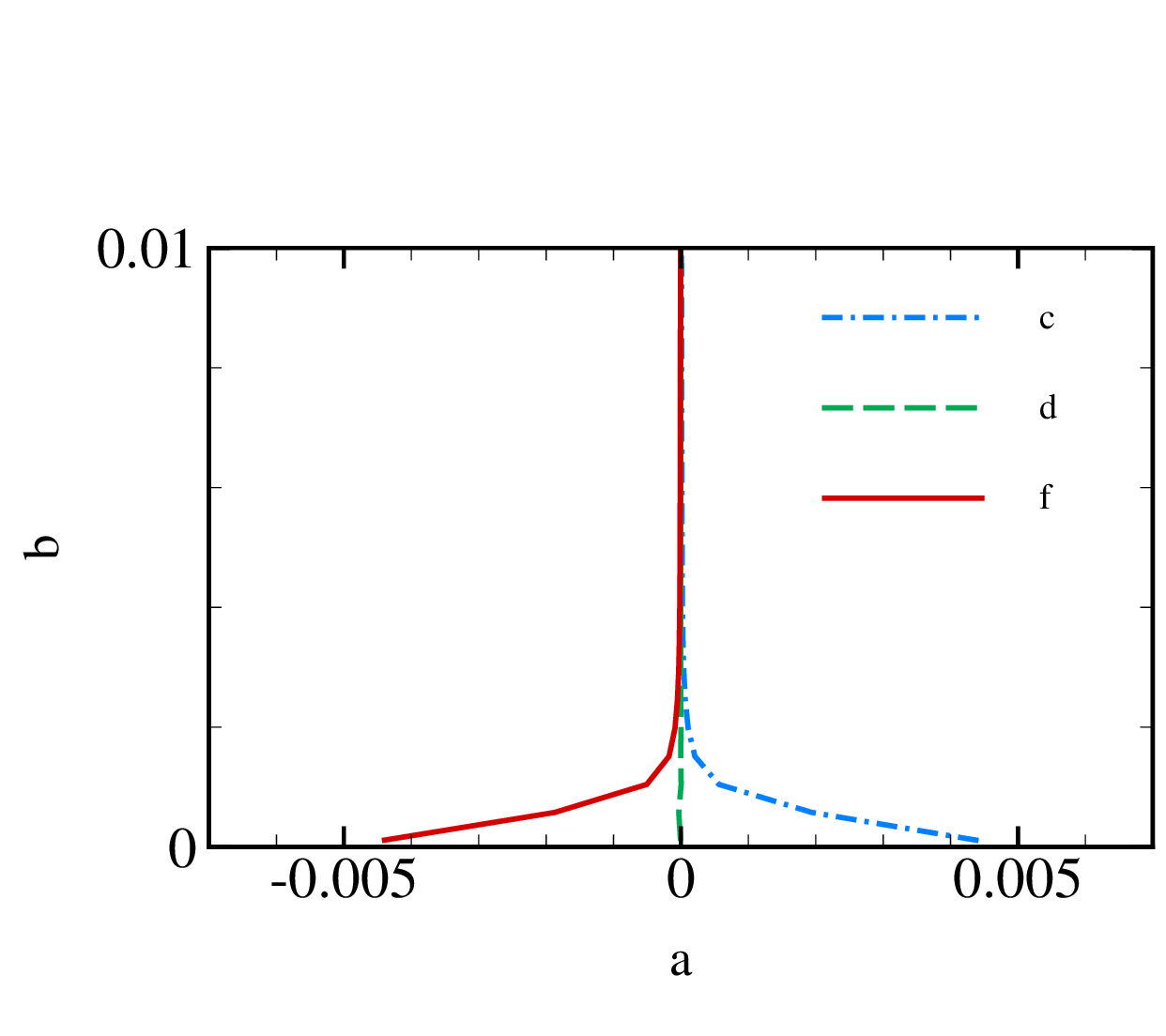}}
}
\caption{Mean isotropic SGS stress tensor components ($\overline{\tau_{11}^{\text{iso}}}$, $\overline{\tau_{12}^{\text{iso}}}$, and $\overline{\tau_{22}^{\text{iso}}}$) (a) and mean anisotropic SGS stress tensor components ($\overline{\tau_{11}^{\text{ani}}}$, $\overline{\tau_{12}^{\text{ani}}}$, and $\overline{\tau_{22}^{\text{ani}}}$) (b) at $x/L=-0.1$ for medium-mesh simulations with the MSM.}
\label{mean_SGS_compare_MSM_part}
\end{figure}

Figure~\ref{mean_SGS_compare_MSM_part} shows the isotropic and anisotropic contributions to the mean SGS stress at $x/L=-0.1$ for the medium-mesh MSM simulation. The mean isotropic stress behaves almost identically to the SM (see figure~\ref{mean_SGS_compare}), indicating that the anisotropic term does not significantly modify the isotropic part. A similar trend is observed in the eddy-viscosity distributions within the FPG region (figure~\ref{nut_mean}), and since SGS dissipation in the MSM arises exclusively from the isotropic term, the SGS kinetic energy dissipation is comparable between the SM and MSM at this location. In contrast, the mean anisotropic stress is dominated by the normal components, with a negligible shear component. Comparison with figure~\ref{mean_SGS_compare} confirms that the normal stresses in the MSM originate almost entirely from the anisotropic term, which is therefore the primary source of the difference in principal directions of the mean SGS stress tensor between the two models.

The differences in mean SGS stress between the SM and MSM thus arise from the normal components $\overline{\tau_{11}^{\text{sgs}}}$ and $\overline{\tau_{22}^{\text{sgs}}}$, which originate from the anisotropic term. However, as shown in the momentum budget analyses of \Cref{sec:mean_budgets} and \Cref{sec:mesh_effect}, these normal stresses contribute little to mean streamwise momentum transport. The mean shear stress $\overline{\tau_{12}^{\text{sgs}}}$, which is most relevant to momentum transport and separation onset at coarse resolutions, is predicted similarly by both models in the critical FPG region. This explains why the coarsest- and coarse-mesh simulations yield similar separation behaviour for both SGS models.

\subsection{SGS stress fluctuations}

\begin{figure}
\centering
\sidesubfloat[]{
{\psfrag{a}[][]{{${\tau_{ij\text{, rms}}^{\text{sgs}}}/(\rho U_\infty^2)$}}
\psfrag{b}[][]{{$x_2/L$}}
\psfrag{c}[][]{{\scriptsize$\tau_{11\text{, rms}}^{\text{sgs}}$}}
\psfrag{d}[][]{{\scriptsize$\tau_{12\text{, rms}}^{\text{sgs}}$}}
\psfrag{e}[][]{{\scriptsize$\tau_{13\text{, rms}}^{\text{sgs}}$}}
\psfrag{f}[][]{{\scriptsize$\tau_{22\text{, rms}}^{\text{sgs}}$}}
\psfrag{g}[][]{{\scriptsize$\tau_{23\text{, rms}}^{\text{sgs}}$}}
\psfrag{h}[][]{{\scriptsize$\tau_{33\text{, rms}}^{\text{sgs}}$}}
\includegraphics[width=.45\textwidth,trim={1cm 1.0cm 0.1cm 3cm},clip]{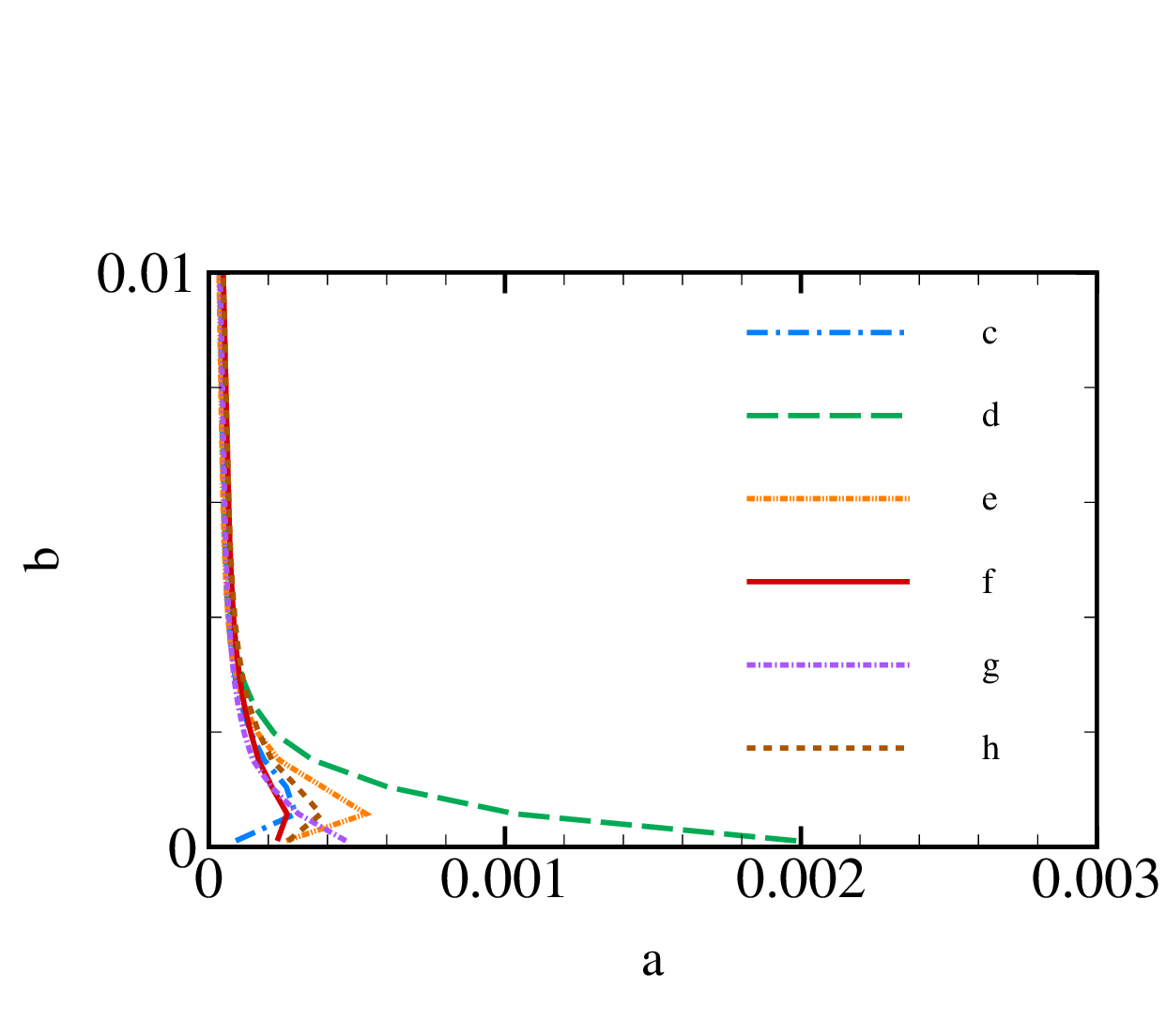}}
}
\sidesubfloat[]{
{\psfrag{a}[][]{{${\tau_{ij\text{, rms}}^{\text{sgs}}}/(\rho U_\infty^2)$}}
\psfrag{b}[][]{{$x_2/L$}}
\includegraphics[width=.45\textwidth,trim={1cm 1.0cm 0.1cm 3cm},clip]{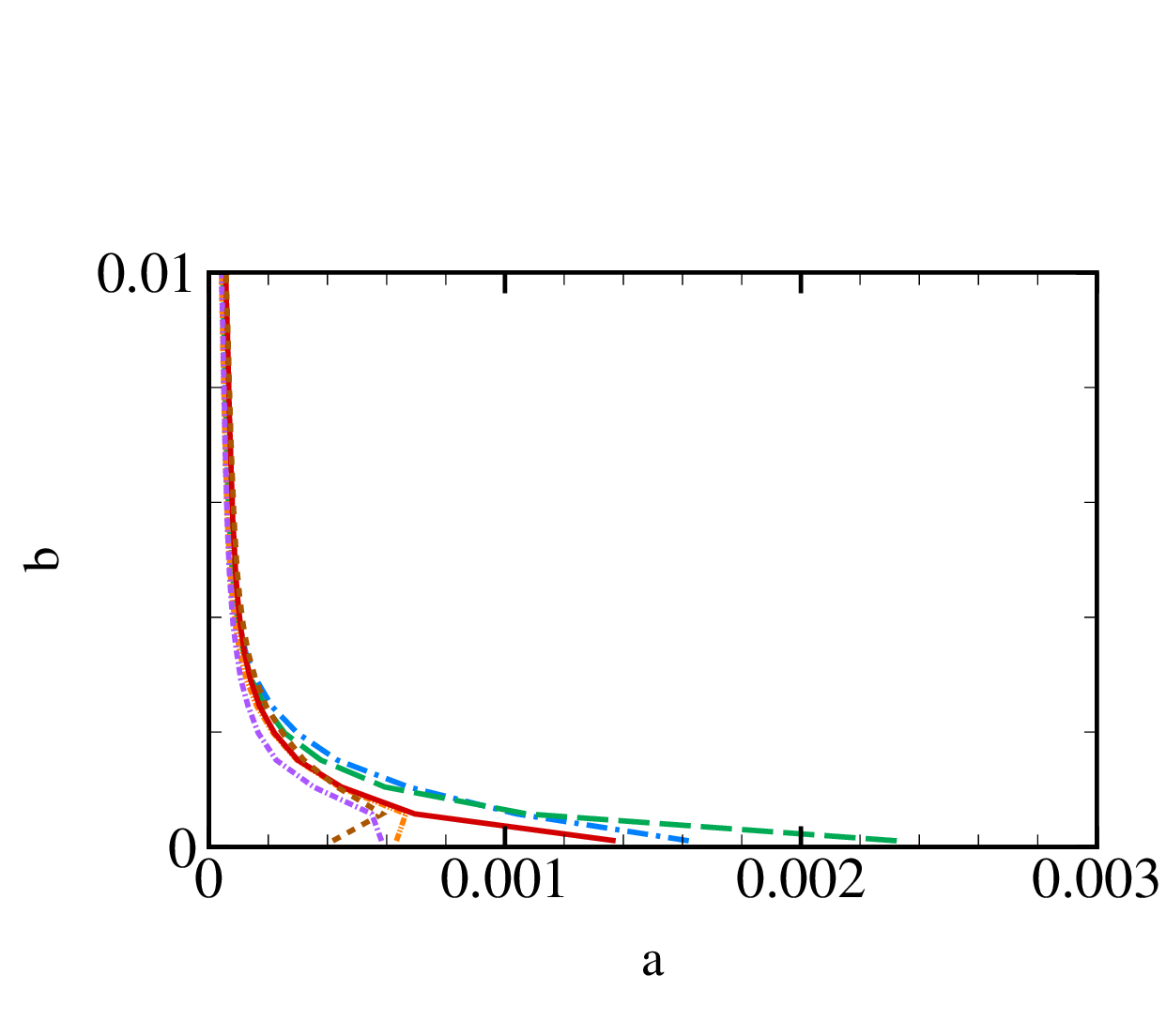}}
}
\caption{SGS stress tensor r.m.s. components ${\tau_{ij\text{, rms}}^{\text{sgs}}}$ at $x/L=-0.1$ for medium-mesh simulations with the SM (a) and MSM (b).}
\label{rms_SGS_compare}
\end{figure}

Figure~\ref{rms_SGS_compare} shows wall-normal distributions of the root-mean-square (r.m.s.) values of SGS stress fluctuations at $x/L=-0.1$ for the medium-mesh simulations. As with the mean SGS stresses, fluctuations are largest near the wall, with $\tau_{12}^{\text{sgs}}$ exhibiting the strongest fluctuations in both simulations. In the MSM, however, the normal components $\tau_{11}^{\text{sgs}}$ and $\tau_{22}^{\text{sgs}}$ also display substantial fluctuations, unlike in the SM. Since fluctuations of the normal stresses contribute directly to SGS dissipation and diffusion in the Reynolds stress transport equation, these enhanced fluctuations explain the distinct SGS dissipation and diffusion behaviour observed for the MSM in \Cref{sec:Reynolds_budgets}. This qualitative behaviour remains consistent throughout the FPG region.

{Across mesh resolutions, the qualitative behaviour of SGS stress fluctuations within each model remains similar upon refinement. In contrast to the mean SGS stresses, which decrease rapidly with mesh refinement, the fluctuations exhibit only a mild reduction, indicating their increasing relative importance on finer grids. The MSM consistently yields larger fluctuations than the SM across all resolutions at this location.}

\begin{figure}
\centering
\sidesubfloat[]{
{\psfrag{a}[][]{{${\tau_{ij\text{, rms}}^{\text{iso}}}/(\rho U_\infty^2)$}}
\psfrag{b}[][]{{$x_2/L$}}
\psfrag{c}[][]{{\scriptsize$\tau_{11\text{, rms}}^{\text{iso}}$}}
\psfrag{d}[][]{{\scriptsize$\tau_{12\text{, rms}}^{\text{iso}}$}}
\psfrag{f}[][]{{\scriptsize$\tau_{22\text{, rms}}^{\text{iso}}$}}
\includegraphics[width=.45\textwidth,trim={1cm 1.0cm 0.1cm 3cm},clip]{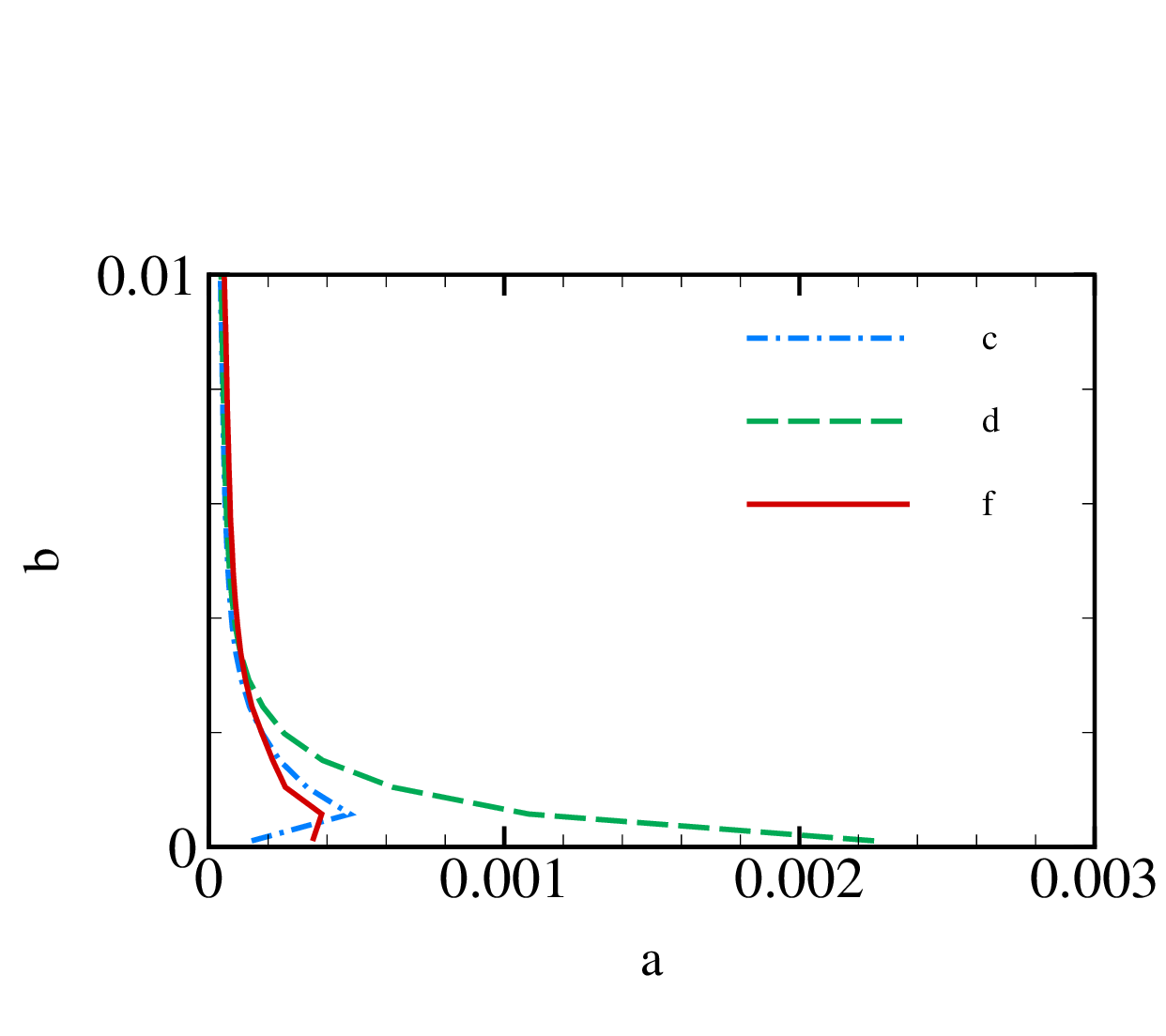}}
}
\sidesubfloat[]{
{\psfrag{a}[][]{{${\tau_{ij\text{, rms}}^{\text{ani}}}/(\rho U_\infty^2)$}}
\psfrag{b}[][]{{$x_2/L$}}
\psfrag{c}[][]{{\scriptsize$\tau_{11\text{, rms}}^{\text{ani}}$}}
\psfrag{d}[][]{{\scriptsize$\tau_{12\text{, rms}}^{\text{ani}}$}}
\psfrag{f}[][]{{\scriptsize$\tau_{22\text{, rms}}^{\text{ani}}$}}
\includegraphics[width=.45\textwidth,trim={1cm 1.0cm 0.1cm 3cm},clip]{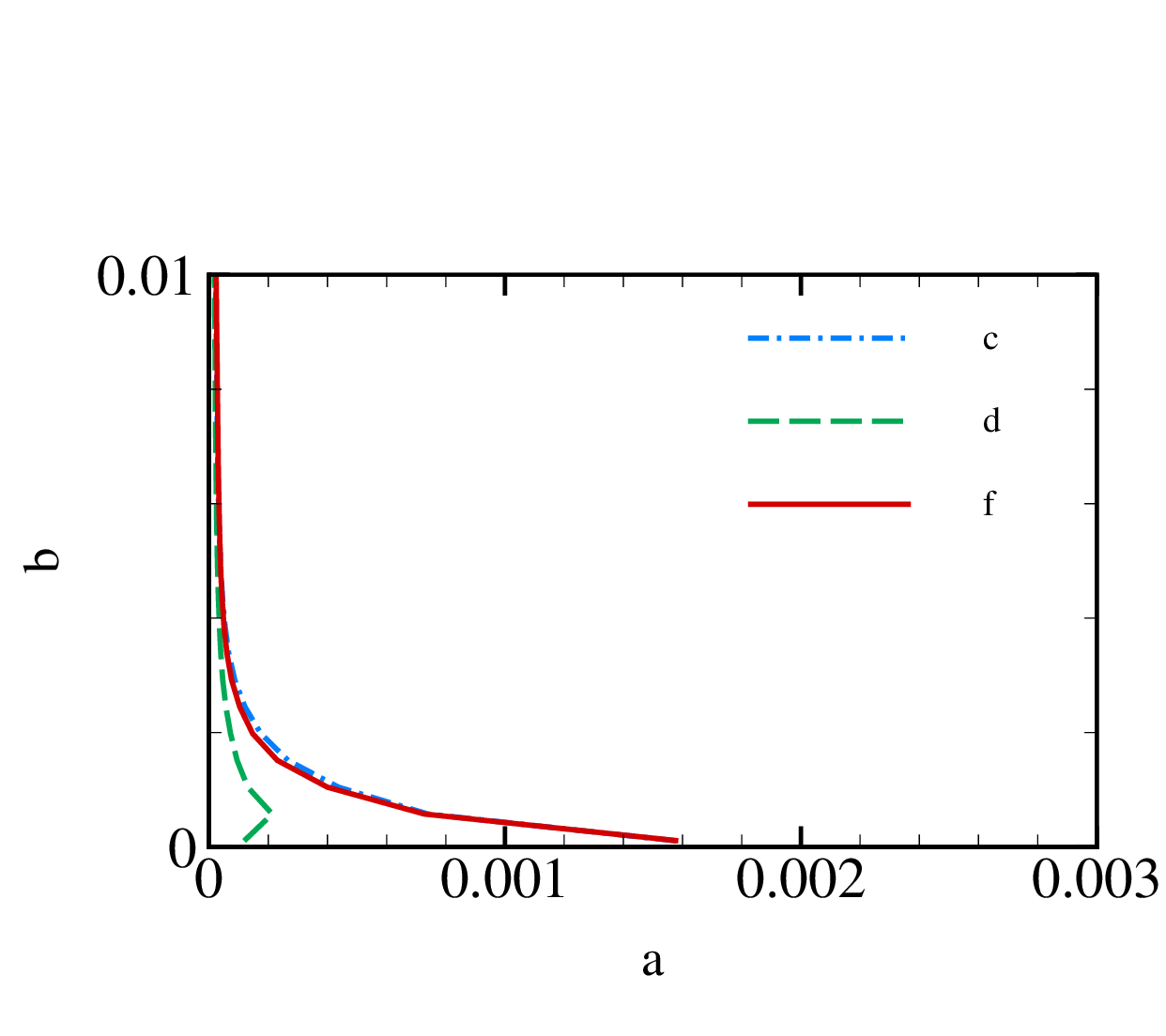}}
}
\caption{Isotropic SGS stress tensor r.m.s. components (${\tau_{11\text{, rms}}^{\text{iso}}}$, ${\tau_{12\text{, rms}}^{\text{iso}}}$, and ${\tau_{22\text{, rms}}^{\text{iso}}}$) (a) and anisotropic SGS stress tensor r.m.s. components (${\tau_{11\text{, rms}}^{\text{ani}}}$, ${\tau_{12\text{, rms}}^{\text{ani}}}$, and ${\tau_{22\text{, rms}}^{\text{ani}}}$) (b) at $x/L=-0.1$ for medium-mesh simulations with the MSM.}
\label{rms_SGS_compare_MSM_part}
\end{figure}

Figure~\ref{rms_SGS_compare_MSM_part} shows the isotropic and anisotropic contributions to the SGS stress fluctuations at $x/L=-0.1$ for the medium-mesh simulation with the MSM. The isotropic fluctuations behave similarly to those in the SM, confirming that the anisotropic term does not substantially alter the isotropic component. The anisotropic fluctuations are dominated by the normal components ${\tau_{11}^{\text{ani}}}$ and ${\tau_{22}^{\text{ani}}}$, consistent with the mean anisotropic stress behaviour, and comparison with figure~\ref{rms_SGS_compare} confirms these as the primary source of the enhanced normal stress fluctuations in the MSM.

These observations confirm that the differences in SGS dissipation and diffusion of Reynolds stresses discussed in \Cref{sec:Reynolds_budgets} stem from the anisotropic stress term in the MSM, not from modifications to the isotropic component. This suggests that optimizing the coefficient and formulation of the anisotropic stress term offers an effective pathway for controlling SGS stress fluctuations and improving Reynolds stress and mean velocity predictions.

\subsection{A priori analysis of filtered DNS}

To further validate the \emph{a posteriori} findings and gain additional insight into SGS stress characteristics in wall-bounded turbulence under FPG, we perform an \emph{a priori} analysis using Gaussian-filtered DNS of turbulent Couette--Poiseuille flow at $Re_H=U_cH/\nu=2{,}500$, where $H$ is the half-channel height and $U_c$ is the wall motion speed. Details of the DNS and filtering procedure are provided in {Appendices~\cref{appB} and \cref{appC}}.

The analysis focuses on the lower half of the channel, where the flow near the bottom wall experiences an FPG and behaves qualitatively similarly to a TBL with FPG. Gaussian filtering is applied only in the streamwise ($x$) and spanwise ($z$) directions, with standard deviations $\sigma_x/\Delta_x=\sigma_z/\Delta_z=1$, 2, and 4, where $\Delta_x$ and $\Delta_z$ are the uniform DNS grid spacings. The SGS stress is computed as $\tau^\text{sgs}_{ij} = \widehat{u_i u_j}-\widehat{u}_i\hspace{1pt}\widehat{u}_j$, where $\widehat{(\cdot)}$ denotes Gaussian filtering, and only the deviatoric part is considered.

\begin{figure}
\centering
\sidesubfloat[]{
{\psfrag{a}[][]{{$\overline{\tau_{ij}^{\text{sgs}}}/(\rho U_c^2)$}}
\psfrag{b}[][]{{$y/H$}}
\psfrag{c}[][]{{\scriptsize$\overline{\tau_{11}^{\text{sgs}}}$}}
\psfrag{d}[][]{{\scriptsize$\overline{\tau_{12}^{\text{sgs}}}$}}
\psfrag{e}[][]{{\scriptsize$\overline{\tau_{13}^{\text{sgs}}}$}}
\psfrag{f}[][]{{\scriptsize$\overline{\tau_{22}^{\text{sgs}}}$}}
\psfrag{g}[][]{{\scriptsize$\overline{\tau_{23}^{\text{sgs}}}$}}
\psfrag{h}[][]{{\scriptsize$\overline{\tau_{33}^{\text{sgs}}}$}}
\includegraphics[width=.46\textwidth,trim={1cm 1cm 0.1cm 3cm},clip]{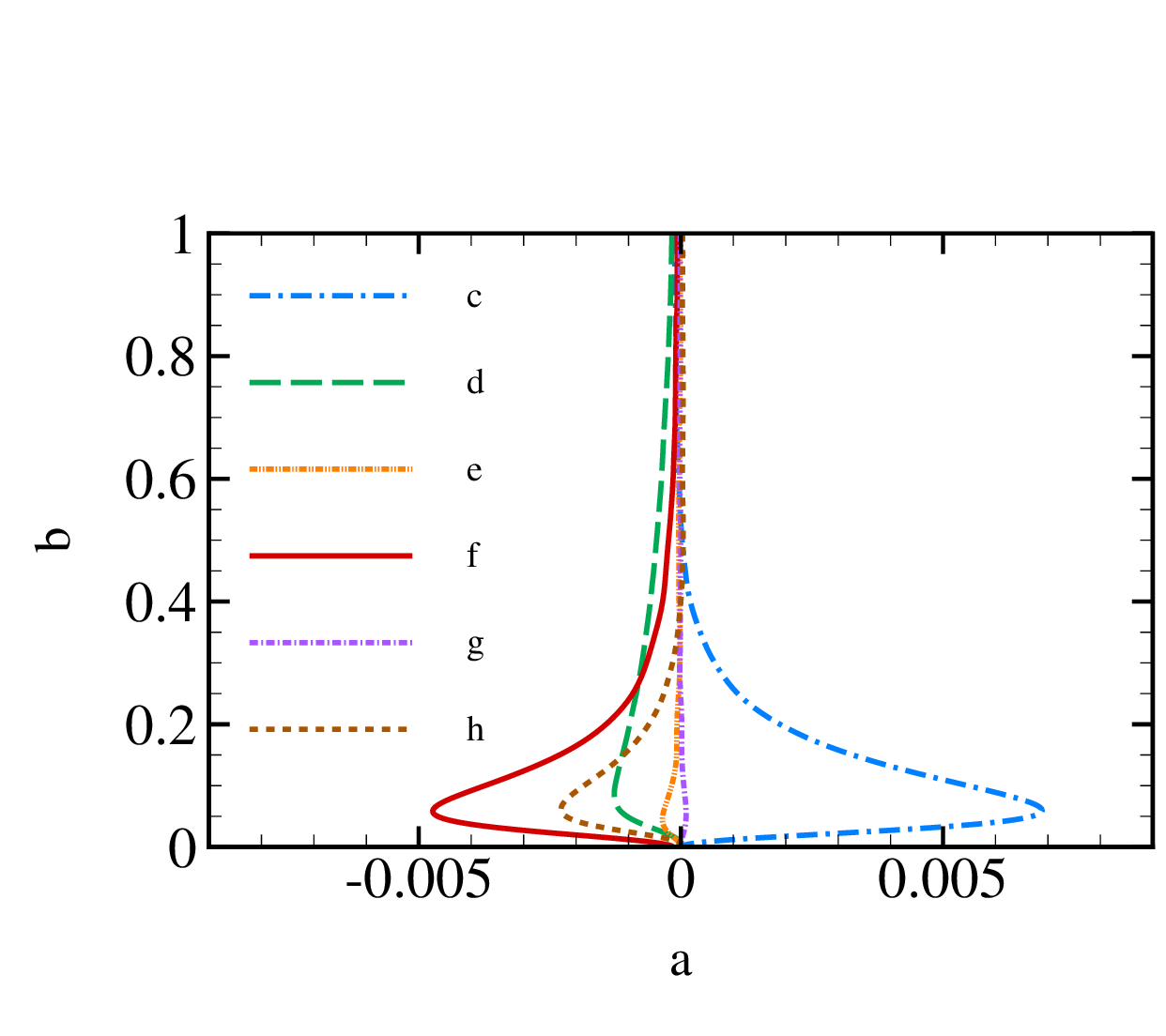}}
}
\sidesubfloat[]{
{\psfrag{a}[][]{{${\tau_{ij\text{, rms}}^{\text{sgs}}}/(\rho U_c^2)$}}
\psfrag{b}[][]{{$y/H$}}
\psfrag{c}[][]{{\scriptsize$\tau_{11\text{, rms}}^{\text{sgs}}$}}
\psfrag{d}[][]{{\scriptsize$\tau_{12\text{, rms}}^{\text{sgs}}$}}
\psfrag{e}[][]{{\scriptsize$\tau_{13\text{, rms}}^{\text{sgs}}$}}
\psfrag{f}[][]{{\scriptsize$\tau_{22\text{, rms}}^{\text{sgs}}$}}
\psfrag{g}[][]{{\scriptsize$\tau_{23\text{, rms}}^{\text{sgs}}$}}
\psfrag{h}[][]{{\scriptsize$\tau_{33\text{, rms}}^{\text{sgs}}$}}
\includegraphics[width=.46\textwidth,trim={1cm 1cm 0.1cm 3cm},clip]{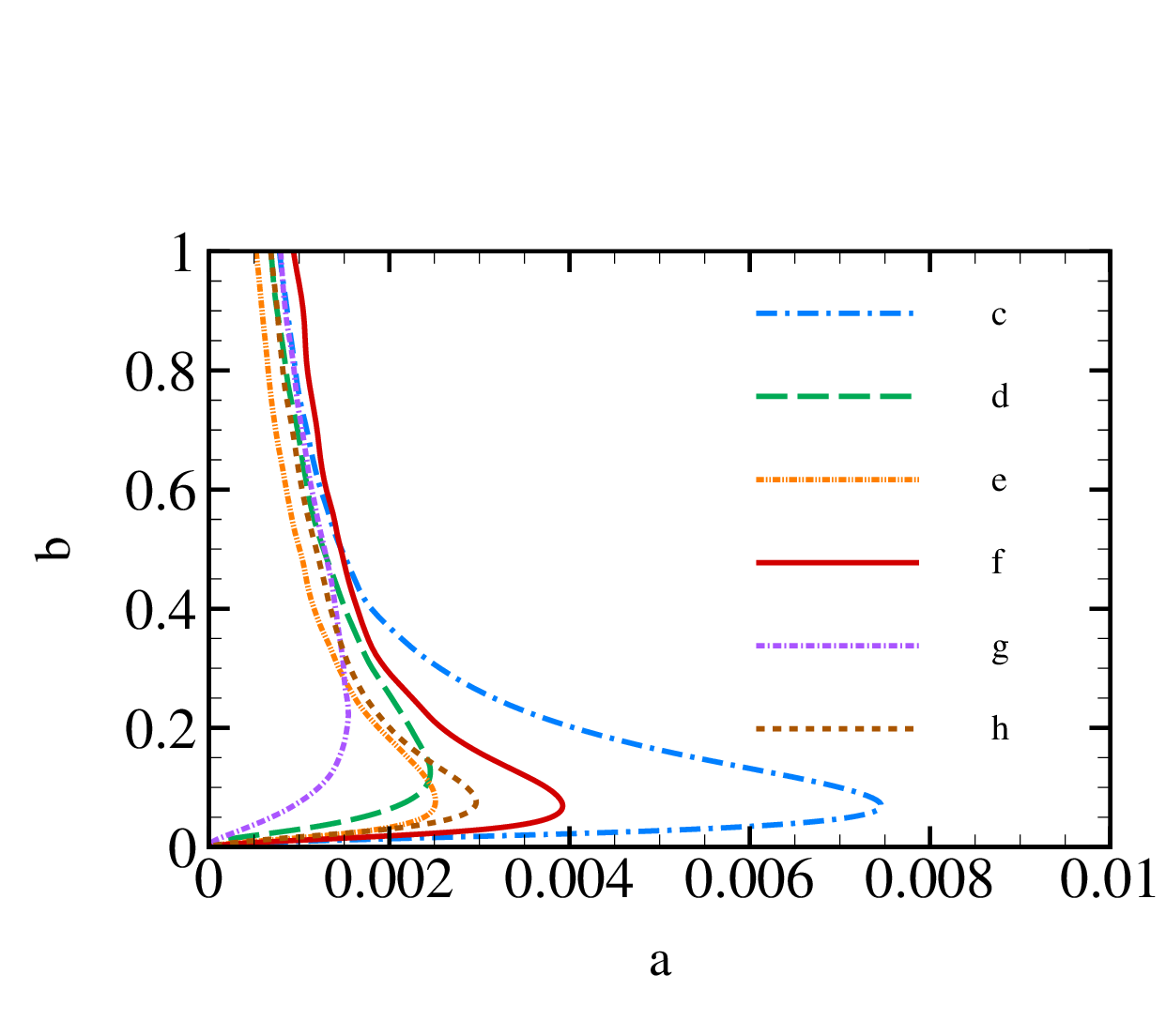}}
}
\caption{Mean SGS stress tensor components $\overline{\tau_{ij}^{\text{sgs}}}$ (a) and SGS stress tensor r.m.s. components ${\tau_{ij\text{, rms}}^{\text{sgs}}}$ (b) obtained from the Gaussian filtering of velocity field from the DNS of turbulent Couette-Poiseuille flow with the standard deviations of the Gaussian kernel $\sigma_x/\Delta_x=\sigma_z/\Delta_z=$2.}
\label{fDNS_SGS_compare}
\end{figure}

Figure~\ref{fDNS_SGS_compare}(a) shows the wall-normal distributions of mean SGS stress components for $\sigma_x/\Delta_x=\sigma_z/\Delta_z=2$. The magnitudes are largest near the wall, peaking at $y/H\approx0.06$ (approximately 20 wall units). The normal components $\tau_{11}^{\text{sgs}}$ and $\tau_{22}^{\text{sgs}}$ are significantly larger than the other components in magnitude, indicating strong SGS anisotropy near the wall. Despite the large mean shear in this region, the small-scale cross-correlation between streamwise and wall-normal velocity fluctuations is weaker than the variance of each component, keeping the mean shear stress $\tau_{12}^{\text{sgs}}$ comparatively small. 

Compared with the mean SGS stress in the TBL within the FPG region of the WMLES, the filtered DNS results are more consistent with the MSM predictions. In both cases, the anisotropic stress introduces dominant normal stress components near the wall with consistent signs for $\tau_{11}^{\text{sgs}}$ and $\tau_{22}^{\text{sgs}}$. The peaks are not fully captured in the WMLES due to the coarse wall-normal resolution.

It is also worth noting that the qualitative behaviour of the mean SGS stresses remains similar across different standard deviations of the Gaussian filter, or equivalently, different effective filter widths. For brevity, results for other filter widths (e.g., $\sigma_x/\Delta_x = \sigma_z/\Delta_z = 1$ or 4) are not shown. Examination of these cases indicates that increasing the filter standard deviation leads to larger mean SGS stress magnitudes, consistent with the fact that wider filters remove more turbulent scales and therefore attribute a greater portion of the momentum transfer to the unresolved motions.

Figure~\ref{fDNS_SGS_compare}(b) shows the r.m.s.\ values of SGS stress fluctuations from the filtered DNS. The fluctuation intensities peak near the wall at approximately the same location as the mean SGS stresses, reflecting the highly intermittent energy transfer between resolved and subgrid scales driven by the bursting and ejection--sweep cycles of near-wall turbulence. The particularly large fluctuations in $\tau_{11}^{\text{sgs}}$ and $\tau_{22}^{\text{sgs}}$ indicate substantial temporal variability in the normal stress components, likely associated with rapid distortion of streaks and vortical structures. These strong normal SGS stress fluctuations are captured by the MSM in the WMLES but are nearly absent in the SM results, reinforcing the role of the anisotropic stress term in reproducing correct near-wall SGS dynamics. The qualitative behaviour is similar across filter widths, with fluctuation intensities increasing with filter width.

From the filtered DNS analysis and its comparison with the WMLES results, it is evident that in wall-bounded turbulence under FPG, SGS dynamics are strongly anisotropic near the wall, with the normal SGS stress components playing a major role. Classical eddy-viscosity models cannot represent these effects adequately. In contrast, the anisotropic SGS model reproduces the near-wall anisotropy much more realistically, consistent with the filtered DNS characteristics. These findings suggest that improving or optimizing the anisotropic stress term and its coefficient in SGS models offers a promising pathway for enhancing model performance in simulations of complex turbulent flows.

\section{Conclusions}\label{sec:conclusion}

This study performed a comprehensive \emph{a posteriori} analysis of the effect of anisotropic SGS stress on WMLES of separated turbulent flow over a spanwise-uniform Gaussian bump. An idealized wall boundary condition prescribing the local mean wall-shear stress from DNS data \citep{uzun2022high} was used to isolate the impact of the SGS model. Two models were compared: the classical Smagorinsky model (SM) and a modified Smagorinsky model (MSM) that includes an additional anisotropic stress term.

The main findings are summarized as follows. First, the predicted flow separation on the leeward side of the bump depends strongly on the SGS model. The isotropic SM exhibits non-monotonic convergence of the mean separation bubble length with mesh refinement, whereas the anisotropic MSM provides consistent predictions across resolutions. Second, the influence of anisotropic SGS stress is found to be most critical upstream of the bump peak, within the region of strong FPG. Changes to the SGS model in this region substantially alter the downstream separation, revealing a pronounced history effect in determining separation onset. Third, inclusion of anisotropic SGS stress improves the prediction of Reynolds shear and normal stress distributions in the FPG region. These modifications propagate downstream and influence the onset and size of the separation bubble. Analysis of the Reynolds stress budget shows that anisotropic SGS stress fluctuations enable both dissipation and backscatter, facilitating the bidirectional energy transfer that isotropic models fail to represent. The dependence of flow-separation prediction on mesh resolution is also clarified. On coarse meshes, the mean SGS shear stress dominates the streamwise momentum balance upstream of the separation point, and both models behave similarly. As the resolution increases, Reynolds stresses become more influential, and the anisotropic MSM better captures their distribution and yields more consistent flow predictions. At fine resolution, model differences diminish as more turbulent scales are resolved. The key physical distinction between the SM and MSM arises from the normal SGS stress components, $\tau_{11}^{\text{sgs}}$ and $\tau_{22}^{\text{sgs}}$, which in the MSM significantly contribute to SGS dissipation and diffusion of Reynolds stresses, particularly under FPG. An \emph{a priori} analysis based on filtered DNS of Couette–Poiseuille flow further confirms that near-wall turbulence under FPG is highly anisotropic and dominated by these normal stress components, which are not captured by isotropic eddy-viscosity models.

Taken together, these findings explain why isotropic and anisotropic SGS models yield qualitatively different predictions of separation behaviour and why the anisotropic model achieves more consistent convergence across mesh resolutions. The results emphasize that accurate WMLES predictions require proper representation of both mean and fluctuating SGS stresses, especially their anisotropy in the near-wall region.

Beyond elucidating the role of anisotropic SGS stress, this study highlights directions for improving WMLES of complex wall-bounded turbulence. Since the unresolved motions in WMLES carry substantial energy and momentum fluxes, the SGS model must account for anisotropic stress dynamics near the wall and under pressure gradients. Developing more advanced anisotropic SGS models, potentially through optimized extensions of eddy-viscosity formulations \citep{marstorp2009explicit, silvis2019nonlinear, agrawal2022non, uzun2025application}, is therefore a promising path forward. Although this work employed an idealized wall model {based on the exact mean wall-shear stress} to isolate SGS effects, {simulations using other idealized wall-model formulations \citep{zhou2024sensitivity}, as well as an equilibrium wall model \citep{wang2002dynamic}, also show improved prediction of flow separation when combined with the anisotropic SGS models considered in the present study. Moreover, to further enhance the overall performance of WMLES, the coupling between realistic wall models and SGS models remains a critical challenge that needs to be addressed. In this context, unified SGS/wall modeling frameworks \citep{ling2022wall, arranz2023wall, arranz2024building, zhou2025wall} represent a promising direction to improve the robustness and predictive accuracy of WMLES for complex turbulent flows.}

\section*{Acknowledgments}
This work was supported by National Science Foundation (NSF) grant No.~2152705. Computer time was provided by the Discover project at Pittsburgh Supercomputing Center through allocation PHY240020 from the Advanced Cyberinfrastructure Coordination Ecosystem: Services \& Support (ACCESS) program, which is supported by NSF grants No.~2138259, No.~2138286, No.~2138307, No.~2137603, and No.~2138296. The authors sincerely thank Dr.~Ali Uzun and Dr.~Mujeeb Malik for generously sharing their DNS data. We also extend our special gratitude to Dr.~Meng Wang for his invaluable assistance.

\section*{Declaration of Interests}
The authors report no conflict of interest.

\appendix

\section{Investigation of the curvature effect on the region upstream of separation}\label{appA_new}

\begin{figure}
\centering
\sidesubfloat[]{
{\psfrag{a}[][]{{$x/L$}}
\psfrag{b}[][]{{$r/L$}}
\includegraphics[width=.47\textwidth,trim={0.8cm 4.5cm 1cm 4.5cm},clip]{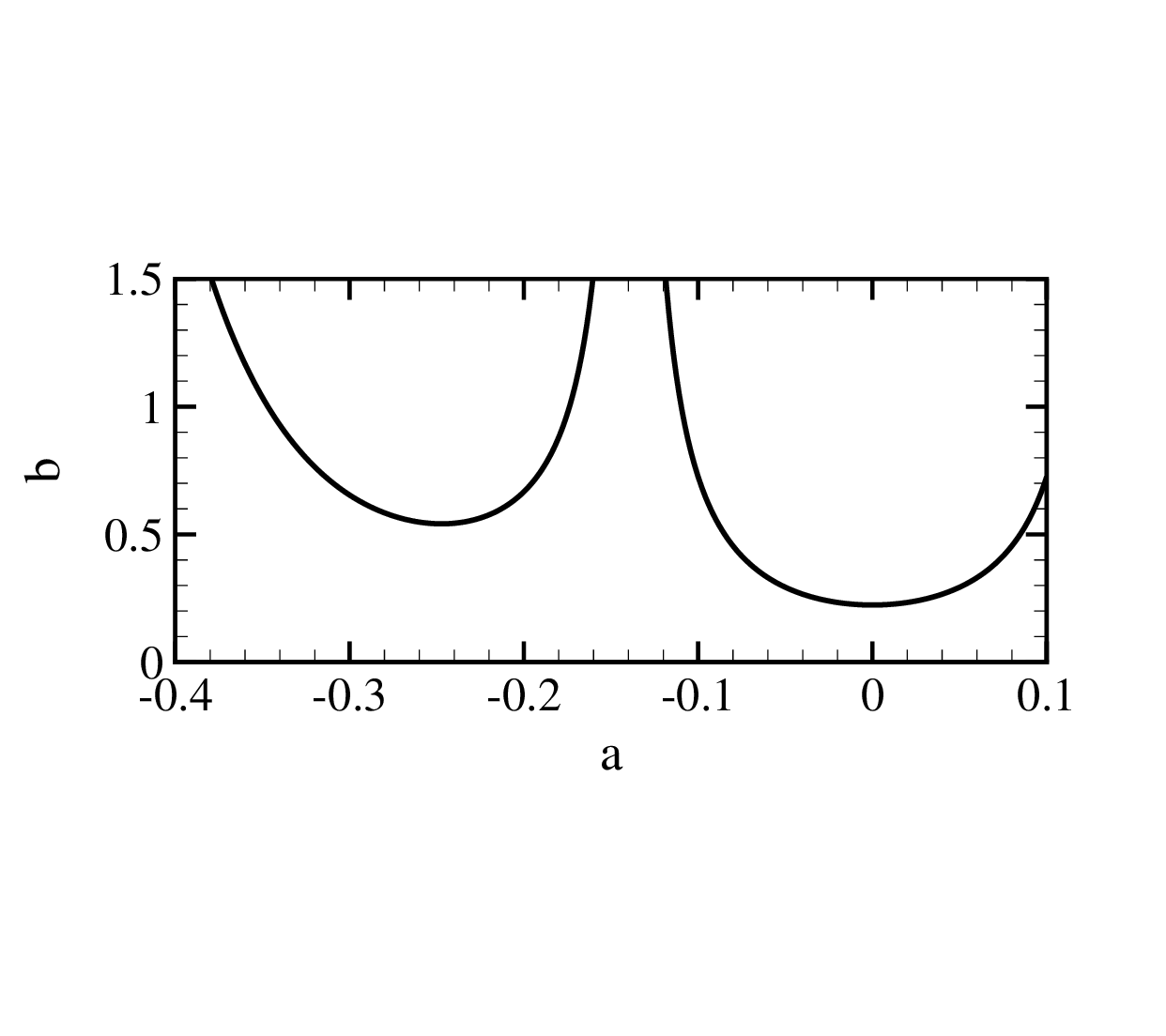}}
}
\sidesubfloat[]{
{\psfrag{a}[][]{{$x/L$}}
\psfrag{b}[][]{{$\delta/L$}}
\includegraphics[width=.47\textwidth,trim={0.8cm 4.5cm 1cm 4.5cm},clip]{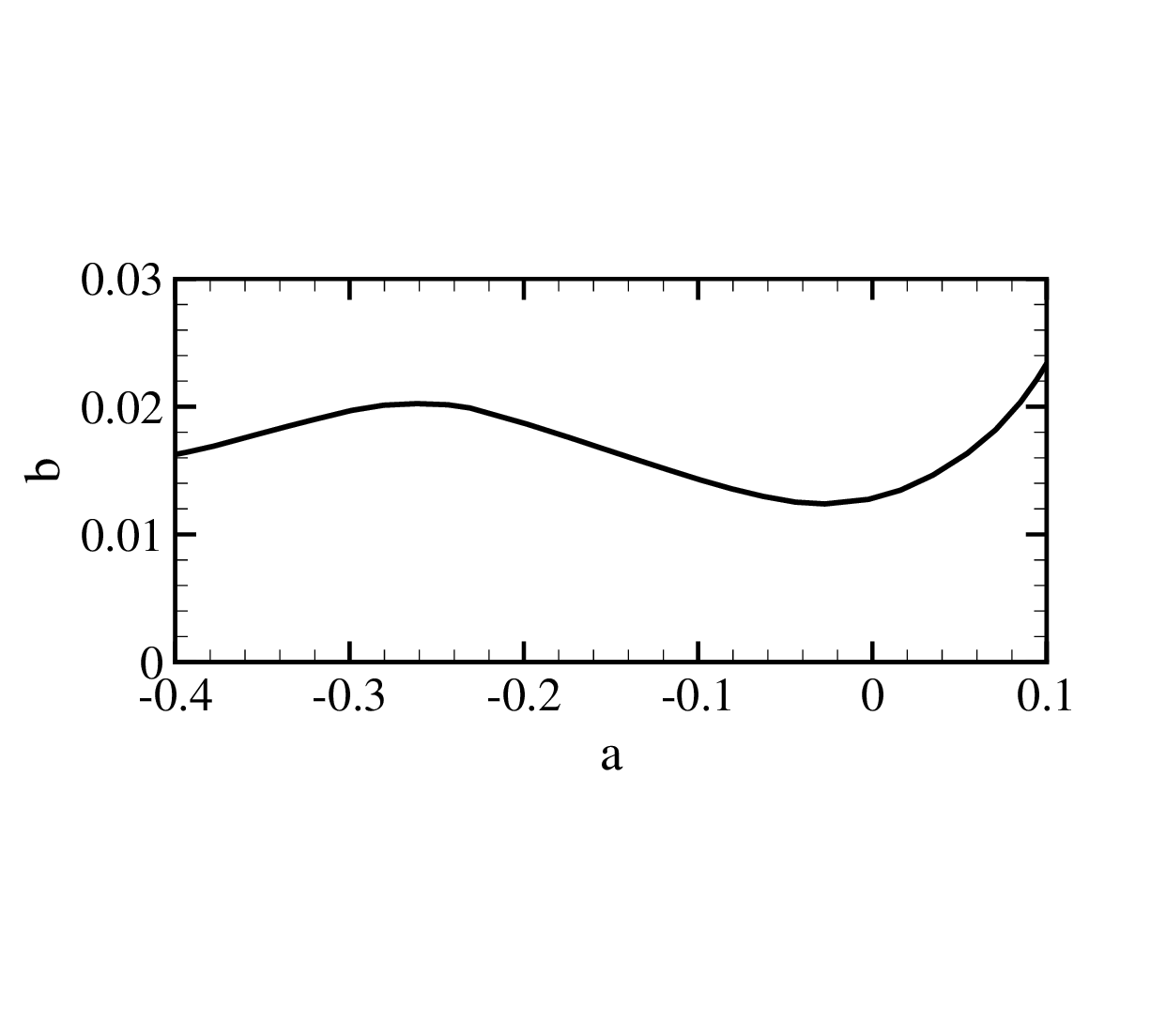}}
}
\caption{Curvature radius (a) of the Gaussian bump surface and local boundary layer thickness (b) for the medium-mesh simulation with the SM in $x/L\in[-0.4, 0.1]$.}
\label{curvature}
\end{figure}

Figure~\ref{curvature} shows the local curvature radius ($r$) of the present Gaussian bump surface and the local boundary layer thickness ($\delta$) from the medium-mesh simulation with the SM in the region $x/L\in[-0.4, 0.1]$. The curvature radius is substantially larger than the local boundary layer thickness throughout, with the ratio exceeding 10 even at the bump peak where curvature is largest. These observations indicate that curvature effects in this region are negligible, consistent with findings of previous studies \citep{prakash2024streamline, spalart2024direct}.

\section{Simulations using the mixed model}\label{appA}
An additional anisotropic SGS model evaluated in the present study is the mixed model (MM) \citep{bardina1983improved, sarghini1999scale}. It combines the SM \citep{smagorinsky1963general} with a scale-similarity term computed using explicit filtering \citep{meneveau2000scale}, which is given by
\begin{equation}\label{MSMeq}
{\tau}^\text{ani}_{ij} = \widehat{u_i\hspace{2pt}u_j}-\hat{u}_i\hspace{2pt}\hat{u}_j\hspace{2pt},
\end{equation} where $\widehat{(\cdot)}$ denotes an explicit filtering operation, chosen here as Gaussian filtering. With the use of Gaussian filtering, the anisotropic term can be approximated \citep{clark1979evaluation} by
\begin{equation}\label{MSMeq_exp}
{\tau}^\text{ani}_{ij} = \widehat{u_i\hspace{2pt}u_j}-\hat{u}_i\hspace{2pt}\hat{u}_j \approx \frac{\Delta^2}{12} \frac{\partial u_{i}}{\partial x_k} \frac{\partial u_{j}}{\partial x_k} = \frac{\Delta^2}{12}\left[S_{ik}S_{kj}-R_{ik} R_{kj}-\left(S_{ik} R_{kj}-R_{ik} S_{kj}\right)\right] \hspace{2pt}.
\end{equation} In addition to introducing anisotropic SGS stress, this term can produce kinetic energy dissipation. In particular, it can also produce negative dissipation, which allows for local backscatter of kinetic energy. The isotropic SGS stress term in the MM is given by the SM with a coefficient of $C_s=0.127$ \citep{bhushan2005large, bhushan2006modeling}, chosen to ensure that the total kinetic energy dissipation for homogeneous isotropic turbulence from both the isotropic and anisotropic stress terms matches that from the SM and MSM used in the present study. The MM simulations are conducted using the same computational meshes and boundary conditions described in \Cref{sec:Computation_method}. 

\begin{figure}
\centering
{\psfrag{a}[][]{{$\Delta_{\text{c}}/L{\times10^3}$}}
\psfrag{b}[][]{{$L_{\text{s}}/L$}}
\includegraphics[width=.65\textwidth,trim={0.1cm 7.5cm 0.1cm 0.1cm},clip]{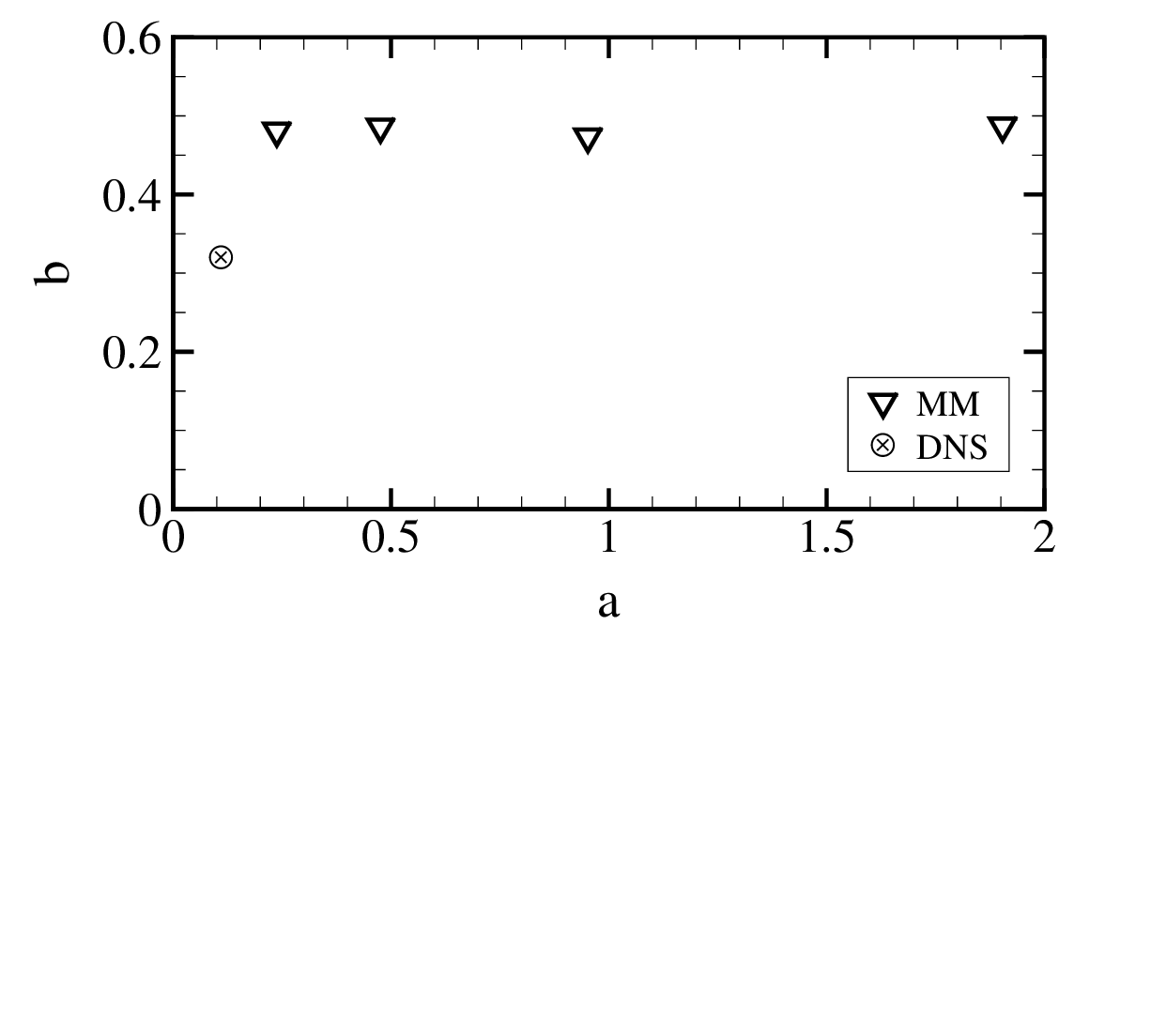}}
\caption{Mean separation bubble length on the leeward side of the bump from the simulations using the MM for different mesh resolutions and the reference DNS \citep{uzun2022high}. Symbols represent data point for each case.}
\label{separation_bubble_size_appendix}
\end{figure}

Figure~\ref{separation_bubble_size_appendix} shows the mean separation bubble length on the leeward side of the bump as a function of characteristic mesh resolution. The MM simulations consistently overpredict the separation bubble length, but the predictions remain nearly insensitive to mesh refinement. This trend mirrors the behaviour observed for the MSM simulations (see figure~\ref{separation_bubble_size}), further underscoring the robustness of anisotropic SGS models for WMLES.

Additional analyses of the mean streamwise momentum, mean pressure, and Reynolds stress transport equations using data from multiple mesh resolutions show that the MM exhibits behaviour qualitatively similar to the MSM, particularly in its representation of anisotropic stress effects. Thus, the conclusions drawn in the main text remain applicable to the MM simulations.

{\section{Budget terms for the transport equation of Reynolds stresses $\overline{u_1'u_2'}$ and $\overline{u_2'u_2'}$}\label{appC_add}}

\begin{figure}
\centering
\sidesubfloat[]{
{\psfrag{b}[][]{{$x_2/L$}}
\psfrag{c}[][]{{$P_{12}$}}
\psfrag{d}[][]{{$\varepsilon_{12}$}}
\psfrag{e}[][]{{$\phi_{12}$}}
\psfrag{f}[][]{{$\xi_{12}$}}
\psfrag{g}[][]{{$\frac{\partial }{\partial  x_k}\zeta_{12k}$}}
\psfrag{h}[][]{{$\frac{\partial }{\partial  x_k} D_{12k}$}}
\psfrag{i}[][]{{$\frac{\partial }{\partial  x_k} T_{12k}$}}
\psfrag{j}[][]{{$\frac{\partial }{\partial  x_k} J_{12k}$}}
\psfrag{k}[][]{$\overline{u}_k\frac{\partial  \overline{ u_1'u_2'}
}{\partial  x_k}$}
\psfrag{n}[][]{{$\text{Res}$}}
\includegraphics[width=.46\textwidth,trim={1.3cm 0.1cm 0.1cm 0.1cm},clip]{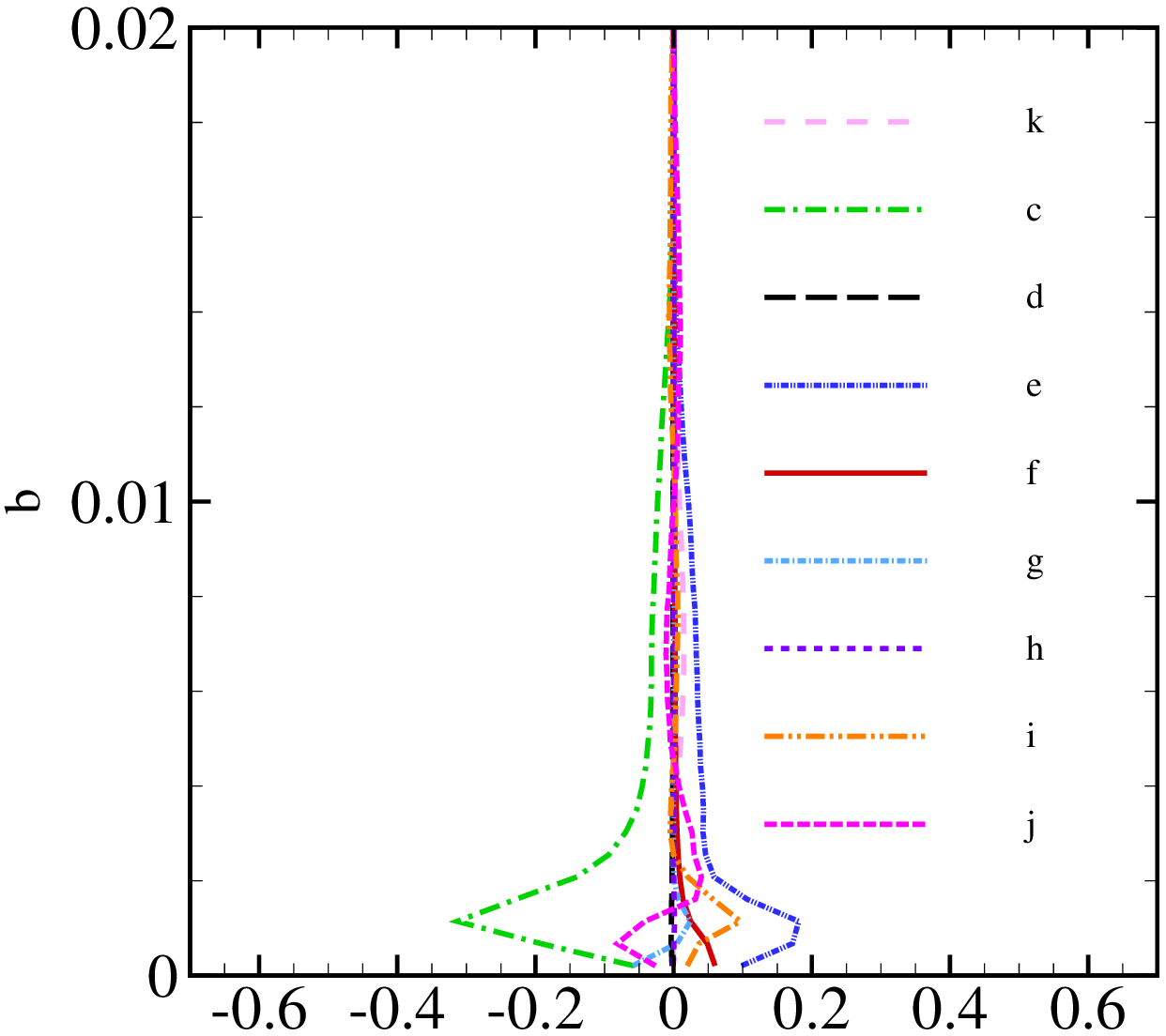}}
}
\sidesubfloat[]{
{\psfrag{b}[][]{{$x_2/L$}}
\psfrag{c}[][]{{$P_{22}$}}
\psfrag{d}[][]{{$\varepsilon_{22}$}}
\psfrag{e}[][]{{$\phi_{22}$}}
\psfrag{f}[][]{{$\xi_{22}$}}
\psfrag{g}[][]{{$\frac{\partial }{\partial  x_k}\zeta_{22k}$}}
\psfrag{h}[][]{{$\frac{\partial }{\partial  x_k} D_{22k}$}}
\psfrag{i}[][]{{$\frac{\partial }{\partial  x_k} T_{22k}$}}
\psfrag{j}[][]{{$\frac{\partial }{\partial  x_k} J_{22k}$}}
\psfrag{k}[][]{$\overline{u}_k\frac{\partial  \overline{ u_2'u_2'}
}{\partial  x_k}$}
\psfrag{n}[][]{{$\text{Res}$}}
\includegraphics[width=.46\textwidth,trim={1.3cm 0.1cm 0.1cm 0.1cm},clip]{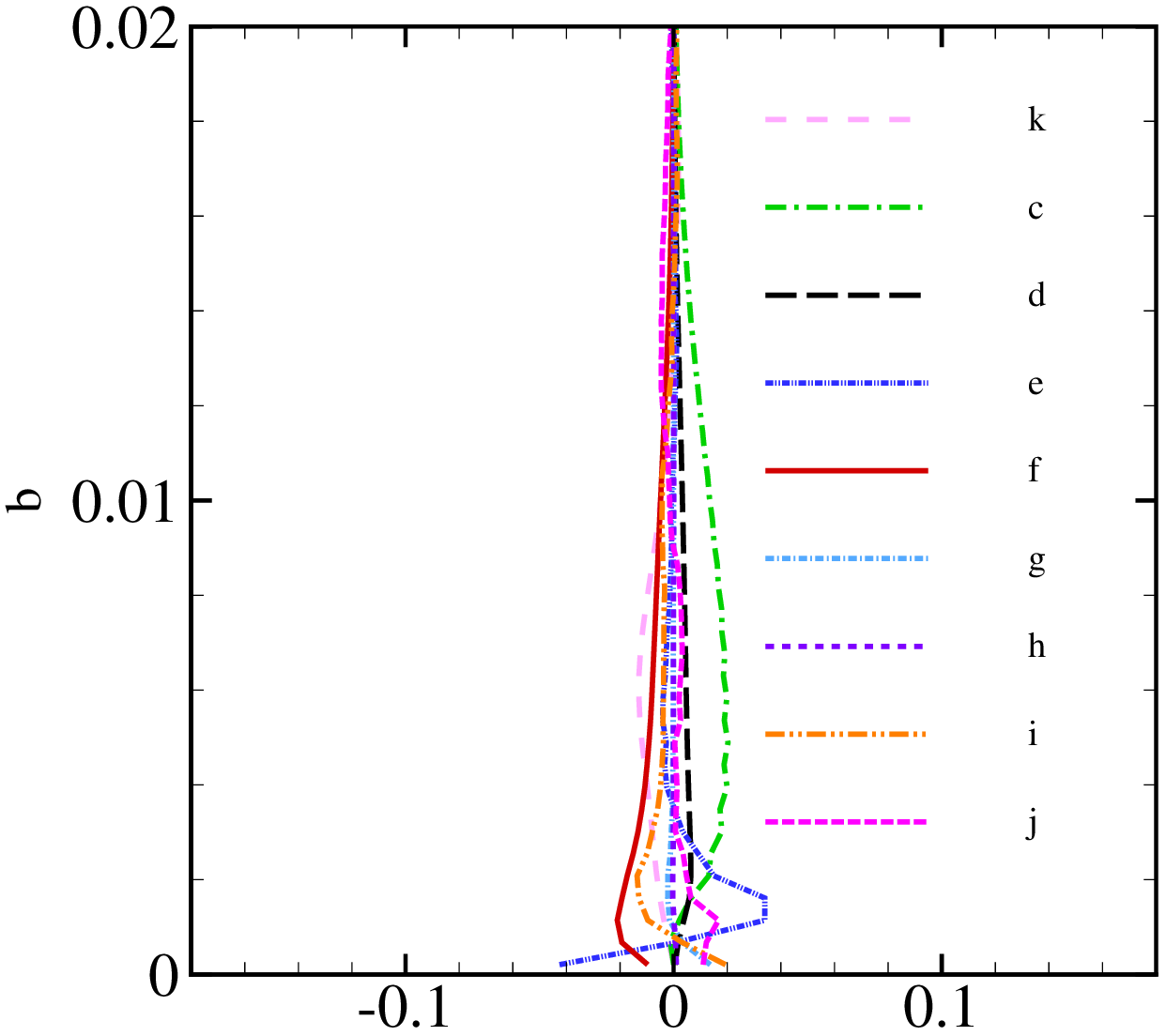}}
}
\vspace{5pt}
\sidesubfloat[]{
{\psfrag{a}[][]{{}}
\psfrag{b}[][]{{$x_2/L$}}
\includegraphics[width=.46\textwidth,trim={1.3cm 0.1cm 0.1cm 0.1cm},clip]{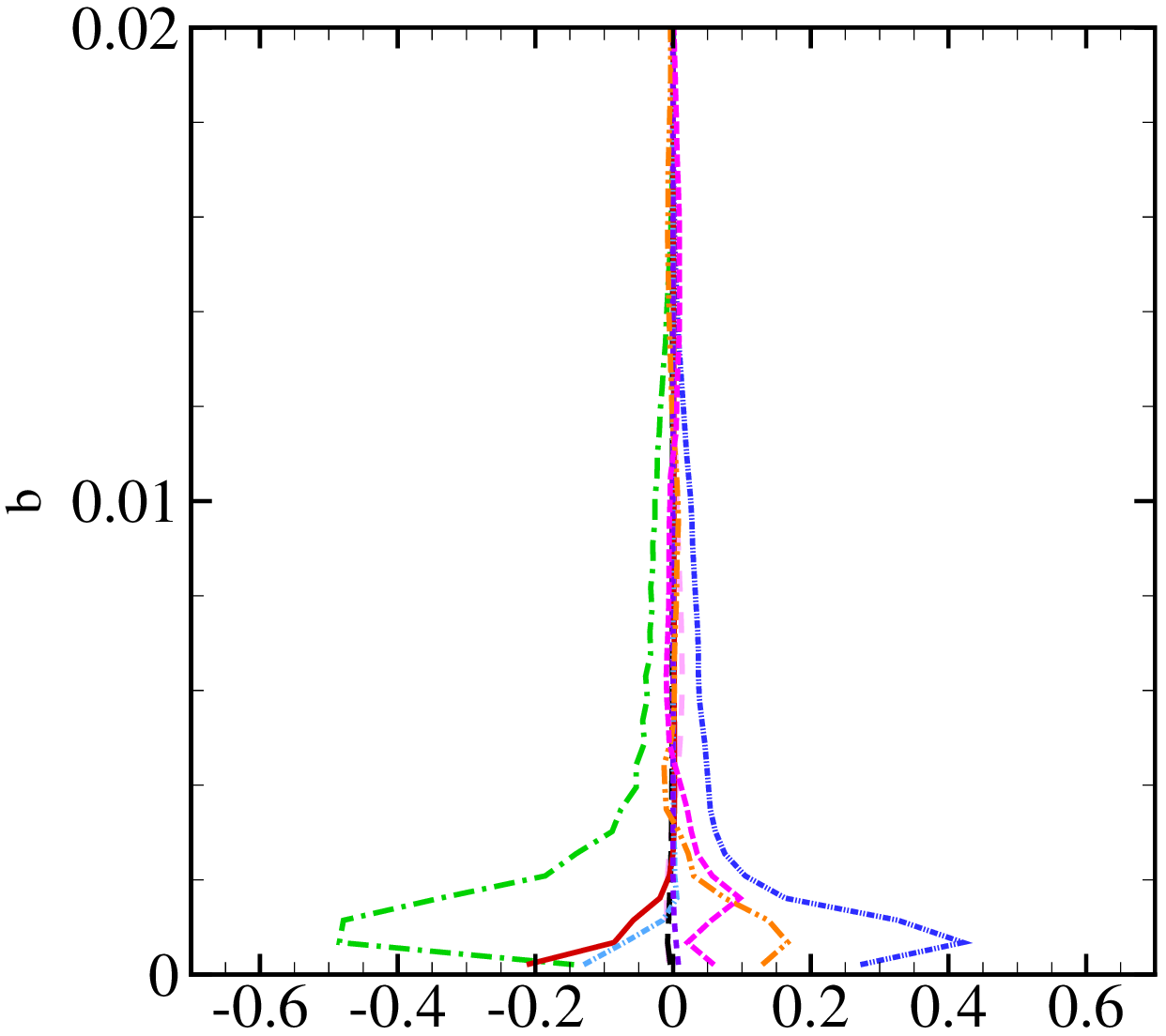}}
}
\sidesubfloat[]{
{\psfrag{a}[][]{{}}
\psfrag{b}[][]{{$x_2/L$}}
\includegraphics[width=.46\textwidth,trim={1.3cm 0.1cm 0.1cm 0.1cm},clip]{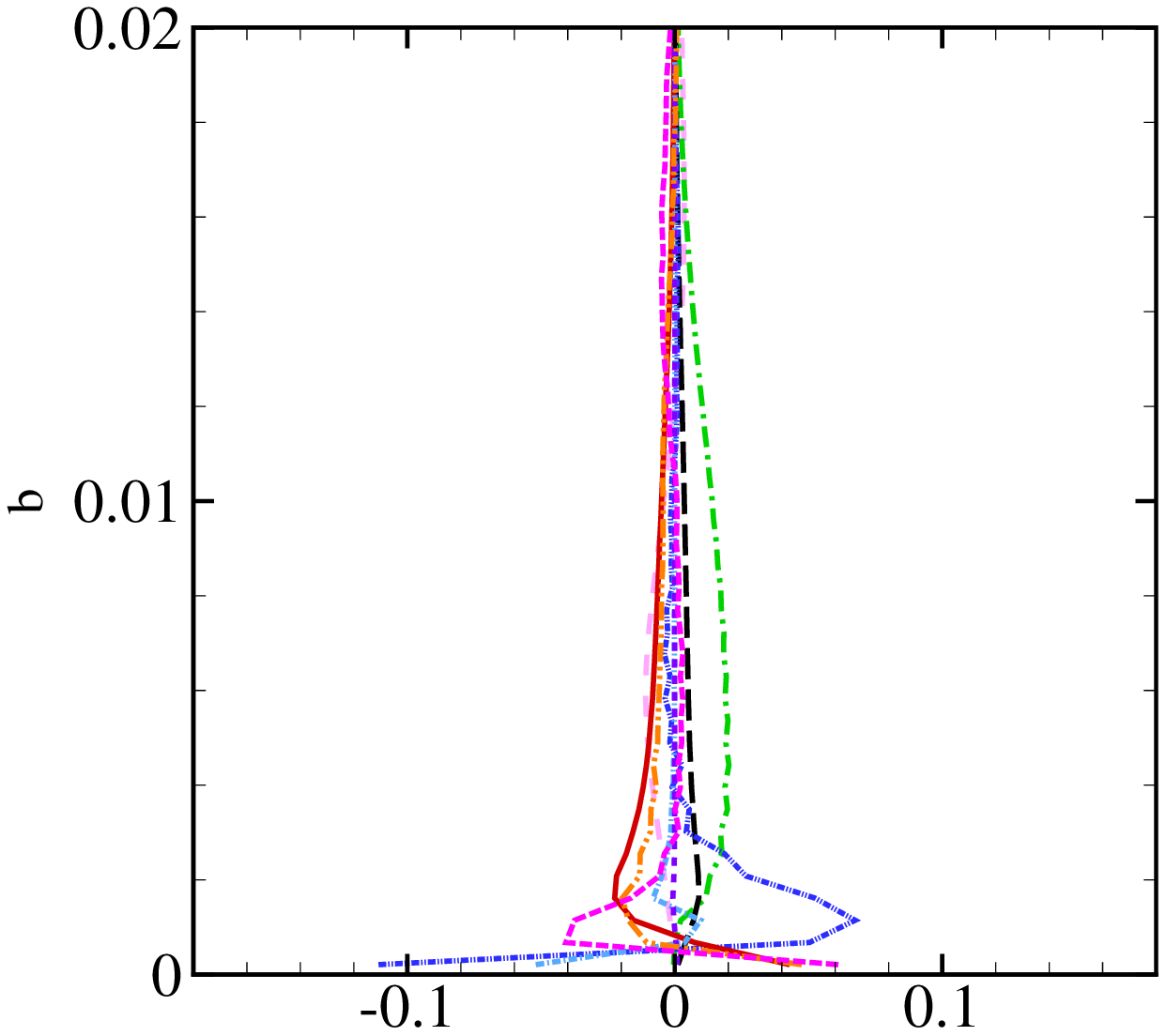}}
}
\caption{{Budget terms for the Reynolds shear stress $\overline{u_1'u_2'}$ (a,c) and Reynolds normal stress $\overline{u_2'u_2'}$ (b,d) transport equations at $x/L=-0.1$ from medium-mesh simulations with the SM (a,b) and MSM (c,d). The line notations correspond to the terms in equation~\eqref{Re_term_all}, excluding the unsteady term that vanishes for the statistically stationary flow, and all terms are nondimensionalized using $U_\infty$, $L$ and $\rho$.}}
\label{Reynolds12_budget_bump}
\end{figure}

Figure~\ref{Reynolds12_budget_bump} shows the wall-normal distributions of all budget terms in the transport equations for the Reynolds shear stress $\overline{u'_1 u'_2}$ and the wall-normal Reynolds normal stress $\overline{u'_2 u'_2}$ at $x/L = -0.1$ from the medium-mesh simulations with the SM and MSM, within the critical FPG region identified in \Cref{sec:identification}. {In addition to the source, sink, and transport terms on the right-hand side of equation~\eqref{Re_term_all}, the convection term on the left-hand side is also shown.}

For $\overline{u_1'u_2'}$, the dominant terms are production $P_{12}$ and pressure strain $\phi_{12}$, with the MSM exhibiting larger magnitudes of both in the near-wall region. Turbulent and pressure diffusion contribute mainly in the region very close to the wall, while viscous diffusion and dissipation remain small throughout. The SGS dissipation $\xi_{12}$ and diffusion $\frac{\partial}{\partial x_k}\zeta_{12k}$ are not the largest terms in absolute magnitude but show clear qualitative differences between the two models, as discussed in \Cref{sec:Reynolds_budgets}.

For $\overline{u_2'u_2'}$, production $P_{22}$ is small near the wall and the dominant source is pressure strain $\phi_{22}$, which redistributes turbulent kinetic energy into the wall-normal stress. The MSM produces more pronounced $\phi_{22}$, consistent with the internal peak in the $\overline{u_2'u_2'}$ profiles within the FPG region (figure~\ref{R12R22_bump}). As with the shear-stress budget, the SGS dissipation and diffusion terms show qualitative differences between the two models.

{The convection term is small throughout the boundary layer for both Reynolds stress components, indicating that the budget is dominated locally by the source, sink, and transport terms on the right-hand side rather than by streamwise mean-flow transport. The residual reflecting the numerical and statistical closure error is small relative to the leading-order terms and smaller than the SGS dissipation and SGS diffusion terms throughout the boundary layer, thus it is not shown for brevity.}

Overall, the budget comparison shows that production and pressure strain are the leading-order terms governing the Reynolds stress evolution, while the SGS-related terms are smaller in magnitude. However, the SGS dissipation and SGS diffusion terms provide the pathway through which the SGS model enters the Reynolds stress transport. Their different behaviour between the SM and MSM, particularly the backscatter and wall-normal redistribution introduced by the anisotropic stress term, modifies the overall budget balance. Through the coupled nature of the governing equations, these changes lead to corresponding adjustments in the other budget terms. The enhanced production and pressure strain observed in the MSM simulation can therefore be interpreted as a response of the resolved flow to the different SGS forcing. Although the SGS dissipation represents the transfer of energy from resolved to unresolved scales and the SGS diffusion corresponds to its spatial redistribution, their importance here lies less in their absolute magnitude than in their influence on the overall balance. In this sense, differences between SGS models do not act in isolation but propagate through the coupled system, ultimately affecting the distribution of Reynolds stresses.\\

\section{DNS of plane Couette-Poiseuille flow}\label{appB}

A DNS of turbulent plane Couette–Poiseuille flow is conducted at $Re_{H} = 2{,}500$, where $H$ denotes the half-channel height. In the simulation, the incompressible Navier–Stokes equations are solved using a staggered finite-difference scheme that is second-order accurate in space and advanced in time with an explicit third-order Runge–Kutta method. The flow solver has been validated in previous studies of turbulent channel flows \citep{bae2018, bae2019dynamic}. In the computational domain, periodic boundary conditions are imposed in the streamwise ($x$) and spanwise ($z$) directions. The top and bottom walls move parallel to each other in opposite directions along the $x$ axis. A Dirichlet boundary condition with constant velocity $u_x=U_c=1$ is applied at the top wall ($y/H=2$), while a constant velocity $u_x=-U_c=-1$ is imposed at the bottom wall ($y/H=0$). A constant streamwise pressure gradient, corresponding to $H/(\rho U_c^2)\frac{dp}{dx} = -0.003$, is applied to drive the flow, indicating that the mean pressure decreases in the positive $x$ direction. The computational domain extends over $L_x/H = 6\pi$, $L_y/H = 2$, and $L_z/H = 3\pi$ in the streamwise, wall-normal, and spanwise directions, respectively. Uniform grids with 512 points are used in both the streamwise and spanwise directions. In the wall-normal direction, 256 non-uniformly spaced points are distributed according to a hyperbolic tangent stretching, yielding $\min(\Delta y)/H = 3.5\times10^{-4}$ and $\max(\Delta y)/H = 2.2\times10^{-2}$. The simulation is first advanced for 100 flow-through times ($6\pi H/U_c$) to eliminate initial transients. After the flow reaches a statistically stationary state, $1{,}500$ temporal snapshots are collected for the present analysis. Moreover, the statistical quantities in the present study are obtained by performing temporal and spatial averaging along the homogeneous streamwise and spanwise directions of the corresponding instantaneous fields.

\begin{figure}
\centering
\includegraphics[width=.85\textwidth,trim={0.0cm 0.0cm 0.1cm 0.0cm},clip]{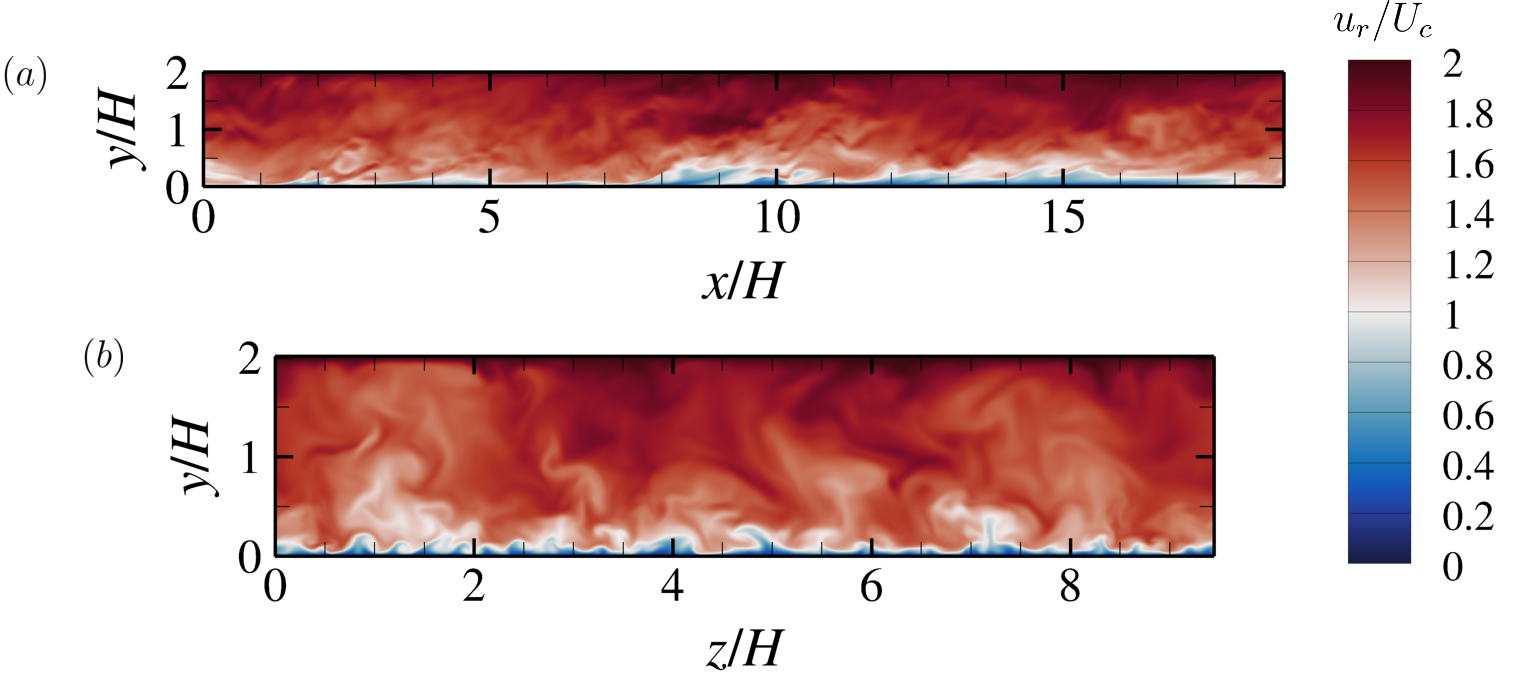}
\caption{Isocontours of the instantaneous streamwise velocity relative to the bottom wall ${u}_r/U_{c}$ in (a) an $x$-$y$ plane and (b) a $z$-$y$ plane from the DNS of turbulent Couette–Poiseuille flow.}
\label{CP_flow_ux_ins}
\end{figure}

\begin{figure}
\centering
{\psfrag{a}[][]{{$y^+$}}
\psfrag{b}[][]{{$\overline{u_r^+}$}}\includegraphics[width=.7\textwidth,trim={0.2cm 5cm 0.2cm 1.5cm},clip]{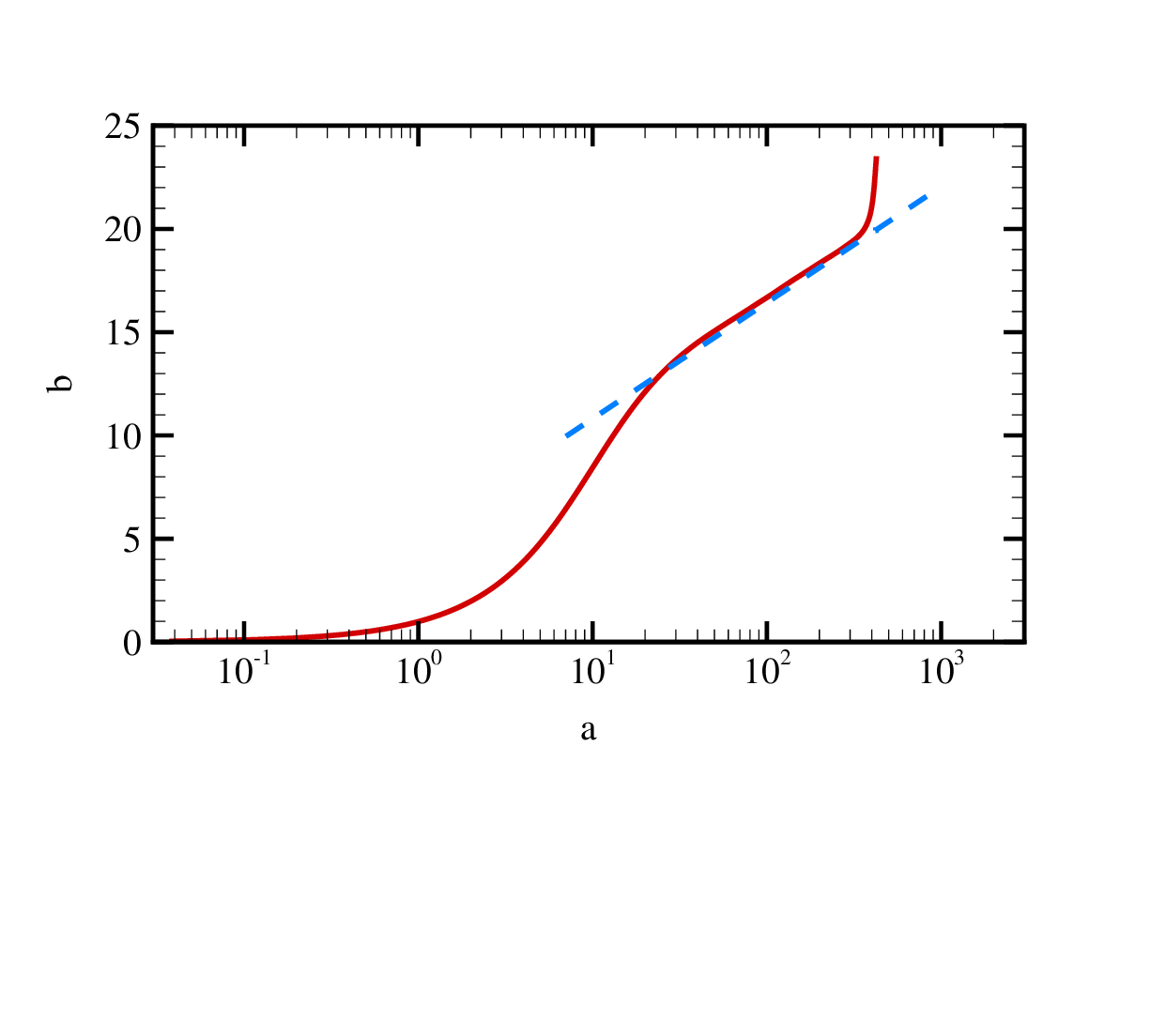}}
\caption{Mean streamwise velocity relative to the bottom wall from the DNS of turbulent Couette-Poiseuille flow. The dashed line represents the classical log law of the wall $\overline{u_r^+}=(1/0.41)\ln(y^+)+5.2$.}
\label{CP_flow_ux_mean}
\end{figure}

Figure~\ref{CP_flow_ux_ins} shows the instantaneous streamwise velocity relative to the bottom wall, $u_r=u_x+U_c$, from the DNS. Due to the motion of the parallel walls and the imposed mean pressure gradient, the wall-bounded turbulence in the upper half of the channel experiences an APG and behaves similarly to an APG TBL containing many large-scale flow structures. In contrast, the turbulence near the bottom wall is subjected to an FPG and qualitatively similar to an FPG TBL. Figure~\ref{CP_flow_ux_mean} presents the profile of inner-scaled mean streamwise velocity relative to the bottom wall, defined as $\overline{u_r^+} = \overline{(u_x+U_c) / u_{\tau,b}} $. Here, the superscript ``$+$'' denotes inner-scaled quantity by wall unit and $u_{\tau,b}$ is the bottom-wall friction velocity. 

\section{Filtering of the DNS velocity field}\label{appC}

To obtain the filtered velocity field from the DNS of the turbulent Couette–Poiseuille flow, a Gaussian filter is applied to the instantaneous velocity field $\boldsymbol{u} = (u_x, u_y, u_z)$. Since the DNS grid is non-uniform in the wall-normal ($y$) direction, the filtering operation is performed only in the streamwise ($x$) and spanwise ($z$) directions to avoid commutation errors. The filtered velocity field $\widehat{\boldsymbol{u}} = (\widehat{u}_x, \widehat{u}_y, \widehat{u}_z)$ is obtained through a two-dimensional convolution in the $x$ and $z$ directions,
\begin{equation}
\widehat{u}_i(x,y,z) = 
\iint G(x - r_x,\, z - r_z)\, u_i(r_x, y, r_z)\, \mathrm{d}r_x\, \mathrm{d}r_z \hspace{2pt},
\end{equation}
where the Gaussian kernel is defined as
\begin{equation}
G(r_x, r_z) = 
\frac{1}{2\pi\, \sigma_x \sigma_z}
\exp\left[-\frac{1}{2}\left(
\frac{r_x^2}{\sigma_x^2} + 
\frac{r_z^2}{\sigma_z^2} \right)\right]\hspace{2pt}.
\end{equation}
Here, $r_x$ and $r_z$ denote spatial separations in the streamwise and spanwise directions, and $\sigma_x$ and $\sigma_z$ are the corresponding standard deviations of the Gaussian kernel. Since the DNS grid is uniform in both directions, the filtering is implemented as a discrete convolution using symmetric one-dimensional Gaussian kernels applied successively in $x$ and $z$. 

The effective filter width $\Delta_{f,i}$ in each direction $i \in \{x, z\}$ is defined by matching the second moment of the Gaussian filter with that of a top-hat filter, giving
\begin{equation}
\Delta_{f,i} = 2\sqrt{3}\,\sigma_i\hspace{2pt}.
\end{equation}
In this study, the standard deviations are set as multiples of the uniform DNS grid spacings such that $\sigma_x/\Delta_x = \sigma_z/\Delta_z = 1$, 2, and 4, corresponding to moderate to coarse filter widths that remove small-scale motions while retaining large-scale flow structures. The resulting effective filter widths $\Delta_{f,x}$ and $\Delta_{f,z}$ are approximately 3.464, 6.928, and 13.856 times the grid spacings $\Delta x$ and $\Delta z$, respectively.

\begin{figure}
\centering
\includegraphics[width=.8\textwidth,trim={0.0cm 0.0cm 0.0cm 0.0cm},clip]{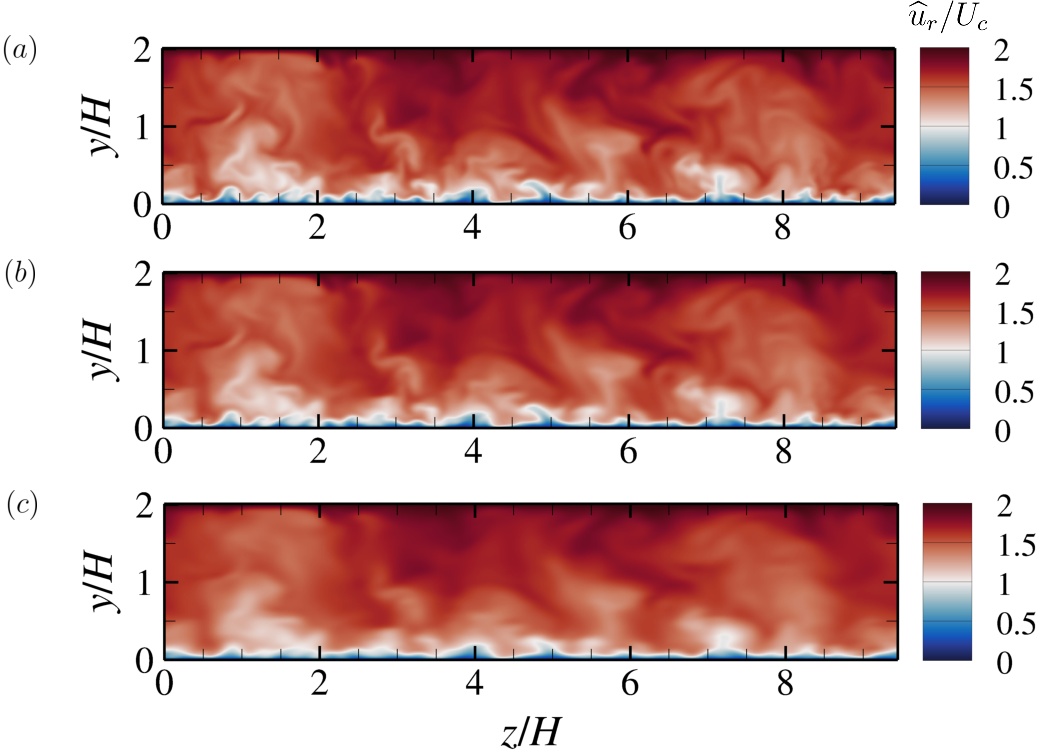}
\caption{Isocontours of the filtered streamwise velocity $\widehat{u}_r/U_{c}$ with $\sigma_x/\Delta{x}=\sigma_z/\Delta{z}=1$ (a), 2 (b), and 4 (c).}
\label{CP_flow_ux_ins_filter}
\end{figure}

Figure~\ref{CP_flow_ux_ins_filter} shows the filtered instantaneous streamwise velocity relative to the bottom wall $\widehat{u}_r/U_{c}$ in an $x$–$y$ plane for these three filter widths, along with the DNS field. As the standard deviations increase, progressively finer structures are removed, demonstrating how the Gaussian filter systematically isolates the larger-scale motions.

\bibliographystyle{jfm}

\end{document}